\documentclass[reprint,superscriptaddress,amsmath,amssymb,aps,prd,longbibliography]{revtex4-2}

\DeclareMathAlphabet{\mathbbold}{U}{bbold}{m}{n}
\usepackage{graphicx}
\usepackage{dcolumn}
\usepackage{mathrsfs,comment}
\usepackage{soul}
\usepackage[normalem]{ulem}

\usepackage{bm}
\usepackage{color}
\usepackage{xcolor}
\definecolor{darkblue}{rgb}{0,0,0.6}
\usepackage{hyperref}
\hypersetup{colorlinks,linkcolor=darkblue,citecolor=darkblue,urlcolor=darkblue}

\newcommand{\dd}{\text{d}}

\begin{document}
\title{Dynamical and structural properties of an absorbing phase transition: a case study from granular systems}
\author{R. Maire}
\affiliation{Universit\'e Paris-Saclay, CNRS, Laboratoire de Physique des Solides, 91405 Orsay, France}
\author{A. Plati}
\affiliation{Universit\'e Paris-Saclay, CNRS, Laboratoire de Physique des Solides, 91405 Orsay, France}
\author{ F. Smallenburg}
\affiliation{Universit\'e Paris-Saclay, CNRS, Laboratoire de Physique des Solides, 91405 Orsay, France}
\author{G. Foffi}
\email{giuseppe.foffi@universite-paris-saclay.fr}
\affiliation{Universit\'e Paris-Saclay, CNRS, Laboratoire de Physique des Solides, 91405 Orsay, France}

\date{\today}

\begin{abstract}

We investigate the dynamical and structural properties of absorbing phase transitions (APTs) within granular systems. Specifically, we examine a model for vibrofluidized systems of spherical grains,  which undergo a transition from a state of purely vertical motion to one characterized by horizontal diffusion as the density increases. Numerical simulations reveal that, depending on the specific system parameters, both continuous and discontinuous transitions can occur, each associated with markedly distinct structural properties at the transition point. We explain this using a theoretical analysis based on kinetic theory applied to an effective 2D model, which elucidates the role of a synchronization effect in determining the nature of the transition. A fluctuating hydrodynamic theory, which quantitatively describes the structural and dynamical properties of the active state such as hyperuniformity is derived from the microscopic dynamics, together with an equilibrium-like assumption concerning the noises on the hydrodynamic fields. This work expands on previous studies by providing a comprehensive examination of the APT characteristics and proposing new theoretical models to interpret the observed behaviors.

\end{abstract}

\maketitle

\section{\label{sec:Intro} Introduction}
In the context of non-equilibrium statistical physics, an \textit{absorbing phase} occurs when a dynamical system is trapped in a stationary or periodic state in which no more evolution takes place.  Upon varying a suitable control parameter, a transition to a diffusive state can occur. Such transitions are termed absorbing phase transitions (APTs).
The characteristics and universality class of these transitions have been extensively discussed in the literature~\cite{hinrichsen2000non, lubeck2004universal}. APTs have been observed in many different contexts, from epidemiology~\cite{Mata2021} to sandpile models~\cite{Dickman1998, hinrichsen1999flowing, hinrichsen2000flowing}, and from quantum systems~\cite{PhysRevResearch.3.013238, PhysRevB.109.L020304, PhysRevLett.130.120402,chertkov2023characterizing,marcuzzi2016absorbing} to chemical reactions~\cite{ziff1986kinetic}.  Since the pioneering work by Pine and coworkers~\cite{pine2005chaos}, particle systems have also become a major area of study for APTs. Specifically, APTs have been observed in low Reynolds number reversible  suspensions~\cite{reichhardt2023reversible, pine2005chaos, corte2008random, nagamanasa2014experimental, jeanneret2014geometrically, franceschini2011transverse},  dense colloidal systems~\cite{keim2014mechanical,tjhung2015hyperuniform,tjhung2016criticality,PhysRevLett.125.148001}, emulsions~\cite{knowltonsm14, Weijs2015}, liquid crystals~\cite{takeuchi2007directed}, glasses~\cite{fiocco2013,regev2013onset, bhaumik2021role}, jammed systems~\cite{PhysRevLett.131.238202,PhysRevLett.127.038002, Ness2020}, granular materials~\cite{Mobius2014,neel2014dynamics}, chemical systems~\cite{vellela2009stochastic, schlogl1972chemical, de2023effects}, ecology~\cite{garcia2024interactions,pascual2005criticality, clar1997phase, chang2020modelling} materials at yielding \cite{PhysRevLett.132.268203}  and turbulent systems~\cite{bertin2024cascade,chantry2017universal, PhysRevLett.132.264002, hof2023directed, lemoult2016directed}.

In several of the cases mentioned above~\cite{jocteur2025random,pine2005chaos, corte2008random, nagamanasa2014experimental, takeuchi2007directed, Ness2020, chantry2017universal}, the transition manifests in a continuous manner. This behavior has been interpreted within the framework of directed percolation ~\cite{Rossi2000} or conserved directed percolation models, notably the Manna models~\cite{LeDoussal2015, Martiniani2019}. These theoretical approaches have proven valuable in understanding the nature of the continuous transitions observed.
However, a contrasting phenomenon has been documented in many other situations~\cite{Kawasaki2016, jeanneret2014geometrically, leishangthem2017yielding, di2016self}. In these instances, researchers have found that the transition exhibits a discontinuous character. To explain this divergence from the continuous behavior, it has been speculated that certain interactions might play a crucial role. Specifically, interactions potentially arising from hydrodynamic effects or material elasticity have been suggested as possible mechanisms responsible for the discontinuous nature of the transition~\cite{mari2022absorbing}. 

Recently, we introduced a simple vibrofluidized granular model in which macroscopic spherical beads, confined within a quasi-2D cell, undergo a transition from a state in which they are locked in a vertical motion to a diffusive and active state in the horizontal direction as the density is increased~\cite{maire2024interplay}. Furthermore, by adjusting parameters like the confinement height or the vibration amplitude, we showed that we can induce either a continuous or discontinuous transition. While continuous to discontinuous absorbing transitions have been observed in various systems~\cite{Fisher97, van1998wilson,pires2023tricritical, assis2009discontinuous, PhysRevE.93.012110, Jagla2014, mari2022absorbing, garcia2024interactions, marcuzzi2016absorbing}, the underlying physical explanations have often remained unclear. In our case, we demonstrated that this phenomenon was related to the synchronization between the vibrating cell and the vertical motion of the beads. The granular system under study was described with two different models: a realistic quasi-2D one which can be numerically integrated through the discrete element method and an effective one in 2D which can be simulated through event-driven molecular dynamics and provides a good starting point for analytical investigations based on kinetic theory and hydrodynamics. 

In this paper, we expand the scope of our previous investigation~\cite{maire2024interplay} in two primary directions. First, we examine the dynamical and structural properties of the absorbing phase transition, aiming to elucidate potential distinctions between continuous and discontinuous cases. Our analysis of the absorbing phase transition's properties focuses on identifying characteristic differences between continuous and discontinuous transitions. This examination of dynamical and structural aspects aims to deepen our understanding of the underlying mechanisms driving these transitions and to draw connections with analogous phenomena in other systems.. Second, we explore theoretical models that can interpret the observed phase transition by extending the kinetic theory framework we proposed in Ref.~\onlinecite{maire2024interplay} and by proposing an approach grounded in fluctuating hydrodynamics to study the active state. These complementary theoretical frameworks offer valuable perspectives on the complex dynamics of the system under study. 

The remainder of this paper is structured as follows. In Sec.~\ref{sec: PRL}, we introduce and simulate the two models used to describe the granular system. In Sec.~\ref{sec: cont vs discont} we compare in greater detail these two models close to the continuous and discontinuous transition by studying their structural and dynamical properties. This sets the stage for the kinetic theory discussed in Sec.~\ref{sec:KinTheo}, which is then extended in Sec.\ref{sec: hydro} using a hydrodynamics theory, allowing us to quantitatively predict the structural properties of the active phase.

\section{Absorbing phase transitions in a granular system}\label{sec:ABSrealistic}
\label{sec: PRL}
In this section, we describe two different numerical approaches used in the present study. First, we consider a realistic quasi-2D model, which is based on a detailed and accurate description of grain interactions and which includes the complete quasi-2D dynamics of the quasi-2D granular system. The advantage of this approach is that it models grain collisions in a more physically accurate manner, incorporating the mechanical properties of the materials involved. However, this approach turns out to be  computationally very demanding.
The second model we present is an effective 2D model that we have developed to simplify the problem \cite{maire2024interplay}. This approach offers two significant benefits. These simulations become significantly less computationally demanding, enabling the exploration of time scales and system sizes increased by three orders of magnitude. Additionally, while our simplified model is approximate, it provides a better understanding of the underlying physical principles behind the observed phenomena and represent a more viable starting for the analytical approach discussed later in the paper.

\subsection{The realistic quasi-2D model}
\label{subsec: numericalSetup}
 
\begin{figure}
\centering
\includegraphics[width=0.99\columnwidth,clip=true]{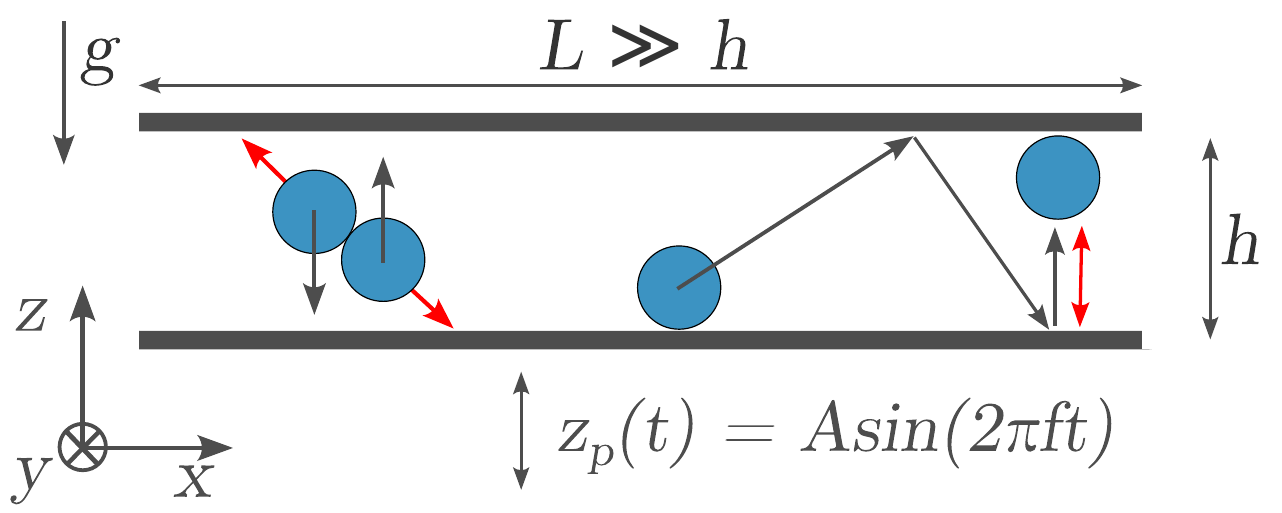}

\caption{Numerical quasi-2D geometry used in realistic simulations. A vertical displacement $z_p(t)$ is imposed to the box in order to provide external energy to the system. Because of tangential frictional forces, the grains lose horizontal energy during collisions with the top and bottom walls. This mechanism introduces an effective dissipation rate $\bar{\gamma}$ for the horizontal dynamics.  Energy transfer between $z$ and $xy$ directions occurs during grain-grain non-planar collisions. This introduces an effective energy gain at collision for the $xy$-motion.} \label{fig:Fig1}
\end{figure}

Our first model consists of a vertically vibrating granular system confined between two plates in a quasi-2D geometry (see Fig.~\ref{fig:Fig1}). The evolution is simulated using molecular dynamics simulations based on discrete element method~\cite{Cundall79,PoeschelBook} using LAMMPS \cite{Plimpton2022, kloss2010granular}. This numerical approach is widely adopted to model realistic in-silico setups \cite{Plati2019,Osullivan2009}, utilizing precise contact mechanics models \cite{Zhang2005,Plati2022,PoeschelBook,PopovBook} for grain-grain and grain-plate interactions. These models encompass normal and tangential forces, capturing both their elastic and dissipative aspects. In particular, in this work, we adopt the Hertz-Mindlin contact model \cite{mindlin1949compliance, LammpsSiteGran}. Particles are also subject to a constant downward gravity field $g$.

In Fig.~\ref{fig:Fig1}, we present a schematic diagram of the numerical setup.
The simulation box has a square base of side $L$ in the $xy$-plane and height $h$, with $L \gg h$. Particles are confined in the $z$-direction by two horizontal parallel plates, and periodic boundary conditions are imposed in the $x$ and $y$ directions. The grain-plate interaction includes tangential friction, which is crucial for the phenomenon under study.

The box vibrates in the $z$-direction following a sinusoidal equation of motion  with frequency $f$ and amplitude $A$: $z_p=A\sin(2\pi f t)$ and is filled with monodisperse spherical grains of spatial coordinates $\{x,y,z\}$, translational velocities $\{v_x,v_y,v_z\}$, angular velocities $\{\Omega_x,\Omega_y,\Omega_z\}$, diameter $\sigma$ and mass $m$. 

Within the LAMMPS simulations, these parameters are implemented in SI units, and we refer to Ref.~\onlinecite{maire2024interplay} and its Supplementary Informations for the numerical values of the physical properties of the beads and additional information concerning the simulations details.

During the simulation, the beads gain vertical ($z$) momentum and energy through the vibration of the plates. This energy is then distributed to the horizontal ($xy$) components of the momentum through collisions between the grains (as represented in the left-hand side of Fig.~\ref{fig:Fig1}). Furthermore, energy is dissipated by tangential friction between the plates and the grains, which tends to slow the particles in between bead-bead collisions and eventually to completely arrest in the $xy$ plane while still oscillating vertically. (see right-hand side of Fig.~\ref{fig:Fig1}).

We consider a system of $N$ particles initialized with random velocities. At low densities, collisions are not frequent enough to convert the continuous supply of vertical energy into horizontal one. Consequently, the system gradually cools due to friction with the plates until it becomes completely arrested in the $xy$ plane, while still bouncing up and down along the $z$ axis. We refer to this state as the absorbing state, as the system cannot escape the $xy$ spatial configurations it has fallen into. At intermediate densities, the system reaches a horizontally active state in the $xy$ plane, where the energy ``injected'' during collisions is sufficient to balance the energy dissipated through tangential collisions with the plates. At very high densities, the system start to crystallize \cite{Reis2006,Castillo2012,Castillo2015,Olafsen98,Komatsu2015} into an hexagonal phase and reaches again an ``absorbing phase'' with 0 diffusivity in the $xy$ plane, consistent with previous studies on absorbing phase transitions in cyclically sheared suspensions~\cite{Ness2020,corte2008random,pine2005chaos, Ness2020}.

\begin{figure}
\includegraphics[width=0.99\columnwidth,clip=true]{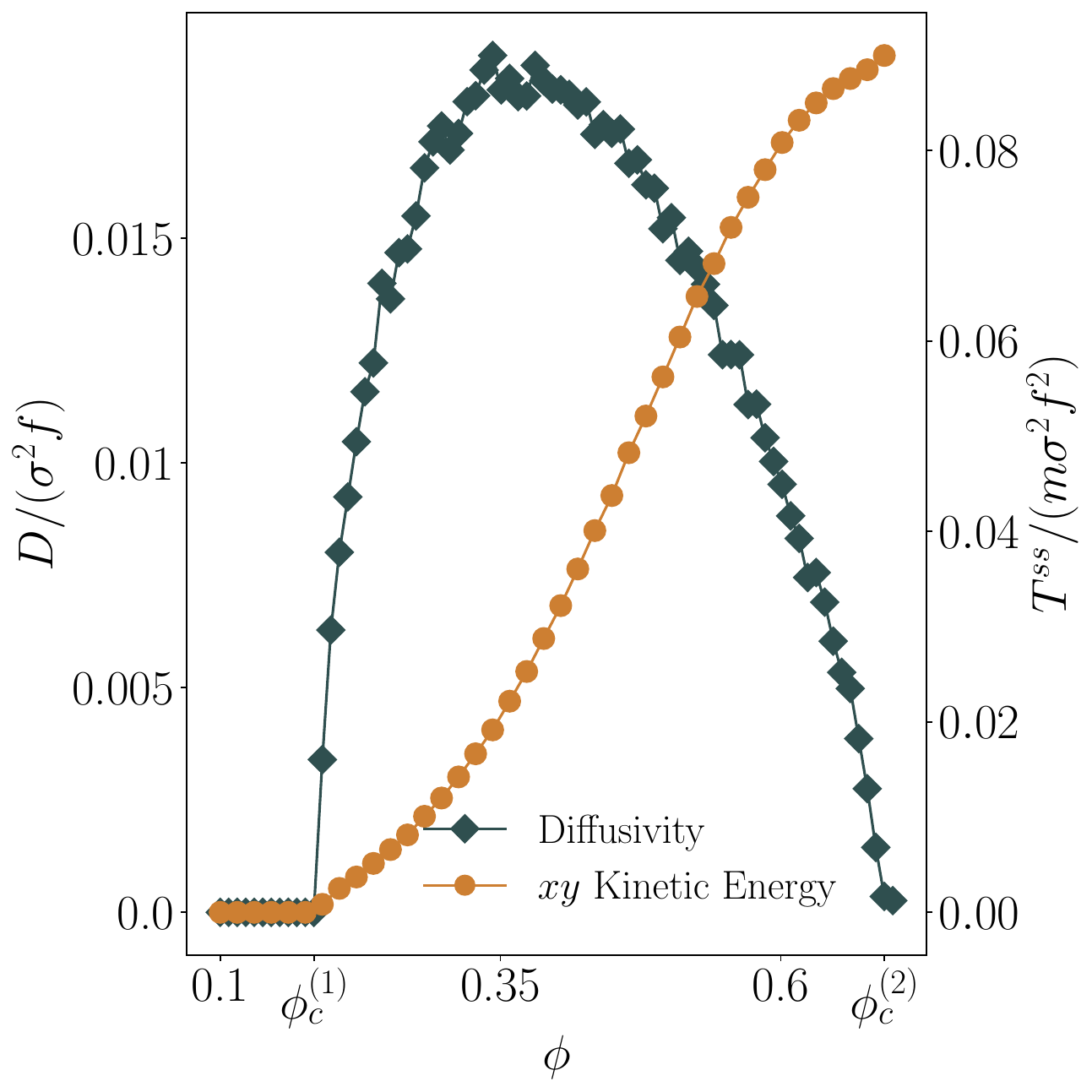}
\centering
\caption{Mean diffusivity (rhombi) and horizontal kinetic energy (circles) as a function of the packing fraction for $h=1.51\sigma$ and $A=0.085\sigma$. Simulations are performed with $N=10^{3}$ grains. } \label{fig:Fig2}
\end{figure}

This qualitative picture is illustrated in Fig.~\ref{fig:Fig2} where the diffusivity in the $xy$ plane is plotted as a function of the area fraction $\phi=\pi N\sigma^2/4L^2$.  The diffusivity is given by $D=\lim_{t\to \infty} \langle (\bm{r}_{xy}(t)-\bm{r}_{xy}(0))^2\rangle/4t $, where $\bm{r}_{xy}$ represents the horizontal coordinates of the position of a given particle, and  $\langle \cdot \rangle$ refers to the average over all the grains. Indeed, at $\phi^{(1)}_c$, we observe the first transition, due to the competition between collision and dissipation, while at $\phi_c^{(2)}$ we observe the second, re-entrant transition driven by cage formation and crystallization. In this paper, we focus entirely on the first transition, whose critical packing fraction will be simply denoted as $\phi_c$.

To accurately characterize this transition, a more suitable order parameter is the mean horizontal kinetic energy of the grains:
\begin{equation}
    T=\frac{1}{\mathcal{T}}\int_{t_0}^{t_0+\mathcal{T}}\frac{m}{2}\langle v_x^2(t) +v_y^2(t) \rangle dt\label{eq:Tgranular}
\end{equation}
where $\langle \cdot \rangle$ refers to the average over all the particles in an instantaneous configuration, $t_0$ is the initial measurement time and $\mathcal{T}$ is the observation time. This plays the role of a macroscopic effective temperature, sometimes called the granular temperature. We define $T_{ss}$ as the steady-state temperature, obtained by choosing $t_0$ to be larger than any relaxation time in the system.

The behavior of $T^{ss}$, is also presented in Fig.~\ref{fig:Fig2}. Its value remains zero up to the absorbing-to-active transition at $\phi_c$.  Following this transition, the order parameter monotonically increases, which is in contrast to the non-monotonic trend observed for diffusivity due to the spatial localization of the particles at higher density~\cite{Ness2020}.  The choice of $T^{ss}$ as the order parameter will prove particularly suitable for the theoretical treatment we propose in Sec.~\ref{sec:KinTheo}.

Interestingly, as discussed in Ref.~\onlinecite{maire2024interplay}, the nature of the transition as the packing fraction varies can be either continuous or discontinuous, depending on the values of physical parameters such as the amplitude of shaking or the height of the plate. This phenomenon was explained by an additional dynamical effect related to the synchronization of the beads with the moving plates. In what follows, we review and discuss this mechanism in detail.

\begin{figure*}[!ht]
\includegraphics[width=0.99\textwidth,clip=true]{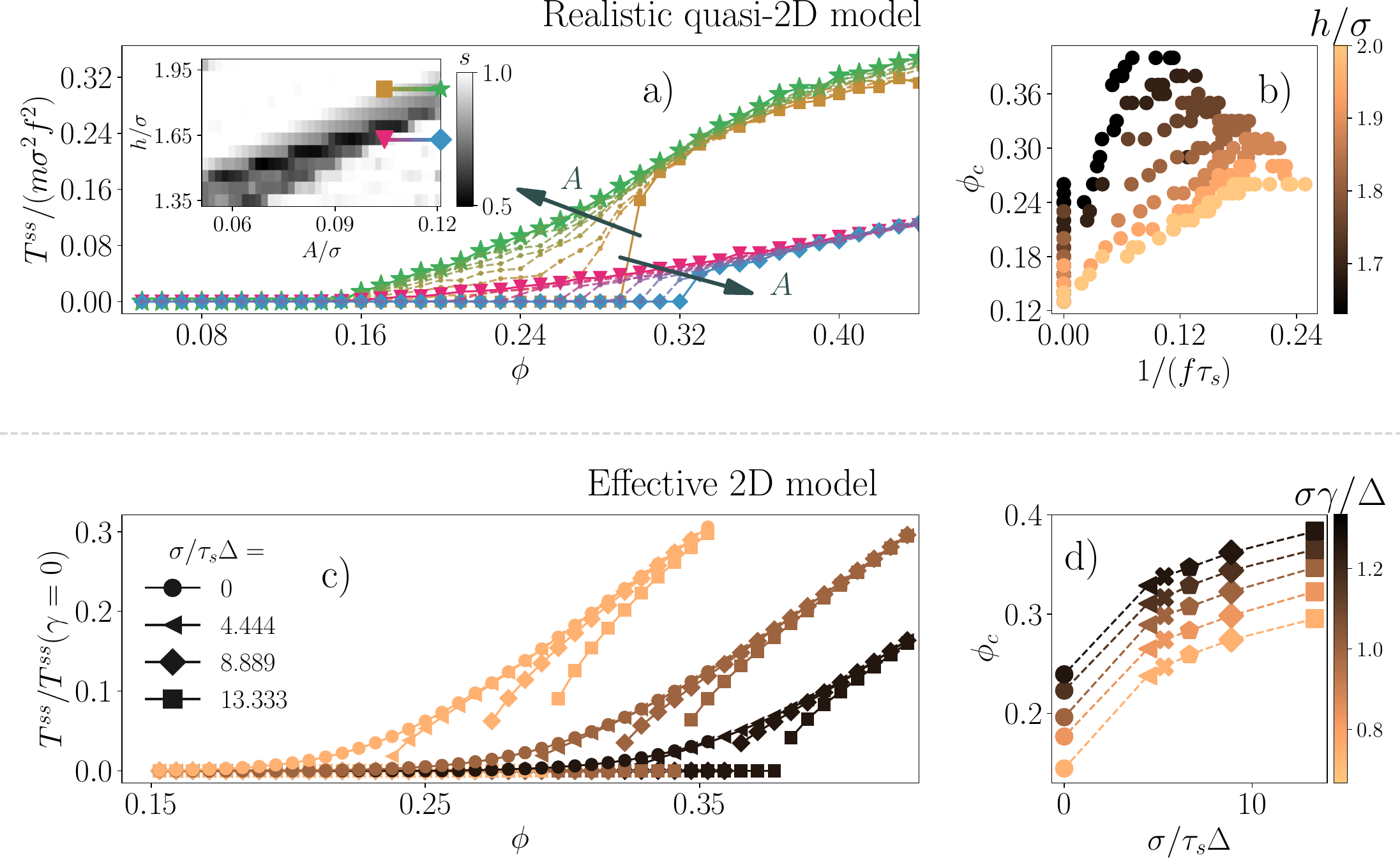}
\centering
\caption{Comparison of the realistic quasi-2D model (a) and b)) with the effective 2D model (c) and d)). a) Evolution of the $xy$ temperature as a function of the density for various amplitude and height in the realistic quasi-2D model. The values of these parameters for each curve are indicated on the inset with corresponding markers. The inset is the synchronization map of the absorbing state with varying $h$ and $A$. When the absorbing state is synchronized, the transition is discontinuous, while the opposite is observed when the absorbing state is chaotic. b) Critical packing fraction as a function of $1/\tau_s$ in the realistic quasi-2D model. The synchronization time controls the packing fraction of the transition. c) Temperature as a function of the packing fraction in the effective 2D model. Without synchronization ($1/\tau_s=0$), the transition is continuous, while at finite $\tau_s$, the transition is discontinuous (see however Sec.~\ref{sec: synchroaa} for a discussion of the existence of a tricritical point at finite $\tau_s$). d) Critical packing fraction as a function of the synchronization time for various damping in the effective 2D model. Consistently with the realistic model, the synchronization time increases the critical packing fraction since it facilitates the transition.}\label{fig:RECAP}
\end{figure*}

From numerical simulations, we observe that an isolated particle, for a given set of parameters, can either begin oscillating in the $z$ direction in phase with the plate (a state we call synchronized) after a characteristic time $\tau_s$, or remain in chaotic motion, not synchronized with the plate. This crossover between periodic and chaotic dynamics for bouncing grains colliding with moving walls has been already observed in different experiments \cite{Pieraski1983,Pieraski1985,Chastaing2015, mujica2016dynamics, schindler2023phase, rivas2011sudden} and understood theoretically by means of simple dynamical models \cite{everson1986chaotic, Metha1990,Luck1993}. In order to quantify the degree of synchronization, we use the following parameter~\cite{acebron2005kuramoto}:
\begin{equation}
    \label{eq: Kuramoto}
    s(t) = \left| \dfrac{1}{N}\sum_{j = 1}^N e^{i\theta^j (t)}\right|,
\end{equation}
where $N$ is the number of particles, and $\theta^j$ is a ``phase'' for particle $j$ defined as:
\begin{equation}
    \theta^j(t) = \frac{z^j(t) - z_+}{z_+ - z_-}\textrm{sign}(v_z^j(t))     
\end{equation}
where $z^j$ and $v_z^j$ are the vertical position and velocity of the particle $j$, $z_+ =h+ A-\sigma/2$ and $z_- = -A + \sigma/2$ respectively the maximum and minimum $z$ position. When all particles jump synchronously, they will coherently contribute to the sum in Eq.~\eqref{eq: Kuramoto} and the synchronization parameter will be equal to one. For an asynchronized system, $s(t)$ will be lowered. We can now introduce the long-time average 
\begin{equation}
   s=\lim_{t_0,t\to \infty}\frac{1}{t}\int_{t_0}^{t_0+t}s(t')dt' 
\end{equation} 
to measure the degree of synchronization as a function of $A$ and $h$. The resulting synchronization map is shown in the inset of Fig.~\ref{fig:RECAP}a. 

This synchronization significantly impacts the energy exchange in the system. If two particles collide at the same height, no energy transfer between the vertical and horizontal degrees of freedom occurs, and the particles will undergo a purely dissipative collision in the $xy$ plane. When most collisions occur between synchronized particles, the energy accumulated in the $z$ direction is not distributed to the $xy$ plane, causing the system to cool down and eventually enter an absorbing state. This has a profound effect on the nature of the transition that is exemplified in Fig.~\ref{fig:RECAP}a. In the region where the particles do not synchronize (green stars) we observe a continuous transition to the absorbing state. However, a slight change in the parameters (in this case the amplitude $A$) yields the curve with the yellow square markers, which reaches an absorbing state at higher densities discontinuously because the particles, for these parameters, can synchronize with the plate (as seen in the inset). Roughly, the collision frequency of the last active steady state corresponds to the average time it takes for particles to synchronize $\tau_s$. If the density is further lowered, particles have enough time to synchronize before colliding with others, leading to purely dissipative collisions in the $xy$ plane and causing the system to cool to the absorbing state. This explains the sudden jump to the absorbing state observed for the curve with square or diamond markers compared to the stars and triangles.

It is important to note that synchronization of the particles requires a typical time $\tau_s$, which plays a key role in determining the critical packing fraction $\phi_c$ where the transition takes place. This timescale can be tuned via the shaking amplitude and the sample height. To illustrate this, we plot in Fig.~3b the behavior of $\phi_c$ as a function of $(f\tau_s)^{-1}$ for a range of different conditions. We find that the critical packing fraction measured for different amplitudes grows with the synchronization frequency $1/\tau_s$ at fixed $h$. As in Ref.~\onlinecite{maire2024interplay}, we measure $\tau_s$ for every $A$ and $h$ using an exponential fit of $s(t)$ starting from an asynchronized and absorbing state.

In summary, the observed phenomenology is attributable to the interplay between two distinct transitions: one between absorbing and active states in the $xy$-motion, and the other between synchronized and asynchronized states in the $z$-dynamics. Without synchronization, the system reaches continuously an absorbing state due to tangential friction with the plate while with synchronization, the transition is discontinuous and driven by purely dissipative collisions. Although it is possible to develop a theory for this system directly \cite{maynar2019understanding, maynar2019homogeneous, brey2019inhomogeneous, brey2020self, brey2020kinetic, mayo2023confined}, we instead recall, in the following subsection, the model introduced in Ref.~\onlinecite{maire2024interplay}, which offers a simplified framework for analyzing the phenomenology observed in the more realistic setting.

\subsection{The effective 2D model}
\label{sec:CGmodel}

In the preceding section, we have accumulated a wealth of information from the realistic quasi-2D model. Here, we will introduce an effective 2D model that incorporates the three primary mechanisms derived from the analysis of the realistic quasi-2D model. These mechanisms are:  an effective dissipation rate for the horizontal motion, an effective parameter for the $z$-to-$xy$ energy transfer at grain-grain collisions and a synchronization time $\tau_s$. This simplified model will allow us to understand the interplay between APT and synchronization, and will rely on a minimal set of physical parameters compared to the realistic quasi-2D model.

The beads are modeled as identical hard disks of mass $m$ and radius $r$ undergoing dissipative collisions characterized by a coefficient of restitution $\alpha$. 
In the realistic quasi-2D model, the energy is injected through vertical vibrations and subsequently converted into $xy$ momentum through off-plane collisions. Brito \textit{et al.} \cite{brito2013hydrodynamic} proposed a purely 2D model effectively replicating the energy transfer mechanism of the realistic quasi-2D model by introducing the following collision rule between particle $i$ and $j$:
\begin{equation}
    \begin{split}
        \bm v_i'&= \bm v_i + \dfrac{1+\alpha}{2}(\bm v_{ij}\cdot \hat{\bm\sigma}_{ij})\hat{\bm\sigma}_{ij} - \Delta_{ij} \hat{\bm\sigma}_{ij} \\
        \bm v_j'&= \bm v_j - \dfrac{1+\alpha}{2}(\bm v_{ij}\cdot \hat{\bm\sigma}_{ij})\hat{\bm\sigma}_{ij} + \Delta_{ij} \hat{\bm\sigma}_{ij},
    \end{split}
    \label{eq: collRule}
\end{equation}
where $0\leq\alpha\leq1$ is the coefficient of restitution, $\bm v_i'$ the post-collision velocity of particle $i$, $\bm v_i$ its pre-collision velocity, and $\hat{\bm\sigma}_{ij}=(\bm r_j-\bm r_i)/|\bm r_j-\bm r_i|$ and $\bm v_{ij}=\bm v_j-\bm v_i$ are the unit vector joining particles $i$ and $j$ and the relative velocity between them, respectively.

The last term in Eqs.~\eqref{eq: collRule} accounts for energy injection, with $\Delta_{ij} > 0$ ensuring that energy input from this term remains positive at each collision, thus reproducing the conversion of vertical motion into horizontal motion in the realistic quasi-2D system. However, the collision can still be dissipative if the relative velocity is large. The  $\Delta_{ij}$  is an effective parameter that takes into account the effects of the height of the plate and the driving on the energy transfer at collision on a coarse level.  We will show that it can be chosen to depend on the degree of synchronization of the particles.
\\
Furthermore, in the realistic quasi-2D model, the beads lose energy upon collisions with the ceiling or bottom of the plate. To incorporate this effect in our effective 2D model, we introduce a viscous drag $\gamma$ during the free flight between collisions:
\begin{equation}
    \dfrac{\mathrm{d} \bm v_i}{\mathrm{d} t} = -\gamma \bm v_i.
\end{equation}
For fixed parameters $\alpha$, $\gamma$, and $\Delta$, the system can exhibit either an absorbing state or an active state, depending on density. 
At low density, the dynamics are primarily governed by viscous drag, causing the system to eventually come to rest in an absorbing state. In contrast, at high density, the dynamics are mainly driven by collisions leading the system to an active steady state. It should be stressed that, up to this point, the dynamics we consider are very close to the ones introduced in Ref.~\onlinecite{lei2019hydrodynamics}.

To model the dynamics observed with the realistic quasi-2D model, we must incorporate synchronization into our effective 2D model as well. In the realistic quasi-2D model, simulations of free (non-colliding) particles with random initial conditions demonstrate a tendency to synchronize with the plate after a randomly distributed time interval $\tau_s$. Moreover, we recall that collisions between the particles disrupt their synchronization, as these collisions effectively randomize the post-collision vertical velocity. Even in-plane collisions between synchronized particles slightly asynchronize them over a timescale $\tau_s'$ which is on average significantly shorter than $\tau_s$. To easily incorporate all these aspects of the realistic quasi-2D model into our model, we make the assumption that a single constant synchronization time, denoted as $\tau_s$, is sufficient to describe qualitatively the synchronization. In the effective 2D model, particles are either labeled as synchronized or asynchronized. Free flight lasting more than $\tau_s$ synchronize the particle, while any collisions desynchronize it.

We recall that this synchronization effect plays a major role in the energy transfer at collision. In the realistic quasi-2D model, since synchronized particles colliding have the same $z$ coordinate, no transfer of vertical kinetic energy to the horizontal degrees of freedom happens. Asynchronized collisions are instead mostly non-planar, causing a significant $z$-to-$xy$ energy transfer. To capture this behavior in our existing effective 2D model, we define a synchronization dependent $\Delta_{ij}$. If both particles involved in a collision have not had a collision for a time period longer than $\tau_s$, the collision does not add energy to the system, i.e. we set $\Delta_{ij} = 0$. For all other collisions, where at least one of the particles involved has had a collision within the last $\tau_s$ time period, we process the collision with a fixed $\Delta_{ij} > 0$, representing the transfer of energy and momentum resulting from asynchronized collisions.

Formally, the value of $\Delta_{ij}$ for a collision between particles $i$ and $j$ takes the form
\begin{equation}
    \Delta_{ij} = \left\{
    \begin{array}{ll}
        \Delta > 0 & \mbox{if  } \delta t_i< \tau_s \mbox{  or  } \delta t_j < \tau_s\\
        0 & \mbox{otherwise,}
    \end{array}
\right.
\label{eq: vdependentDelta}
\end{equation}
with $\delta t_i$ the time since the last collision of particle $i$.
Clearly, in the limit  $\tau_s\rightarrow\infty$, no synchronization takes place.

Event driven molecular dynamics simulations \cite{smallenburg2022efficient} of this 2D model are performed with up to $1.5\times 10^7$ particles in a square box with periodic boundary conditions. We ensure that simulations are run long enough to let the system reach a steady-state temperature $T^{ss}$. 
The system-wide granular temperature $T$ of the 2D model, defined as in Eq.~\eqref{eq:Tgranular}, will play the role of a mean field order parameter and is the equivalent of the $xy$ temperature defined for the realistic quasi-2D model. We define again the steady state temperature $T^{ss}$ as the temperature measured after the system has relaxed. As shown in Fig.~\ref{fig:RECAP}c, we observe that by setting $1/\tau_s = 0$ (i.e. no synchronization, circular plot markers) the system undergoes a continuous transition as the packing fraction of particles is varied. However, once synchronization is introduced by setting $1/\tau_s > 0$ a discontinuous transition is observed, in agreement with the results found in the realistic quasi-2D model. For fixed values of $\Delta$ and $\alpha$, two timescales are at play, as seen in Fig.~\ref{fig:RECAP}d.  On the one hand, for a fixed value of $\tau_s$, the critical packing fraction increases with increasing drag coefficient $\gamma$. Indeed, higher drag implies that a denser system is needed to ensure sufficiently frequent collisions to sustain an active steady state. On the other hand, at a fixed value of $\gamma$, the critical packing fraction increases upon decreasing $\tau_s$, as this makes it easier to synchronize and hence lose energy. Another way to look at this is through the frequency of collision. When synchronization is at play in our system, we expect that at the critical packing fraction, the rate at which a particle synchronizes $1/\tau_s$ is of the same order of the frequency of collision which is itself, at a given temperature, an increasing function of $\phi$.

\begin{figure*}[!ht]
\centering
\includegraphics[width=1.98\columnwidth, clip=true]{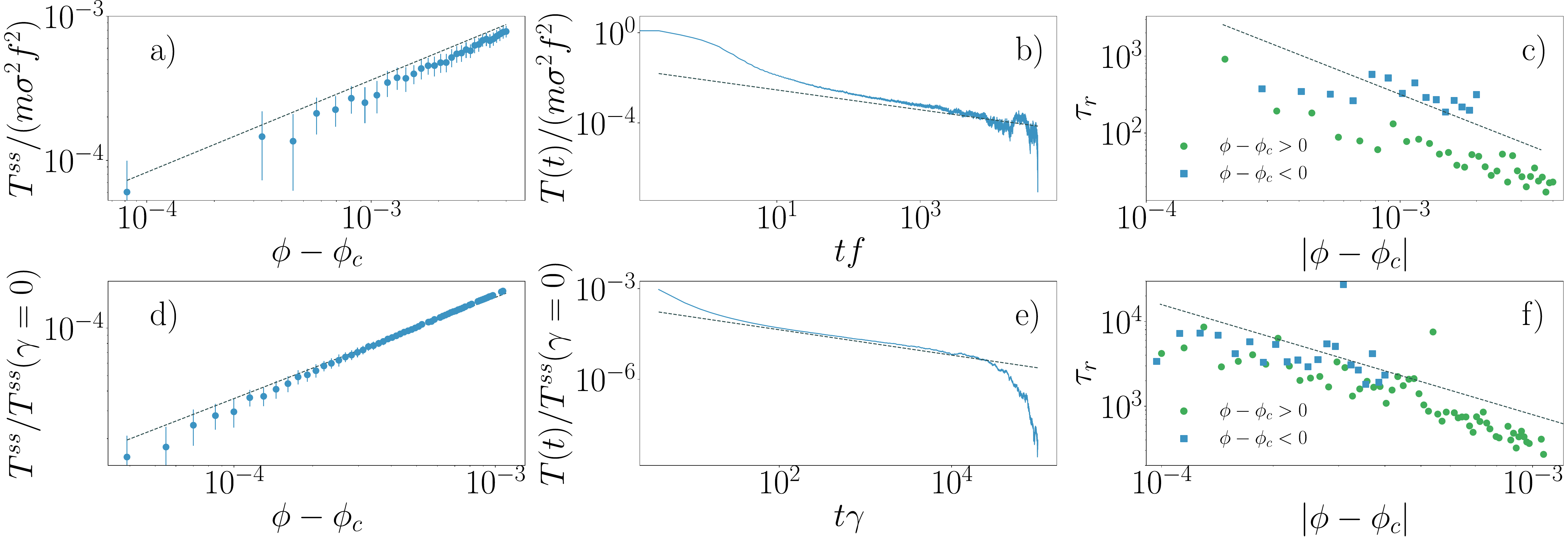}
    \caption{ a, b, c) Power law behavior of observables for the realistic quasi-2D model with parameters $A = 0.085\sigma$, $h = 1.5077\sigma$ and $N = 30000$ leading to $\phi_c \simeq 0.186$. d, e, f) Power law behavior of observables for the effective 2D model with parameters $\Delta/(\sigma \gamma) = 0.15$, $\alpha = 0.95$, $N = 10^6$ leading to $\phi_c\simeq  0.1509$. The dashed lines represent the power law expected from a $2D$ system belonging to the conserved directed percolation universality class \cite{tjhung2016criticality,  odor2004universality}. a, d) Order parameter as a function of $\phi - \phi_c$. Expected exponent $\beta \simeq 0.64$ with $T^{ss}\sim (\phi-\phi_c)^{\beta}$. b, e) Average over 100 runs of the evolution of the order parameter at $\phi_c$. Expected exponent $a \simeq 0.42$ with $T(t)\sim t^{-a}$. c, f) Time to relax to the steady state as a function of $|\phi - \phi_c|$. Expected exponent $\gamma \simeq 1.3$ with $\tau_r\sim |\phi-\phi_c|^{-\gamma}$. }
    \label{fig:manna}
\end{figure*}

\section{Dynamical and structural properties of the phase transition}
\label{sec: cont vs discont}

In this section, we examine the behavior of our system near the transition points. In particular, we highlight the differences between continuous and discontinuous transitions. Whenever possible, we also compare our findings with the behavior observed in similar systems reported in the literature.

\subsection{Nature of the transitions}
\label{sec: nature of the transition}

Thus far, we have discussed the APT in terms of a discontinuous or continuous transition, simply based on whether there is an obvious jump in kinetic energy as a function of packing fraction. In this section, we take a closer look at the behavior around the transition point in order to examine the nature of the transition in more detail, for both the realistic quasi-2D and effective 2D model.

For continuous transitions, in systems characterized by an infinite number of absorbing states and a conserved number of particles without additional symmetries, the associated phase transition is believed to belong to the conserved directed percolation universality class \cite{PhysRevLett.89.190602, Ness2020, henkel2008non, lei2019hydrodynamics}.  Evidence of this universality class, from the analysis of power law scaling, is given in Fig.~\ref{fig:manna} where  simulation results for both the realistic quasi-2D and the effective 2D models are compared to the expected result for systems belonging to the conserved directed percolation universality class.  In particular, panels a), b) and c) present data of the realistic quasi-2D model while d), e) and f) present those of the effective 2D model.
In panels a) and d), we show that the order parameter variation, as a function of the distance from the critical point, obeys a  power-law scaling  characterized by an exponent very close to what is expected for conserved directed percolation universality class. In panels (b) and (e), we give the time evolution of the order parameter at the critical packing fraction $\phi_c$. For the effective 2D model, the expected scaling closely matches the evolution of the order parameter. For the realistic quasi-2D model, however, it is more challenging to discern the correct scaling due to the limited simulation time and the difficulty to accurately pinpoint the transition point. In panels d) and f), we provide data for the characteristic time $\tau_r$ to reach the steady state. Above $\phi_c$, $\tau_r$ is defined as the time at which the system first reaches its steady-state value within an arbitrary precision, which we set to 10 percent of the average steady-state temperature. These curves match well the expected scaling. Below $\phi_c$, $\tau_r$ is defined as the first time the system reaches an arbitrarily small value. We find significant variations in the curves depending on the chosen threshold, since the energy decreases as an exponential after the last collision. Here, we define the threshold as the lowest metastable steady-state temperature observed in a system that eventually decayed to the absorbing state. 
Note that the measured exponent is known to be dependent on the initial conditions \cite{odor2004universality}.  From the overall agreement with the expected scaling laws,  we conclude that the phase transition is consistent with the conserved directed percolation universality class, although a definitive confirmation would require a finite size analysis.

It is worth noting that our focus has been on systems exhibiting their phase transition at relatively high critical packing fraction ($\phi_c > 0.1$).  \cite{lei2019hydrodynamics}\cite{lei2019hydrodynamics} Ref.~\onlinecite{lubeck2004universal}. In contrast, when the transition occurs at very low $\phi_c$, the system may exhibit mean-field conserved directed percolation critical exponents due to effectively long-range interactions, as discussed in Refs.~\onlinecite{lei2019hydrodynamics, lubeck2004universal}.

\begin{figure*}
\centering
\includegraphics[width=2\columnwidth,clip=true]{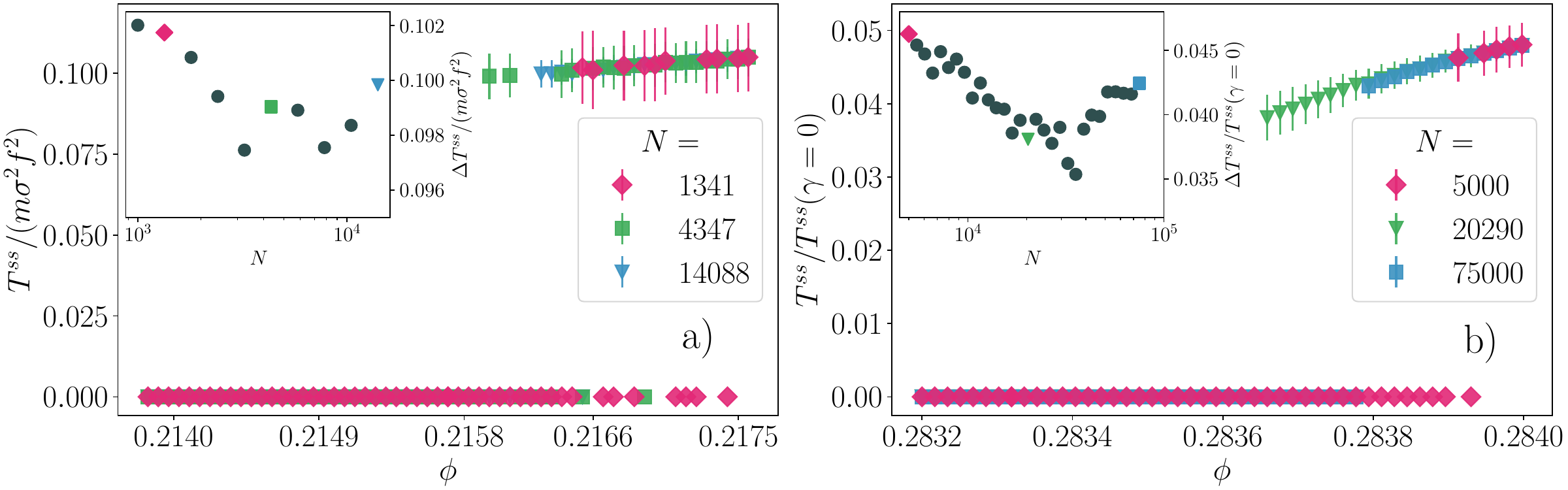}
\caption{ Comparison of the order parameter evolution as a function of packing fraction in the realistic quasi-2D a) and effective 2D model b) for different system sizes. The main figures represent the evolution of the steady state temperature in the discontinuous case (we recall that the temperature is the $xy$-kinetic energy for the realistic quasi-2D model and the usual kinetic energy for the effective 2D model) for different system sizes and insets show the evolution of the order parameter of the last (the most dilute) active steady state ($\Delta T_{ss}$) as a function of the system size a) realistic quasi-2D model: $A = 0.0626\sigma$ and $h = 1.95\sigma$. b) Effective 2D model: $\Delta/\gamma\sigma = 1.5$, $\alpha = 0.95$, and $1/\gamma\tau_s=16.67$.} 
\label{fig: discontinuous}
\end{figure*}

Having clarified the nature of the continuous transition, we now turn to the discontinuous one and perform a finite size analysis to prove that  the finite discontinuity does persist in the infinite system size limit.  
We define $\Delta T_{ss}$ which corresponds to the observed value of the order parameter of the most dilute system in the active state at a given system size and simulation time. 
For a continuous transition, we expect that at finite times, $\Delta T_{ss}$ decreases with $N$ and reaches 0 at infinite system size since the order parameter must vanish continuously. On the contrary, for discontinuous transition, $\Delta T_{ss}$ should not vanish.
Evolution of the order parameter as a function of $\phi$ close to the discontinuous transition as well as $\Delta T_{ss}$ as a function of system size in inset are given, for the realistic quasi-2D and effective 2D models in the insets of Fig \ref{fig: discontinuous} a) and b) respectively. Starting from small system sizes, we see that increasing its size allows us to obtain a transition at a lower density, as would be expected as well for a continuous transition. However, beyond a given system size, $\Delta T_{ss}$ ceases to decrease and instead increases with size, indicating a discontinuous jump in the order parameter in the thermodynamic limit and confirming the first-order nature of the transition.

\begin{figure}[t]
\centering
\includegraphics[width=0.99\columnwidth,clip=true]{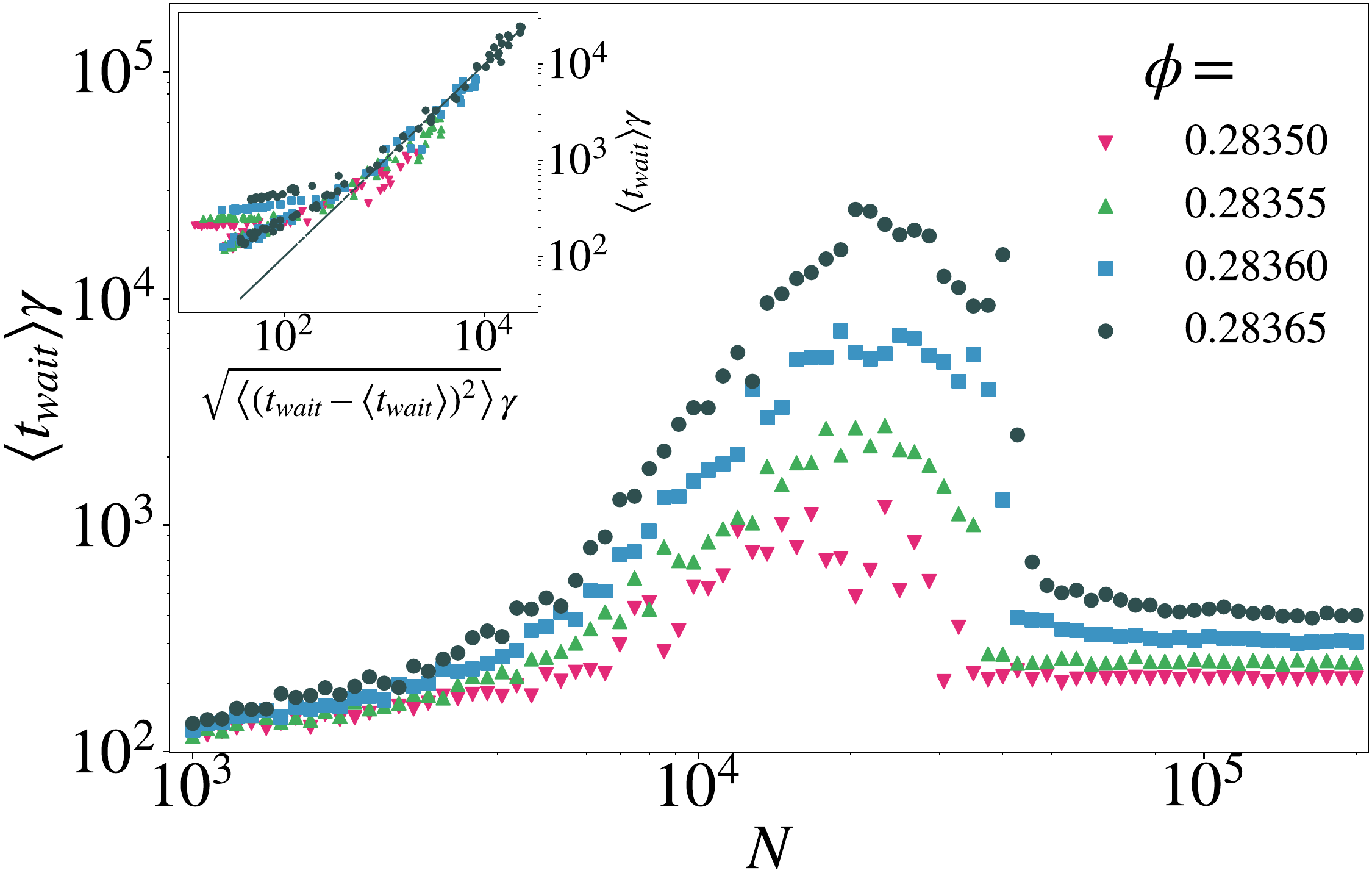}
\caption{Waiting time in the effective 2D model for a system initialized in an active state to reach an absorbing state for different densities in the metastable zone; $\alpha = 0.95$, $1/(\tau_s\gamma) = 16.67$ and $\Delta/(\gamma\sigma) = 1.5$. The inset shows the average waiting time as a function of its standard deviation. The dashed line is a linear scaling, showing that the process is Poissonian. Each point corresponds to an average over 50 independent runs.} \label{fig: waiting Time}
\end{figure}

We interpret such non-monotonic behavior of $\Delta T_{ss}$ as a nucleation dynamics -- characteristic of discontinuous transitions -- and better understood directly from the average time necessary for the system to reach an absorbing state at a given packing  fraction in the metastable zone. We denote this time $t_\mathrm{wait}$ and study its behavior as a function of the number of particles for the effective 2D model in Fig.~\ref{fig: waiting Time}. The time $t_\mathrm{wait}$ acts as a proxy for the nucleation time, since the growth of the nucleus after it has reached its critical size should be fast compared to $t_\mathrm{wait}$.  We only perform the analysis for the effective 2D model due to the prohibitive simulation time of the realistic quasi-2D model. For small system sizes, the waiting time increases with $N$ since the fluctuation required for the system to reach an absorbing state is of the order of the size of the system. Then, the waiting time reaches a maximum, which corresponds to a system size close to the size of the critical nucleus. It then decays due to the increase of available nucleation sites, until reaching a constant at larger system sizes due to the finite cooling rate imposed by $\gamma$ and the frequency of collision. Note that the decrease of the waiting time for system sizes larger than the critical nucleus is very sharp, while in an equilibrium we would expect a decay proportional to $1/N$. This might be the result of hydrodynamical instabilities or non-equilibrium structural properties at small $k$, inaccessible to small system sizes, which are common in granular systems \cite{garzo2005instabilities, mitrano2011instabilities, mcnamara1994inelastic}. Note also that contrary to equilibrium systems, the nucleation dynamics, and especially the growth of the nucleus, exhibit an exotic behavior, both for the effective 2D and the realistic quasi-2D model. Since the order parameter is the activity or the temperature of the system, the nucleus cannot be stationary because, during its growth, active particles around it will quickly convert its inactive grains into active ones. We instead observe damped traveling waves of energy and density, slowly decreasing in amplitude until the absorbing state is reached (see Video 1 for the Realistic quasi-2D model and Video 2 for the effective 2D model). 
Moreover, contrary to Ref.~\onlinecite{neel2014dynamics} we do not find a strongly diverging radius of the critical nucleus as we approach the  stable fluid region but merely a strong increase of the effective energy barrier of nucleation. In the inset of Fig.~\ref{fig: waiting Time} we confirm that the statistics of the waiting time are Poissonian when the reaching of the absorbing state is controlled by nucleation \cite{khain2011hydrodynamics}. 

Interestingly, a very similar system has been studied by Lei, Hu, and Ni \cite{lei2021barrier,lei2023does}.\onlinecite{lei2021barrier, lei2023does} \cite{di2016self, PhysRevResearch.2.013318}. ~\onlinecite{neel2014dynamics} They however found a different phenomenology. We provide additional discussions on this matter in Appendix.~\ref{sec: appendix ran ni}.

\subsection{Structural properties of the transitions}

\begin{figure*}[!t]
\includegraphics[width=0.98\textwidth,clip=true]{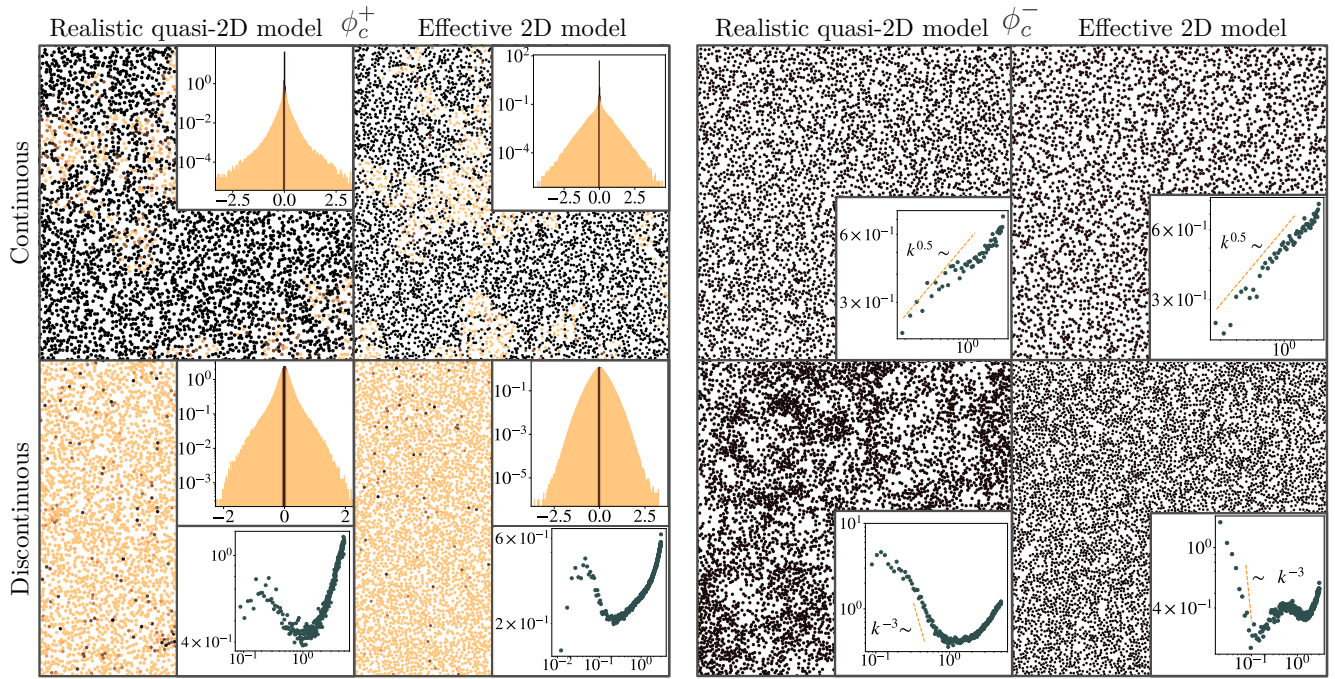}
\centering
\caption{Zoomed snapshots of the systems above ($\phi_c^+$, left) and below ($\phi_c^-$, right) the critical packing fraction for the realistic quasi-2D and effective 2D model. The particles with small velocities are drawn in black. The insets above $\phi_c$ represent the velocity distribution of particles corresponding to the snapshots in units of $\Delta$ for the effective 2D model and $\sigma f$ for the realistic quasi-2D model. The colors of the histograms of velocities match those of the particles in the snapshots (inactive are in black and active are in orange). The other insets represent the structure factor $\bm S_{\rho\rho}(k)$ of the corresponding snapshots as a function of wavenumber in units of $1/\sigma$.} \label{fig:Fig9}
\end{figure*}

Having clarified the nature of the transition, we now focus on dynamical and structural properties of the system close to it. In Fig.~\ref{fig:Fig9},  we present, for both the effective 2D model and the realistic quasi-2D one, snapshots above and below $\phi_c$ for both type of transition, and the probability distribution of the velocities in the diffusive phase, and the long-wavelength behavior of the structure factor in the absorbing phase. 

In the case of a continuous transition, the system reaches the absorbing state due to viscous drag forces. As commonly observed in similar situations, the critical point is characterized by avalanches of active particles that span the whole system \cite{bak1991self, pruessner2013average, manna1990cascades}. This behavior is exemplified in the snapshots and leads to a distinctly identifiable population of inactive particles, discernible through the peak observed in the velocity distribution at the origin.

To examine the structure of our systems, we calculate the density structure factor, given by
\begin{equation}  
S_{\rho\rho}(\bm {k} )={\frac {1}{N}}\left\langle \sum _{j=1}^{N}\sum _{k=1}^{N}\mathrm {e} ^{-i\bm {k} \cdot \bm {r}_{jk}}\right\rangle.
\end{equation}
Close to the transition point, $S_{\rho\rho}$ exhibits a scaling behavior of the form $S_{\rho\rho}(|\bm k|\rightarrow 0) \sim k^{\nu}$, with $\nu\simeq 0.5$ \cite{wiese2024hyperuniformity}. The vanishing of the structure factor at $k=0$ is called hyperuniformity \cite{torquato2018hyperuniform, torquato2016hyperuniformity, salvalaglio2024persistent} and is directly related to a suppression of density fluctuations on large length scales. This trend is consistent with the typical hyperuniform behavior observed at the critical point associated with the conserved directed percolation universality class \cite{PhysRevLett.114.110602,  lei2019hydrodynamics}. It has been observed that this scaling is expected to hold exactly at the transition point and to be lost below or above it \cite{hexner2017enhanced}. In our models, however, system sizes are presumably not large enough for this hyperuniform scaling to break. Very recent works have successfully predicted hyperuniform scaling of the structure factor from a Renormalization Group approach by mapping the conserved directed percolation problem to the interface position at depinning for an elastic manifold \cite{wiese2024hyperuniformity}. A direct analysis of the hyperuniformity from the Renormalization Group of the conserved directed percolation field theory was also performed in Ref.~\cite{ma2023theory} at first loop. 
\\
Finally, we report new  hyperuniform scaling for the longitudinal and transverse velocity fields at criticality in Appendix.~\ref{sec: new hyperuniform scaling}.

The velocity distributions in the active state exhibit peculiar scaling. As already noted, a finite portion of the system is almost completely inactive, hence there is a sharp peak at $v=0$, both for the realistic quasi-2D model and the effective 2D model to the velocity distribution. Moreover, at small velocity, we observe an exponential distribution with a nontrivial power: $\exp(-a|v|^\delta)$ with $\delta\leq1$, which then morphs into a simple exponential scaling: $\exp(-b|v|)$ at $v\simeq \Delta$. This effect is not surprising as non-Gaussian scalings are common in granular systems and, notably, it is known that the homogeneous cooling state exhibits a simple exponential tail while the homogeneously heated granular gas has an exponential tail with exponent $\delta=3/2$ \cite{yu2020velocity, esipov1997granular, van1998velocity}. In particular,  a simple exponential tail  appears frequently \cite{arnarson1998thermal, jarzynski1993universal} and is conjectured to be universal for systems with adiabatic energy change. However, this is not the case in our system, since the energy change during a collision is large compared to the energy of the particles.
In our case, the pronounced non-Gaussian behavior is most likely due to the finite jump processes affecting the velocities, effectively transforming the dynamics into a generalized Ornstein-Uhlenbeck process with discrete jumps of size $\pm \Delta$. This makes the system analogous to a stochastic resetting problem, where an overdamped particle in a harmonic potential is reset away from the origin. Such processes are known to asymptotically yield stationary distributions with exponential or power-law tails \cite{singh2020resetting, pal2015diffusion}. In Appendix~\ref{sec: appendix brilliantov}, we provide an  explanation of this peculiar velocity distribution using an approximation for the Boltzmann equation.

In the discontinuous case, the system primarily transitions into the absorbing phase in a manner similar to the cooling behavior observed in a granular gas. This is because most of the energy is dissipated through collisions between synchronized particles, with the drag playing a secondary role in this process. 
Interestingly, the presence of viscous friction is not a prerequisite for observing a transition. While it helps the system to reach sooner an absorbing state, the dissipative character of synchronized collisions alone is sufficient to deplete the entire energy of the system in the infinite time limit. This implies the formation of clusters similar to those observed in the inhomogeneous cooling state \cite{goldhirsch1993clustering, garzo2005instabilities, brey1998hydrodynamics} for the realistic quasi-2D model, as can be seen in the snapshots and from the upturn of the structure factor at $k\rightarrow 0$. A scaling close to $k^{-3}$, but less steep is observed in the intermediate wavenumber region, which is roughly consistent with Porod's law \cite{bray1994theory, paul2017ballistic}. The inhomogeneous cooling is however arrested at a finite time due to the drag. In the effective 2D model,  as collisions lead to a complete asynchronization of particles, the formation of clusters due to dissipative collisions is less prominent because high density regions of synchronized particles will quickly be shattered by any collision. However, even if the growth of the nucleus cannot proceed as the beginning of the inhomogeneous cooling in the effective 2D model, Porod's law is approximately verified.

The active state of the discontinuous transition does not exhibit a population of inactive particles but still has a marked non-Gaussian behavior, that, contrary to the case of the continuous transition, can be in part attributed to the dissipative collisions. We also provide the structure factor of the fluid close to the metastable region, which is in itself interesting. Lei and Ni \cite{lei2023does} observed a hyperuniform scaling $S\sim k^{1.2}$ in a metastable fluid at very small wavenumbers. As already mentioned before, due to the dissipative collisions between synchronized particles, we instead observe a peak in the structure factor corresponding to the inverse average cluster size.   We show in Appendix~\ref{sec: appendix ran ni} that without dissipative collisions ($\alpha=1$) close to the metastable fluid, we recover a structure factor similar to the one found by Lei and Ni.

\section{Kinetic theory}\label{sec:KinTheo}

In this section we develop the kinetic theory of the homogeneous state for the effective 2D model. We predict quantitatively the transition points and the steady state temperature of our systems. We relate our findings to results already obtained in the literature.

\subsection{\texorpdfstring{Simple $\Delta$ model}{Simple Delta model}}

\begin{figure*}[!t]
\includegraphics[width=0.98\textwidth,clip=true]{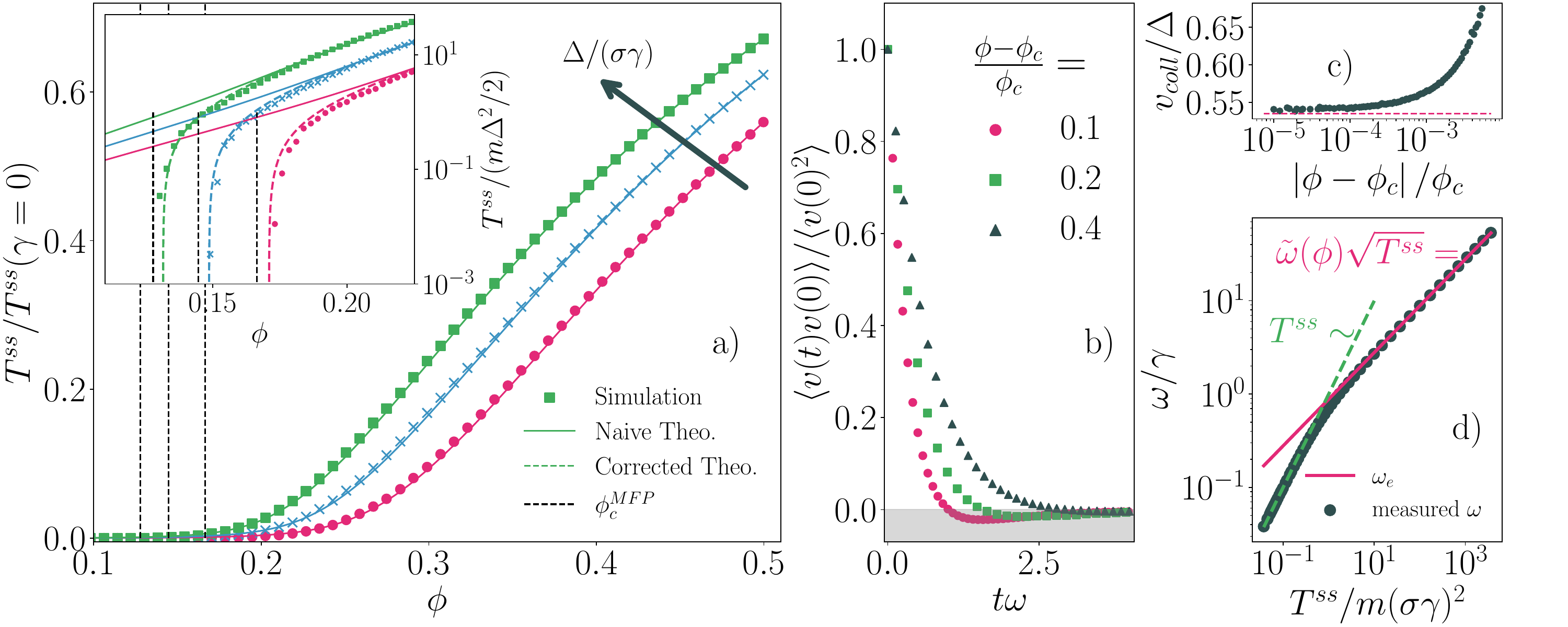}
\centering
\caption{Comparison between the different theories applied to the simple $\Delta$ model without synchronization and simulations. The analysis is accompanied by examples of the breakdown of the assumptions used. a) Comparison of the theory of the simple model with the simulations. ``Naive theo.'' is Eq.~\eqref{eq: solution}. ``Corrected theo.'' is Eqs.~\eqref{eq: correct temperature} and \eqref{eq: corrected theory}. Dark dashed lines are the theoretical critical packing fractions derived from the mean free path argument. The inset is in a semi-log scale close to the transition and in units of $m\Delta^2/2$. We see that when the naive theory predicts a temperature of the order $m\Delta^2/2$, the system reaches an absorbing state in simulation. $N = 5000$ and $\alpha = 0.95$. The increasing values of $\Delta/\gamma\sigma$ are: 1.25, 1.5 and 1.75.  b) Evolution of the velocity autocorrelation function as the critical density of the continuous transition is approached. The system starts to exhibit backscattering and the prediction from the Enskog theory with molecular chaos cannot be trusted. c) Evolution of the \textit{pre}-colliding velocity $v_{coll}$ as a function of the proximity to the absorbing state. $v_{coll}$ tends to saturate, implying that the active particles have a finite energy at the transition. d) Evolution of the measured frequency of collision as the critical point is approached ($T^{ss}\rightarrow 0$) by varying $\Delta$ from $\Delta_c$, which gives a vanishing $T^{ss}$ to a large value where the measured frequency of collision matches the one predicted by the kinetic theory $\omega_e$. At small temperatures, the frequency of collision increases linearly with $T^{ss}$ because it becomes proportional to the number of active particles. }  \label{fig: new}
\end{figure*}

Before introducing the theory of the full effective 2D model, it is useful to discuss the scenario where particles never synchronize which corresponds to the case $\tau_s\rightarrow\infty$ and, consequently,  $\Delta_{ij} = \Delta$ is constant, see Eq~\eqref{eq: vdependentDelta}. If the theory is accurate it should predict, in this situation,  a continuous transition. We first recall the theory developed in Ref.~\onlinecite{maire2024interplay} and introduced in Ref.~\onlinecite{brito2013hydrodynamic}.

The temperature change of the system, assuming homogeneity, is only determined by the energy injection and dissipation following the relation
\begin{equation}
    \dfrac{\partial T}{\partial t} = G(\phi, T),
    \label{eq: Tevolution}
\end{equation}
where $G(\phi, T)$ represents the rate of energy change. The stable steady state temperature $T^{ss}$ is thus given by:

\begin{equation}
    \begin{split} 
        G(\phi, T^{ss}) &= 0,\\
        \left.\dfrac{\partial G(\phi, T)}{\partial T}\right|_{T^{ss}}&<0.        
    \end{split}
    \label{eq: condition}
\end{equation}

$G$ has two contributions: energy change due to collisions and drag. It can be written as 
\begin{equation}
    G(\phi, T) = \dfrac{\omega(\phi, T)}{2}\langle E' - E\rangle_{\text{coll}} - 2\gamma T.
    \label{eq: G}
\end{equation}
Here, the first term represents the change of energy due to collisions, calculated as the product of the frequency of collision $\omega(\phi, T)$ of a single particle, divided by 2 to avoid double-counting, with the average energy change during a collision, while the second one represents the drag.

Using the collision rule of the simple $\Delta$ model (Eq.~\eqref{eq: collRule}) and averaging over collisions, the energy change per collision can be estimated explicitly \cite{brito2013hydrodynamic}:
\begin{equation}
    G(\phi, T) = \frac{\omega_e(\phi, T)}{2}(m\Delta^2+\alpha\Delta\sqrt{\pi m T}-T(1-\alpha^2))-2\gamma T.
    \label{eq: GsimpleDelta}
\end{equation}
where $m$ is the mass of a particle. This result is based on the assumption of uncorrelated velocities  (molecular chaos) and Gaussianity of the velocity distribution.  The detailed calculations required to arrive at Eq.~\eqref{eq: GsimpleDelta} can be found in Ref.~\onlinecite{maire2024interplay}. These assumptions immediately imply that the frequency of collision is given by the so-called Enskog frequency of collision $\omega_e$ \cite{pagonabarraga2001randomly}:

\begin{equation}
    \omega(\phi, T)=\omega_e(\phi, T)=\dfrac{\langle |\boldsymbol v| \rangle}{l(\phi)} = \frac{ 8 \phi \chi(\phi)}{\sigma\sqrt{\pi m}}\sqrt{T}=\tilde \omega(\phi)\sqrt{T},
    \label{eq: freq coll main text}
\end{equation}
where $l(\phi)$ is the mean free path and $\chi$ is the Enskog factor taken to be the radial pair distribution function at contact of an equilibrium system at the same density. It arises from the Enskog's correction to the molecular chaos assumption \cite{pagonabarraga2001randomly}. It can be approximated for $\phi <0.6$ by \cite{mulero2009equation}:
\begin{equation}
    \chi(\phi) = \dfrac{1 - 7\phi/16 - \phi^3/20}{(1-\phi)^2}
\end{equation}
Imposing stationarity to Eq.~\eqref{eq: Tevolution}, we find the analytic steady state temperature:

\begin{equation}
    T^{ss}(\phi) = \left(\dfrac{\epsilon(\phi)  + \sqrt{\epsilon(\phi)^2 +4m\Delta^2(1-\alpha^2)}}{2(1-\alpha^2)}\right) ^2.
    \label{eq: solution}
\end{equation}
with $\epsilon(\phi)= \alpha\Delta \sqrt{\pi m}-4\gamma/\tilde\omega(\phi)$.

In Fig.~\ref{fig: new}a, we compare this theoretical prediction with measurements done with simulations of the effective 2D model at fixed $\Delta$ (i.e. no synchronization) and indeed the theory seems to work very well. This result, however, must be taken with care since the simulations reach an absorbing state at a finite packing fraction and this behavior is not predicted by the theory, which predicts zero temperature only at zero packing fraction, see the inset of Fig.~\ref{fig: new}a. The assumptions of homogeneity, molecular chaos and Gaussian distribution of velocity do not hold close to or below the critical packing fraction due to the significant effect of viscous drag. As can be seen in Fig.~\ref{fig: new}b,  close to the transition, molecular chaos is broken since the velocity autocorrelation function develops a negative minimum at a time roughly equal to the mean collision time that is a clear indication of a backscattering effect. This is incompatible with the assumption of uncorrelated or Markovian collisions underlying the molecular chaos assumption. Indeed, such a theory can only predict exponentially decaying velocity autocorrelation \cite{van2001kinetic, gardiner2009stochastic, sarracino2010irreversible}. This backscattering is induced by the considerable slowdown of the particles before colliding with their neighbors and gaining velocities in a direction opposite to the one they came from. Treating this effect would require going beyond molecular chaos; an approach that is made possible, by some tedious calculation~\cite{van1998ring}. Moreover, we observed in Fig.~\ref{fig:Fig9} that close to the critical packing fraction, the dynamics is dominated by avalanches of active particles within a population of inactive particles. The transition could then occur via two different scenarios. Either the temperature of the active particles reaches 0 at $\phi_c$, or their number vanishes while they still possess a finite amount of energy. The second scenario is the one observed in our system, as illustrated in Fig.~\ref{fig: new}c. Indeed, as the critical point is approached, the \textit{pre}-colliding velocity $v_0$ (the velocity just before a collision) tends to saturate. This indicated that, at the transition, the active particles have finite energy, roughly equal to the minimal excitation possible: $m\Delta^2/2$, but their number vanishes.

This can also be inferred from the measured frequency of collision in the system as the critical point is approached. In Fig.~\ref{fig: new}d we compare the measured frequency of collision $\omega$ with the one used in our theory $\omega_e$ (Eq.~\eqref{eq: freq coll main text}). At high temperature, the measured frequency of collision $\omega$ is well approximated by the equilibrium Enskog $\omega_e$. However, at lower energy, closer to the critical point, when the dynamics is dominated by free flight and avalanches, a large portion of the particles becomes inactive and the scaling of $\omega(\phi, T)$ changes from the equilibrium $\tilde \omega(\phi)\sqrt{T}$ to $\bar\omega(\phi)T$ with $\bar\omega(\phi)$ a prefactor independent of $T$. As discussed above, the collision frequency becomes proportional to the fraction of active particles because these particles carry finite energy and the measured temperature also scales with their fraction, leading to a linear dependence of $\omega$  with $T$. As a result, the ratio $\omega/\omega_e$ provides a rough estimate of the fraction of active particles near the critical point.

This evidence indicates that close to the transition point, the kinetic theory breaks completely and would be hard to fix, assuming no spatial correlations in the distribution of active particles over the system. As a consequence of the significant increase in inactive particles, we adopted a more phenomenological approach that couples the population dynamics of active and inactive particles with the kinetic temperature field.\\
We denote $P_a$ the probability of finding a particle in the active state, characterized by its own finite temperature $T_a$. The probability of being in the absorbing state is $1-P_a$. We propose the following evolution equation for $P_a$: 
\begin{equation}
    \dfrac{\partial P_a}{\partial t}=\dfrac{\omega(\phi, T_a)}{2}P_a(1-P_a) - \dfrac{e^{-\omega(\phi, T_a)\tau_a}}{\tau_a}P_a.
    \label{eq: population dynamics}
\end{equation}
The first term accounts for the conversion of inactive particles to active ones due to collisions. The second term is a rate of inactivation modulated by the probability for an active particle of not colliding in a time $\tau_a$ assuming that collisions are Poissonian \cite{visco2008non}. It accounts for particles reaching the absorbing state due to an absence of collision between them and a neighbor in a time $\tau_a$. This $\tau_a$ is acting as a cutoff with a value that will be derived below and should be inversely proportional to the drag $\gamma$. We recall that $\omega/2$ is the number of collision events per particle happening in the system per unit of time, while $\omega$ is the average frequency at which \textit{one} particle collides with others.

In order to close the problem, we need to specify the temperature and the frequency of collision of the active particles.  We will assume that $T_a$ is simply given by the same Eq.~\eqref{eq: G} used for a homogeneous system far away from the critical region:
\begin{equation}
    \dfrac{\partial T_a}{\partial t} =\dfrac{\omega(\phi, T_a)}{2}\langle E' - E\rangle_{\text{active coll}} - 2\gamma T_a.
    \label{eq: GactiveDelta}
\end{equation}
Since active particles evolve in an avalanche-like fashion, their neighborhood is mostly homogeneous and made of active particles. Therefore, we make the approximation that an average over collisions for active particles in the critical zone is the same as the one performed in the non-critical region. In the same spirit, we chose as well the frequency of collision between active particles to be the equilibrium one: $\omega =\omega_e$ (Eq.~\eqref{eq: freq coll main text}). Eq.~\eqref{eq: GactiveDelta} is thus exactly the same as Eq.~\eqref{eq: G}. We now understand that the temperature prediction from the naive theory was correct but only as a description of a subset of the system:  the active particles; for which the equilibrium frequency of collision is a good approximation. We immediately obtain the temperature of the whole system:
\begin{equation}
    \begin{split}
        T^{ss}_a &= \left(\dfrac{\epsilon  + \sqrt{\epsilon^2 +4m\Delta^2(1-\alpha^2)}}{2(1-\alpha^2)}\right) ^2\\
        T^{ss} &= P_a^{ss} T_a^{ss}.
    \end{split}
    \label{eq: correct temperature}
\end{equation}
Far away from the critical point, $P_a^{ss}\to1$ and Eq.~\eqref{eq: solution} is recovered. Moreover, as argued above, at the critical point $T_a^{ss}$ is finite and the criticality is provided solely by the fraction of active particles going to 0 and not by the activity of the active particles.

With these assumptions, since $T_a$ is decoupled from $P_a$ and varies slowly close to the transition, we can assume it as a constant around the critical point, hence Eq.~\eqref{eq: population dynamics} reads:
\begin{equation}
    \tau_a\dfrac{\partial P_a(t)}{\partial t} = \left(\frac{\tau_a \omega_e}{2}-e^{-\tau_a\omega_e}\right)P_a-\frac{\tau_a\omega_e}{2} P_a^2,
    \label{eq: CDP}
\end{equation}
this has the following stable non-negative stationary solution:
\begin{equation}
    P_{a}^{ss}= \left\{
    \begin{array}{ll}
         \dfrac{\tau_a \omega_e(\phi, T^{ss}_a)-2e^{-\tau_a\omega_e(\phi, T^{ss}_a)}}{\tau_a\omega_e(\phi, T^{ss}_a)} & \mbox{ if }  \tau_a\omega_e> W_0(2) \\
        0 & \mbox{ else},
    \end{array}
\right.
\label{eq: corrected theory}
\end{equation}
with $W_0$ the principal branch of the Lambert $W$ function. $P_a^{ss}$ varies continuously, and the transition happens when the mean free time $1/\omega_e$ of the active particle is of the same order ($W_0(2)\simeq0.85)$ as the stopping time $\tau_a$. It is gratifying to observe that Eq.~\eqref{eq: CDP} is the mean field equation describing the conserved directed percolation universality class, which is the correct equation to describe the vicinity of our phase transition \cite{ma2023theory}. At this point, it is interesting to make a parallel with models on lattice. As said before, the criticality in our model comes from the vanishing of $P_a$ while the temperature of the active particles is fixed. This behavior closely resembles lattice models in the conserved directed percolation universality class, where the order parameter is the number of active particles and for which the activity kick allowing a particle to move on the lattice is fixed and finite \cite{manna1990cascades, manna_sandpile_1999}.

To close our equations, we must determine a suitable value for $\tau_a$. This can be achieved,  for example, by finding an independent prediction for $\phi_c$. An argument presented in Ref.~\cite{lei2019hydrodynamics} allows us to find a lower bound for the critical packing fraction of the continuous transition. If we assume spatial homogeneity, at the critical packing fraction, particles must collide on average with zero velocity. Below this density, particles do not have sufficient momentum after a collision to reach their neighbors, which ultimately leads to the system reaching an absorbing state. Hence, after a collision, two particles  will simply have a velocity $\Delta$ which on average is dissipated to zero over a distance $\Delta/\gamma$. This distance must be equal to the mean free path,  which lets us derive the equation of the critical packing fraction:
\begin{equation}
    l(\phi_c) = \Delta/\gamma,
    \label{eq: MFP}
\end{equation}
with $l(\phi)$ the mean free path. Surprisingly, this simple reasoning, whose results are presented as a dashed vertical bar in the inset of Fig.~\ref{fig: new},  predicts a critical packing fraction in good agreement with the simulations. Note that with this argument, it is assumed that the system transitions to the absorbing state uniformly, and that the pre-collisional velocity, $v_{coll}$, vanishes at the transition point. This assumption would suggest $T_a(\phi_c)=0$, which is not consistent with the way the transition occurs in the simulations.

We can inject this critical packing fraction into our theory. We expect the transition to happen when $\tau_a \omega_e(\phi_c, T_a)= W_0(2)$. Using the Gaussian relation: $\omega_e = \sqrt{\pi T_a(\phi)/(2m)}/l(\phi)$ and supposing that $T_a$ is of order $m\Delta^2/2$ at the transition, we finally obtain an expression for $\tau_a$ from the mean free path argument: \begin{equation}
    \tau_a^{MFP}= \dfrac{2W_0(2)}{\sqrt{\pi}}\dfrac{1}{\gamma}.
    \label{eq: mfp}
\end{equation}
As expected, $\tau_a$ is proportional to $1/\gamma$.
Our equations are now devoid of adjustable parameters. In the inset of Fig.~\ref{fig: new}a, we compare this new theory (Eqs.~\eqref{eq: correct temperature}, \eqref{eq: corrected theory} and \eqref{eq: mfp}) with the simulations. The agreement is very satisfactory, especially in view of the assumption made (Gaussianity of the active particles, homogeneity, absence of spatial and temporal correlation, mean field approach, etc.). The temperatures in the inset are rescaled by $m\Delta^2/2$ to emphasize the fact that the system reaches an absorbing state when the energy of the active particles is close to the minimal excitation $m\Delta^2/2$. Indeed, the corrected theory predicts the transition when the naive theory --now representing the theoretical energy of the active particles only-- is close to $m\Delta^2/2$. Three points however warrant attention. (i) The equations are solved in the mean field approximation, hence they cannot be exact close to the transition and this would be reflected, for example, in the discrepancy of critical exponents. (ii) With the new theory, the transition is not predicted to happen exactly at the critical packing fraction obtained from the mean free path argument $\phi_c^{MFP}$ even if it is used to find $\tau_a$. This is because the temperature of the active particles predicted Eq.~\eqref{eq: correct temperature} at $\phi_c^{MFP}$ is obviously not strictly equal to $m\Delta^2/2$. (iii) The coefficient of restitution $\alpha$ is close to 1, hence, the probability distribution is expected to be almost Gaussian in the active region where the drag is not playing a major role \cite{garzo2018enskog, brey2013homogeneous}. The predictions would thus not be as good for systems with stronger dissipation at collision.

\subsection{\texorpdfstring{Synchronization dependent $\Delta$ model}{Synchronization dependent Delta model}}
\label{sec: synchroaa}

\begin{figure*}[!t]
\centering
\includegraphics[width=0.97\textwidth,clip=true]{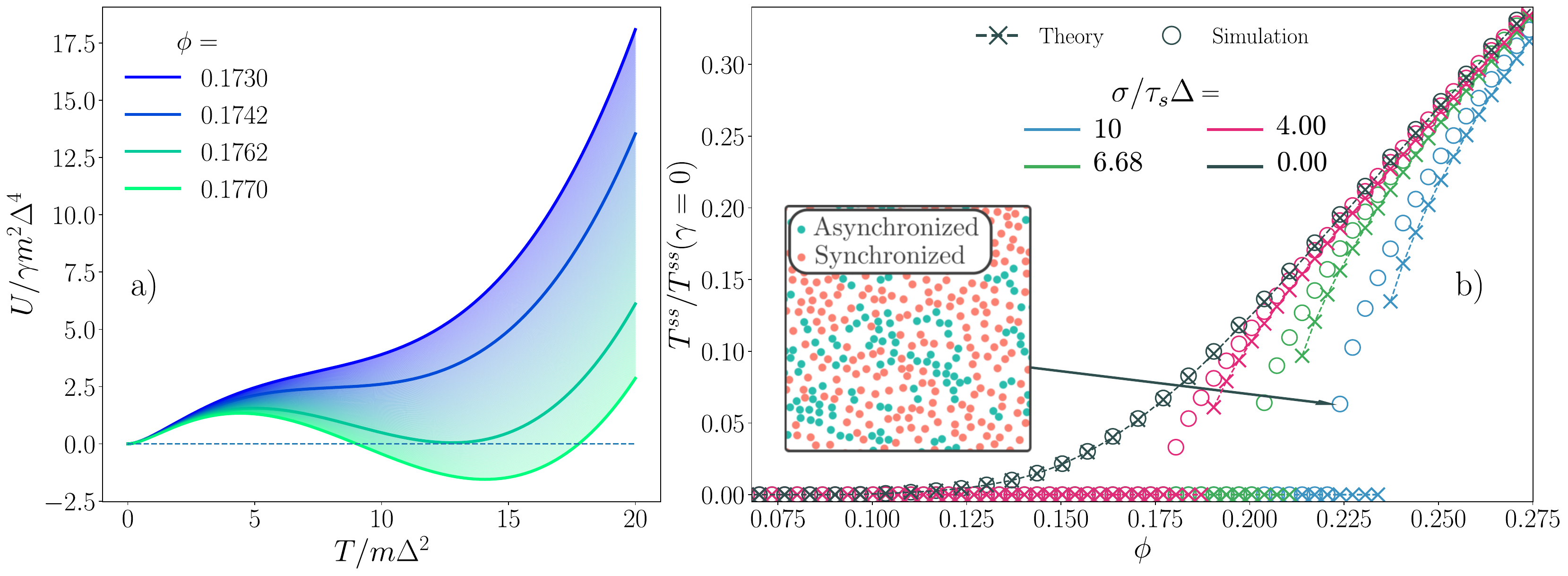}
\caption{a) Theoretical non-equilibrium potential $U$ as a function of $T$ for different $\phi$ and $1/(\gamma\tau_s) = 10$, $\Delta/(\gamma\sigma) = 2.5$ and $\alpha = 0.95$. The potential behave exactly as an equilibrium one for a thermodynamic first order phase transition as the order parameter is changed. b) Effect of synchronization over the nature of transition and the critical packing fraction: $N = 20000$, $\Delta/(\gamma\sigma) = 2.5$ and $\alpha = 0.95$. Inset is a typical snapshot of the system close to the transition point.} \label{fig:theoryDeltaSynchro}
\end{figure*}

Here, we extend the theory outlined in the previous section to incorporate synchronization. At first, we assume that the discontinuous transition happens far away from the zone where the continuous transition happens, thus neglecting the inactivation dynamics described by the population dynamics Eq.~\eqref{eq: population dynamics} and effectively setting $P_a = 1$ which implies $T = T_a$.

The model Eq.~\eqref{eq: vdependentDelta} implies that the value of $\Delta_{ij}$ depends on the time since the last collision for both particles involved $i$ and $j$. If at least one particle has traveled for a time smaller than $\tau_s$, the value of $\Delta_{ij}$ is $\Delta$ otherwise it is set to 0. In this case an exact calculation of the collisional average  in Eq.~\eqref{eq: G}  is challenging, however we can make a reasonable estimation  by assuming that the synchronization dependent $\Delta_{ij}$ can be replaced by an effective one: $\overline \Delta$. Here, we propose:  
\begin{equation}
    \overline{\Delta}(\phi, T_a, \tau_s)\equiv\overline{\Delta}_{ij} = \Delta \left(1 - e^{-2\omega_{e}(\phi, T_a)\tau_s}\right) 
    \label{eq: effective delta}
\end{equation}
where the term in parentheses represents the probability that at least one of the two particles involved in the collision has collided at a time smaller than $\tau_s$ in the past, assuming again uncorrelated particles, Poissonian collisions and an equilibrium frequency of collision. With this effective $\bar{\Delta}$, it becomes evident that the synchronization mechanism results from the competition between the collision frequency $\omega$ and the synchronization rate $1/\tau_s$. At low frequencies of collision compared to the synchronization rate, $\overline{\Delta}$ is small, while the opposite is true for high values of $\omega \tau_s$.

One final assumption is needed,  we simplify the averages over collisions which include a $\Delta_{ij}$ by doing the following approximation when calculating the average of an arbitrary function $h(\bm v_i, \bm v_j)$:
\begin{equation}
    \langle h(\bm v_i, \bm v_j)\Delta_{ij}\rangle_{\text{coll}} \simeq \langle h(\bm v_i, \bm v_j)\rangle_{\text{coll}}\overline{\Delta}.
\end{equation}
This approximation essentially neglects any correlations between particle velocities and their synchronization state.

This simplification allows us to derive an analytical expression for  $G(\phi, T_a, \tau_s)$:
\begin{equation}
        G(\phi, T_a, \tau_s) = \frac{\omega}{2}(m\overline\Delta^2+\alpha\overline\Delta\sqrt{\pi m T_a}-T_a(1-\alpha^2))-2\gamma T_a,
    \label{eq: GsynchroDelta}
\end{equation}
which has the same form as the one found for the simple $\Delta$ model without synchronization except that $\Delta$ is replaced by the effective $\overline\Delta$ which depends on the packing fraction, temperature and synchronization time. This effective $\Delta$ radically changes the energy landscape of the transition compared to the continuous one. To see this, it is useful to define an effective non-equilibrium thermodynamic potential:
\begin{equation}
    U(T_a, \phi, \tau_s) = -\int_0^{T_a} G(T'_a, \phi,\tau_s) dT'_a,
    \label{eq: potential U main}
\end{equation}
such that Eq.~\eqref{eq: Tevolution} can be rewritten as a gradient descent:
\begin{equation}
    \dfrac{\partial T_a}{\partial t}=-\dfrac{\partial U}{\partial T_a},
    \label{eq: gradientDescent}
\end{equation}
and stable steady states are hence given by local minima of the effective non-equilibrium potential.

As a qualitative check, when $1/\tau_s > 0$, the non-equilibrium potential given by the integral of Eq.~\eqref{eq: GsynchroDelta} has the same functional shape as an equilibrium Landau free energy \cite{goldenfeld2018lectures, cardy1996scaling} of a system undergoing a first-order phase transition as seen in Fig.~\ref{fig:theoryDeltaSynchro}a. A quantitative comparison between the simulation and the theory is depicted in Fig.~\ref{fig:theoryDeltaSynchro}b. The theory, corresponding to the local minima of $U$ or equivalently the root of $G$ (found numerically), consistently predicts the observed transition. We note that the mean-field nature of the solution limits our ability to accurately determine the critical packing fraction. Near the transition point, density and spatial synchronization inhomogeneities arise due to the local nature of synchronization and due to the dissipative nature of the collisions between synchronized particles, as is depicted on the snapshot in the inset of Fig.~\ref{fig:theoryDeltaSynchro}b and in Fig.~\ref{fig:Fig9}. We can see clusters of synchronized and asynchronized particles, which can reasonably explain the inaccuracies of the theories (especially close to the transition) together with the effective $\Delta$ approximation.

It is interesting to further explore the similarities between our effective potential and the Landau free energy of an equilibrium system. This resemblance is for example highlighted by the Taylor expansion of the non-equilibrium potential:
\begin{multline}
    U = \gamma T_a^2  - \dfrac{\tilde\omega}{5}\left[2\tilde\omega\tau_s\Delta(\sqrt{\pi m}\alpha + 2 m\tilde\omega\tau_s\Delta) -\vphantom{\frac12}\right.\\ \left. \vphantom{\frac12}(1-\alpha^2)\right]T_a^{5/2}  + \sum_{n = 6}^{\infty}(-1)^{n} K_{n/2}T_a^{n/2},
    \label{eq: taylor}
\end{multline}
where $K_{n/2}$ are positive numbers for $n\geq 6$.  \cite{mari2022absorbing, PhysRevResearch.2.043390, 10.21468/SciPostPhys.8.5.074, PhysRevLett.130.207102,ptaszynski2025nonanalyticlandaufunctionalsshaping}

For infinite system sizes, the stability of the phases and their robustness to small perturbations are controlled by the depths of the minima corresponding to these phases. Specifically, for small damping $\gamma$,  the stability of the absorbing state with respect to small perturbations is dictated  by the sign of $K_{5/2}$:
\begin{equation}
    K_{5/2} = \tilde\omega(2\tilde\omega\tau_s\Delta(\sqrt{\pi m}\alpha + 2 m\tilde\omega\tau_s\Delta) - (1-\alpha^2))/5,
    \label{eq: K52}
\end{equation}
playing a role akin to the $\epsilon$ of Ref.~\onlinecite{neel2014dynamics}. When $K_{5/2} < 0$, the dissipation exceeds the energy injection, leading to an easily unstable absorbing state with respect to small perturbations. Conversely, when $K_{5/2} > 0$, the absorbing state is stable. We conclude that the critical packing fraction $\phi_c$ approximately satisfies $K_{5/2}(\phi_c) = 0$ when the drag can be neglected. Importantly, we see that for this transition, the critical packing fraction is approximately given by a competition between the average energy injected and dissipated at collision. Instead, for the continuous transition, the critical packing fraction was dictated by the competition between energy injection at collision and dissipation during the free flight. Notably, the mean free path argument (Eq.~\eqref{eq: MFP}) predicts a critical packing fraction for the continuous transition independent on $\alpha$. This is consistent with our understanding that synchronization allows the system to reach the absorbing state for a range of packing fractions that would have corresponded to the active phase in the asynchronized case. Increasing $\phi$, and in turn the collision rate $\omega(T,\phi)$, gives the grains less time to synchronize between collisions, which happens over a characteristic time $\tau_s$. For the effective 2D model, we have numerically confirmed that the typical time between collisions $1/\omega$ is always on the same order as (but slightly below) $\tau_s$ in the first active state after the discontinuous transition.

\cite{hinrichsen2000non, dornic2005integration}\cite{van1992stochastic, canet2011general, honkonen2011ito}:
\begin{equation}
    \dfrac{\partial T_a}{\partial t}=-\dfrac{\partial U}{\partial T_a}+\sqrt{2 K T_a}\eta(t),
\end{equation}

\cite{zakine2023minimum, elgart2004rare, meerson2011extinction, assaf2017wkb}

Thus far, in our theoretical model with synchronization, we have neglected the possible presence of inactive particles by setting $P_a = 1$, effectively limiting ourselves to cases where $T_a(\phi_c^+) \gg m\Delta^2/2$. In these cases, any finite synchronization time would lead to a discontinuous transition, however, in cases where our theory with $P_a=1$ predicts a discontinuous transition with a theoretical $\phi_c$ such that $T_{a}(\phi_c^+)< m\Delta^2/2$, the expected observed transition is in fact continuous.
This is because, in these cases, as the density is decreased, the population of inactive particles would begin to grow due to the drag, eventually leading the system to an absorbing state before the dissipative collisions can provoke the theoretically predicted discontinuous phase transition.  To correctly account for this effect, we simply need to restore in the theory the evolution equation for the proportion of active particle instead of assuming $P_a = 1$.

This situation occurs, for example, when $\tau_s$ is very large, as particles cannot synchronize before coming to a complete stop. Indeed, the synchronization timescale is not competing against the diverging timescale linked to the criticality of the continuous transition but against the mean collision time of the active particles, which is finite. Hence, while the population of inactive particles may synchronize, the active particle population does not and the transition follows the same route as without synchronization. We provide more details on this in Appendix~\ref{sec: appendix tricritical}.

Finally, we note that the theoretical treatment of the model with synchronization could have been approached by introducing an additional dynamical variable: the fraction of synchronized particles, instead of relying on an effective $\Delta$. This method, analogous to the population dynamics introduced for the active/inactive populations of particles near the continuous transition, offers slightly better results compared to the effective $\Delta$ approach, albeit at the expense of a less clear physical picture and a more complex theoretical treatment since the description of the system would then require at least 3 fields: the temperature of the active particles, the fraction of active particles and the fraction of synchronized particles. 

\section{Hydrodynamics of the active state}
\label{sec: hydro}
We now derive the hydrodynamic theory of the effective 2D model in its active state. In doing so, we will neglect synchronization effects and any heterogeneity in the population of active particles. This approach allows us to quantitatively predict the structural properties of the active states.

\subsection{Derivation of the hydrodynamics of the model}
Up to this point, the theory we use is neglecting all spatial dependence. Starting from the Boltzmann equation of our system and the singe-particle microscopic velocity and position distribution, we can derive the corresponding hydrodynamical equations through its velocity moments \cite{pitaevskii2012physical, mazenko2006nonequilibrium, kardar2007statistical}. Following Refs.~\onlinecite{garzo2018enskog} and \onlinecite{brey2015hydrodynamics}, the Boltzmann equation dictating the evolution of the one particle distribution function $f(\bm r, \bm v, t)$ in our system is given by:
\begin{equation}
    \dfrac{\partial f(\bm r, \bm v, t)}{\partial t}  + \bm v\cdot \dfrac{\partial f}{\partial \bm r}-\gamma   \dfrac{\partial}{\partial \bm v} \cdot(\bm v f)=J(\bm r, \bm v|f, f),
    \label{eq: boltzmann}
\end{equation}
where $J$ is the collision kernel:
\begin{widetext}
        \begin{multline}  
        J(\bm r_i, \bm {v}_i|f, f) = \chi \sigma_{ij}\int \dd \bm v_j\int \dd\hat{\bm\sigma}_{ij}\left[\Theta(-\bm {v}_{ij}\cdot \hat{ \bm\sigma}_{ij} + 2\Delta_{ij})(-\bm {v}_{ij}\cdot \hat{ \bm\sigma}_{ij} + 2\Delta_{ij})\dfrac{f(\bm r_j, \bm v_j'')f(\bm r_i+\bm \sigma_{ij}, \bm v_i'')}{\alpha^2}- \right.\\\left.\vphantom{\frac12} \Theta(-\bm {v}_{ij}\cdot \hat{ \bm\sigma}_{ij})(-\bm {v}_{ij}\cdot \hat{ \bm\sigma}_{ij})f(\bm r_j, \bm v_j)f(\bm r_i+\bm \sigma_{ij}, \bm v_i)\right].
             \label{eq: collisional part boltzmann}
        \end{multline}
\end{widetext}
Additionally, $\bm v_{i}''$ and $\bm v_j''$ are the precollisional velocities that would lead to postcollisional velocities $\bm v_{i}$ and $\bm v_j$. By inverting Eq.~\eqref{eq: collRule}, we find:
\begin{equation}
    \begin{split}
        \bm v_i''&=\bm v_i + \dfrac{1+\alpha^{-1}}{2}(\bm {v}_{ij}\cdot \hat{ \bm\sigma}_{ij})\hat{ \bm\sigma}_{ij} - \Delta_{ij}\alpha^{-1}\hat{ \bm\sigma}_{ij}, \\
        \bm v_j''&=\bm v_j - \dfrac{1+\alpha^{-1}}{2}(\bm {v}_{ij}\cdot \hat{ \bm\sigma}_{ij})\hat{ \bm\sigma}_{ij} + \Delta_{ij}\alpha^{-1}\hat{ \bm\sigma}_{ij}.
    \end{split}
\end{equation}
Molecular chaos was again assumed in Eq.~\eqref{eq: collisional part boltzmann} to close the equation for the single velocity distribution.
The collisional contribution $J$ takes the usual form of a Master equation  \cite{brilliantov2004kinetic}. While the first term accounts for collisions increasing the number of particles with velocity $\bm v_i$, the second term is a loss term for $\bm v_i$ due to collisions. Both terms are modulated by the relative velocities of the particles along the collision since faster particles collide more often. $\Theta$ is the Heaviside function ensuring that collisions are physical and particles are headed toward each other before colliding.  Finally, the factor $1/\alpha^2$ comes from the Jacobian associated with the change of variable $\bm v''\to \bm v$ and the change of normal velocity at collision \cite{brilliantov2004kinetic}.

The macroscopic hydrodynamic (mass) density $\rho$, velocity $\bm u$ and temperature $T$ fields can be defined through the moment of the single particle distribution:
\begin{equation}
    \begin{split}
        \rho(\bm r, t)&= m\int \dd\bm v f(\bm r, \bm v, t),\\
        \rho(\bm r, t)\bm u(\bm r, t)&=m\int \dd\bm v \bm v f(\bm r, \bm v, t),\\
        \rho(\bm r, t)T(\bm r, t)&=m\int \dd\bm v (\bm v - \bm u(\bm r, t))^2 f(\bm r, \bm v, t).
    \end{split}
    \label{eq: definition field}
\end{equation}
We also define the number density $n$ and packing fraction field $\phi$; useful quantities related to the (mass) density through $\rho(\bm r, t) = mn(\bm r, t) = 4m\phi(\bm r, t)/\pi\sigma^2$. By multiplying the Boltzmann equation (Eq.~\eqref{eq: boltzmann}) with different power of $v$ or $u - v$, the evolution equations for the hydrodynamical fields can be obtained. Assuming at first a constant $\Delta$ (no synchronization), we obtain:
\begin{equation}
    \begin{split}
        \dfrac{\partial \rho}{\partial t} + \bm u \cdot \bm\nabla\rho &= - \rho  \bm\nabla \cdot \bm u,\\
        \dfrac{\partial \bm u}{\partial t}+ \bm u\cdot \bm\nabla\bm u &=-\rho^{-1}\bm\nabla\cdot\bm\Pi -\gamma \bm u,\\
        \dfrac{\partial T}{\partial t} + \bm u\cdot \bm\nabla T&=- n^{-1}\left[\bm\nabla \cdot \bm q +\bm\Pi : \left(\bm\nabla\bm u\right)\right] + \\&~~~~~\tilde G(\bm r, t),
    \end{split}
    \label{eq: navier stokes}
\end{equation}
where some dependencies on $\bm r$ and $t$ have been omitted for ease of reading. Now that the microscopic velocity has been integrated, we will simply denote $\bm\nabla$ as the spatial gradient. ``$:$'' is the dyadic dot product understood as: $(\bm A : \bm B)=\sum_{i, j}A_{ij}B_{ij}$. $\bm \Pi$ and $\bm q$ are respectively the \textit{microscopic} stress tensors and the \textit{microscopic} heat flux. 

Eqs.~\eqref{eq: navier stokes} are currently undetermined since the currents and $G$ are defined from averages with respect to the microscopic probability distribution. To obtain a useful hydrodynamic description, we must assume or derive a closure in order to relate the currents to the macroscopic fields. 
One widely used approach to approximate the velocity distribution involves expanding it in terms of the gradient of the hydrodynamical fields, a method commonly known as the Chapman-Enskog expansion \cite{chapman1990mathematical,pinto2025hydrodynamic,garzo2018enskog, chapman1912xii, enskog1917kinetische}. The important assumption being that the spatial and temporal dependence of the distribution of velocity at a given point can be solely described by a functional dependence on the hydrodynamic fields \cite{dorfman2021contemporary}: $f(\bm r, \bm v, t) \equiv f[\bm v| \rho(\bm r, t), \bm u(\bm r, t), T(\bm r, t)]$. The Boltzmann equation is then solved pertubatively order by order, by using the following expansion:
\begin{equation}
    \begin{split}
        \bm \nabla &\to \varepsilon \bm \nabla\\
        f &= f^{(0)} + \varepsilon f^{(1)} +\dots\\
    \end{split}
    \label{eq: expansion}
\end{equation}
Where $\varepsilon$ is a dummy expansion parameter keeping track of the order of the gradient for the velocity distribution and set to 1 at the end of the procedure. The solutions are expected to be valid only for almost homogeneous system with weak contributions coming from the gradients. 

Performing the Chapman-Enskog expansion at first order leads to the following constitutive relations \cite{garzo2018enskog}:
\begin{equation}
    \begin{split}
        \bm \Pi&=p\bm{1} + \eta\left(\bm\nabla\bm u+(\bm\nabla\bm u)^T-(\bm\nabla\cdot\bm u)\bm{1}\right) + \zeta(\bm\nabla\cdot\bm u)\bm{1}\\
        \bm q &= -\kappa \bm\nabla T - \mu \bm\nabla n\\
        \tilde G &= G + \nu \bm \nabla \cdot \bm u,
    \end{split}
    \label{eq: constitutive relation}
\end{equation}
with $\bm 1$ the unit tensor, $p$ the hydrostatic pressure, $\eta$ and $\zeta$ respectively the shear and bulk viscosity and $\kappa$ and $\mu$ the thermal and diffusive heat conductivity. The latter is 0 at equilibrium and arises in our model from the coupling between density gradient and temperature through the inelastic nature of the collisions \cite{brey1996model}. $G \equiv G(\rho(\bm r, t), T(\bm r, t))$ has formally the same functional shape as Eq.~\eqref{eq: GsimpleDelta}, while $\nu$ is an additional non-equilibrium transport coefficient \cite{brey1998hydrodynamics}. Other transport coefficients technically arise for $\tilde{G}$ at the hydrodynamic order $k^2$ in the hydrodynamic equations. However, these are omitted here as their contributions are negligible.

Given the typical smallness of the correction due to the presence of a $\Delta$ or $\gamma$ \cite{garzo2018enskog, gomez2022enskog} over a wide range of values, we will use the equilibrium hard disk transport coefficients in what follows. We give additional details on the chosen expressions for all the numerical values of these transport coefficients in Appendix~\ref{sec: appendix transport}.

\vspace{1em}

\subsection{Hydrodynamic matrix and stability analysis}

Eqs.~\eqref{eq: constitutive relation} together with the constitutive relations Eqs.~\eqref{eq: navier stokes} are non-linear and difficult to analyze. However, their linearization with respect to the homogeneous solution found in Sec.~\ref{sec:KinTheo} still provides useful information, notably concerning the stability of this reference homogeneous state.

We expand the density, velocity and temperature field around the homogeneous steady state $(\rho, \bm u, T) = (\rho_0, 0, T_0)$:

\begin{equation}
    \begin{split}
        \rho(\bm r, t) &= \rho_0 + \delta \rho(\bm r, t)\\
        \bm u(\bm r, t) &= 0 + \delta \bm u(\bm r, t)\\
        T(\bm r, t) &= T_0+\delta T(\bm r, t).
    \end{split}
    \label{eq: definition around}
\end{equation}

 $\rho_0$ corresponds to the global density imposed in the system and $T_0$ is given by the solution of the homogeneous equation introduced in Sec.~\ref{sec:KinTheo}: $G(\phi_0, T_0\equiv T^{ss}) = 0$ (Eq.~\eqref{eq: solution} without synchronization and Eq.~\eqref{eq: gradientDescent} with synchronization). 

Following Ref.~\onlinecite{hansen2013theory, brito2013hydrodynamic}, we perform a Fourier transform, with $k$ the momentum variable and define the longitudinal $\delta {u}_\parallel(\bm k, t)=\hat {\bm k}\cdot {\delta \bm u}(\bm k, t)$ and transversal velocity field  $\delta {u}_\parallel(\bm k, t)=\hat {\bm {k}}_\perp\cdot {\delta \bm u}(\bm k, t)$ with $\hat {\bm k}$ and $\hat{\bm {k}}_\perp$ unit vector in the direction of $\bm k$ and perpendicular to $\bm k$,  respectively. We define $\bm \Psi(\bm k, t)=(\delta \rho, \delta u_\parallel, \delta u_\perp, \delta T)$ a vector containing the linearized hydrodynamic field. At first order, Eqs.~\eqref{eq: navier stokes} then read:

\begin{equation}
\dfrac{\partial \bm {\Psi}(\bm k, t)}{\partial t}  = \bm M(|\bm k|)\bm\Psi(\bm k, t),  
\label{eq: system of eq}
\end{equation}

with $\bm M(|\bm k|)$ the so-called hydrodynamical matrix:

\begin{widetext}
    \begin{equation}
    \bm M(k) = \begin{pmatrix}
0 & -i\rho_0  k & 0 &0\\
-i k\dfrac{\left .p_\rho\right|_{\rho_0, T_0}}{\rho_0} & -\gamma - \dfrac{\eta_0^\parallel }{\rho_0}  k^2 & 0&-i k\dfrac{\left.p_T\right|_{\rho_0, T_0}}{\rho_0} \\
0&0&-\gamma-\dfrac{\eta_0^\perp}{\rho_0}k^2&0\\
\left.G_\rho\right|_{\rho_0, T_0}-\dfrac{\mu_0}{n_0} k^2 ~~& ~~-i k\dfrac{p_0 + \nu_0}{n_0 } ~~& ~~ 0~~& ~~\left. G_T\right|_{\rho_0, T_0} -\dfrac{\kappa_0}{n_0 }k^2
\end{pmatrix}.
\label{eq: hydro matrix}
\end{equation}
\end{widetext}

The subscript $0$ on the pressure and each transport coefficient indicates their evaluation at the homogeneous state. For conciseness, the derivatives are denoted by a subscript: $\partial F/\partial X \equiv F_X$. $\eta_0^\parallel = \eta_0 + \zeta_0$ is the longitudinal viscosity and $\eta_0^\perp = \eta_0$ is the transverse viscosity. As usual, the transverse velocity field decouples from the other modes at linear order. The results obtained are consistent with related ones found in the granular system literature \cite{mcnamara1993hydrodynamic, van1999randomly, brilliantov2004kinetic, brito2013hydrodynamic} and correctly reduce to equilibrium hydrodynamics in the absence of energy change ($G=0$) and damping \cite{hansen2013theory}.

From Eq.~\eqref{eq: system of eq}, it is clear that to first order, the time evolution of the perturbation around the steady state is characterized by a sum of exponentials with exponents corresponding to the eigenvalues $z(k)$ of $\bm{M}(k)$. Indeed, the evolution of the hydrodynamic fields can be expressed as:
\begin{equation}
\bm{\Psi}(t)=e^{\bm M t}\bm\Psi(0)=\sum_{\lambda}e^{z_\lambda t}\bm{\psi}_\lambda(\bm{\varphi}_\lambda\cdot \bm{\Psi}(0) ),
\label{eq: basis expansion}
\end{equation}
where $\bm{\varphi}_\lambda$ and $\bm{\psi}_\lambda$ are the left and right normalized eigenvectors corresponding to the $\lambda$-th eigenvalue, respectively, and we used the completeness relation $\sum_{\lambda} \bm{\psi}_\lambda \bm{\varphi}_\lambda^T = \bm{1}$.

From Eq.~\eqref{eq: basis expansion}, we understand that the stability of the homogeneous state is related to the negativity of the real part of the eigenvalues of $\bm M$. An analysis of the positivity of the eigenvalues utilizing the Routh–Hurwitz stability criterion \cite{nise2020control} demonstrates, as in previous studies of related systems \cite{brito2013hydrodynamic}, that the active state is stable. Consequently, we conclude that the homogeneous state predicted by the kinetic theory in Sec.~\ref{sec:KinTheo} remains stable against spatial inhomogeneity.

This means that the hydrodynamic theory, which goes beyond the mean field approach, also fails to predict a phase transition since the non-zero homogeneous state is always stable, even at low $\phi$. This justifies \textit{a posteriori}  the introduction of an inactive particle population in the kinetic theory of Sec.~\ref{sec:KinTheo} in order to achieve a continuous transition.

\subsection{Hydrodynamic modes, dynamic and static correlation functions}

The eigenvalues of the hydrodynamic matrix are interesting because they inform us about the process at play, on different length scales, in our system. In Fig.~\ref{fig:hydrodynamic}a, we computed numerically the eigenvalues of $\bm M$ for typical values of $\gamma$ and $\Delta$. We omitted the eigenvalue related to the transverse velocity field since it is trivial. The imaginary part of $z$ corresponds to propagating waves while the real part corresponds to damped or diffusive excitations. We see that only at intermediate values of $k$, in region II, propagating waves are present. Notably, in the inset, we provide a zoom on the region I, where it is clear that at very low wavenumber, there is a regime without propagating waves. We observe as well that only one eigenvalue vanishes at $k=0$ while the others do not. This is directly related to the fact that only the density is conserved in this system. 

While the direct measurement of the evolution of hydrodynamic modes is challenging both experimentally and numerically, spatio-temporal correlations of fluctuations are more accessible \cite{hansen2013theory}. We define the correlation functions we will employ as follows (with a slight abuse of notation):

\begin{equation}
    \begin{split}
    \bm F_{ab}(\bm k, t)&\equiv\lim_{t'\to\infty}\langle\bm\Psi_a(\bm k, t')\bm \Psi^T_b(-\bm k, t' +t)\rangle\\
    &\equiv\lim_{t'\to\infty}\langle \delta a(\bm k, t')\delta b(-\bm k, t' + t)\rangle\\
     \bm S_{ab}(\bm k) &\equiv \bm F_{ab}(\bm k, 0)\\
    \bm S_{ab}(\bm k, w) &\equiv \int_{-\infty}^\infty \dd t e^{-iwt}\bm F_{ab}(\bm k, |t|),
    \end{split}
    \label{eq: correlation function definition}
\end{equation}

where $a$ and $b$ are hydrodynamic fields. If $a$ and $b$ are the density $\rho$ (or $n$ up to a mass factor), the correlation functions defined above will be called respectively the intermediate scattering function, the structure factor and the dynamic structure factor. Note that here the subscripts correspond to the element of the matrix and not to derivative.

In order to perform the averages $\langle \cdot \rangle$ over different steady state realizations, we need to either introduce a probability distribution for the steady state values of the fields or a dynamical noise to the hydrodynamic equations Eq.~\eqref{eq: system of eq}. We chose the latter approach. When the steady state reached by the hydrodynamical fields is an equilibrium one, the probability distribution of the noises for the linearized equations, assuming Gaussianity, is equivalently found through  the Mori-Zwanzig projection method \cite{duran2017general}, Einstein's fluctuation theory  or through the fluctuation dissipation theorem \cite{de2006hydrodynamic, vazquez2001fluctuating, landau1980vol}. Out of equilibrium, the choice of an appropriate probability distribution remains ambiguous and, in principle, must be derived from the microscopic dynamics \cite{manacorda2017lattice,manacorda2018lattice,lasanta2015fluctuating,  bixon1969boltzmann, brey2009fluctuating,maynar2009fluctuating,  bixon1989hard, bouchet2020boltzmann}. However, in our work, we will simply assume that the noise has the same form as in equilibrium, a hypothesis often used with success in granular systems \cite{gradenigo2011fluctuating, pagonabarraga2001randomly}. Therefore, the fluctuating hydrodynamic equations take the form:
\begin{equation}
    \dfrac{\partial \bm {\Psi}(\bm k, t)}{\partial t}  = \bm M(|\bm k|)\bm\Psi(\bm k, t)+ \bm \Xi(\bm k, t),  
    \label{eq: fluctuating hydrodynamic}
\end{equation}
where $\bm \Xi$ is a Gaussian process with the following statistics:
\begin{equation}
    \begin{split}
         \langle \bm\Xi(\bm k,t)\rangle &= 0,\\
        \langle \bm \Xi(\bm k, t)\bm \Xi^T(\bm k', t')\rangle&\equiv (2\pi)^2\bm C(\bm k)\delta(t-t')\delta(\bm k + \bm k'),
    \end{split}
    \label{eq: correlation useless}
\end{equation}
with
\begin{equation}
    \bm C(\bm k)= 2\bm{k}^2T_0\begin{pmatrix}
            0 & 0 & 0 & 0 \\
            0 & \eta_0^\parallel/\rho_0^2 & 0 & 0 \\
            0 & 0 & \eta_0^\perp/\rho_0^2 & 0 \\
            0 & 0 & 0 & \kappa_0 T_0/n_0^2 
        \end{pmatrix}.
\label{eq: noise correlation}
\end{equation} 
The dependence on $\bm{k}^2$ of the correlation $\bm C$ ensures local conservation of momentum and energy by the noise. At equilibrium, this dependence is necessary because hydrodynamic fields must be locally conserved, implying that noises must be a divergence of a random current. In this non-equilibrium context, where neither momentum nor temperature is conserved, this requirement is less immediate. Nevertheless, the source of stochasticity remains the discrete collisions that the particles making up the fluid undergo, which do conserve momentum. The temperature field, is however not conserved by collisions, and a term in $k^0$ is therefore expected. However, such term is usually found to be proportional to the non-gaussianity of the velocity distribution~\cite{brey2009fluctuating}, which we found to be negligible. Finally, we note that in the linear regime, the noises are non-multiplicative since their coefficients are evaluated at the homogeneous field values.

With a way to perform averages, we can now compute the correlation functions. From Eq.~\eqref{eq: basis expansion}, it is clear that the intermediate scattering function is a sum of exponentials, which implies that the dynamic structure factor is a sum of Lorentzian, one for each hydrodynamic mode entering into the correlation function. Since Eq.~\eqref{eq: fluctuating hydrodynamic} is a generalized Ornstein-Uhlenbeck process, we can immediately obtain $\bm S(\bm k, w)$ from a time Fourier transform of Eq.~\eqref{eq: fluctuating hydrodynamic} and the Wiener-Khintchine-Einstein theorem:
\begin{equation}
    \bm S(\bm k, w) = (\bm M(\bm k)-i w \bm 1)^{-1}\bm C(\bm k)(\bm M^T( -\bm k)+i w\bm 1 )^{-1}.
    \label{eq: dynamic corr}
\end{equation}
It can be integrated to obtain the static correlation function $\bm S(\bm k)$. However, it is usually easier to derive it from the equal-time relation \cite{gardiner2009stochastic}:
\begin{equation}
    \bm M(\bm k)\bm S(\bm k) + \bm S(-\bm k)\bm M^T(-\bm k)=-\bm C(\bm k)
    \label{eq: static corr}
\end{equation}
From these equations, it is possible to theoretically derive the static and dynamic correlation functions. However, the expressions are lengthy ratios of sixth-order polynomials and thus not given explicitly. Note however that by adiabatically integrating the temperature field, simpler expressions can be obtained (see Appendix.~\ref{sec: appendix adiabatic slaving delta model}).

It is insightful to provide some asymptotic analysis in the case where $\gamma \neq 0$:

\begin{align}
\frac{\bm{S}_{\rho\rho}(k)}{\rho_0} &\sim
\begin{cases}
\dfrac{G_T T_0 \eta_0^\parallel}{\gamma \rho_0 (G_T p_\rho - G_\rho p_T)} k^2 & \text{as } k \to 0 \\
\dfrac{T_0}{p_\rho} & \text{as } k \to \infty
\end{cases}
\label{eq:pp}\\
\bm{S}_{u_{\parallel}u_{\parallel}}(k) \rho_0 &\sim
\begin{cases}
\dfrac{T_0 \eta_0^\parallel}{\gamma \rho_0} k^2 & \text{as } k \to 0 \\
T_0 & \text{as } k \to \infty
\end{cases}
\label{eq:ss}
\\
\bm{S}_{u_{\perp}u_{\perp}}(k) \rho_0 &\sim
\begin{cases}
\dfrac{T_0 \eta_0^\perp}{\gamma \rho_0} k^2 & \text{as } k \to 0 \\
T_0 & \text{as } k \to \infty
\end{cases}
\label{eq:tt}
\\
\bm{S}_{TT}(k) \rho_0 &\sim
\begin{cases}
g(T_0, \gamma, \dots) k^2 & \text{as } k \to 0 \\
T_0^2 & \text{as } k \to \infty
\end{cases}
\label{eq:TT}
\end{align}

with $g$ a complicated function. We recover the equilibrium results in the limit $k\to \infty$, however, at low $k$, we depart from the equilibrium result, notably by having expressions with transport coefficients appearing. Moreover, as seen in Eq.~\eqref{eq:pp}, we find strong hyperuniformity \cite{lei2024non, kuroda2023microscopic,lei2019hydrodynamics, mukherjee2024anomalous, hazra2024hyperuniformitymasstransportprocesses, PhysRevMaterials.8.104002, anand2025emergent}: $\bm S_{\rho\rho}(k)\sim k^2$. This is directly linked to the fact that the noise locally conserves the momentum while the damping does not, and is reminiscent of long-range correlations induced by bulk conservation law and boundary dissipation in self-organized criticality \cite{grinstein1990conservation, garrido1990long, van1999randomly, plati2021long, simha2002hydrodynamic, kundu2016long, spohn1983long, dorfman1994generic, grinstein1991generic, bak1992self, giusfredi2024localization}. See Ref.~\onlinecite{bonachela2009self} for a recent review. This hyperuniformity in a crystal, allows for the breakdown of the Mermin-Wagner theorem \cite{maire2024enhancing,kuroda2024long,galliano2023two, keta2024longrangeordertwodimensionalsystems}.

For the dynamic correlation functions, asymptotic behaviors are less interesting since most of the relevant physics occurs at intermediate frequencies, where sound propagates. However, we will make use of the asymptotic relation:

\begin{equation}
    \bm S_{\rho\rho}(k, w)=2 T_0\eta_0^\parallel \left(k/w\right)^{4}\text{as } w \to \infty,
    \label{eq:viscosity from dynamic structure factor}
\end{equation}
which can be derived from Eq.~\eqref{eq: dynamic corr}. Another interesting limit is given by $w\to 0$:

\begin{equation}
    \bm S_{\rho\rho}(k, 0)\simeq\dfrac{2 T_0}{\rho_0^2}\dfrac{\kappa_0 T_0 p_T^2 n_0^2k^2+G_T\eta_0^\parallel n_0 \rho_0^2(G_Tn_0-2\kappa_0k^2)}{(G_Tn_0p_\rho-G_\rho n_0p_T-p_\rho\kappa_0k^2)^2}.
    \label{eq: asymptotic w = 0 lei article}
\end{equation}

At equilibrium, when $G=0$, we recover $\bm S_{\rho\rho}(k, 0)\sim k^{-2}$ as $k\to 0$ \cite{chaikin1995principles}. However, in our non-equilibrium setting, this scaling is not verified and at small $k$, $\bm S_{\rho\rho}(k, 0)$ reaches a plateau.  It was postulated in Ref.~\cite{lei2019hydrodynamics} that this plateau could be used to facilitate the experimental detection of random-organizing hyperuniform fluid. As illustrated by the scaling in Eq.~\eqref{eq: asymptotic w = 0 lei article}, this phenomenon is not exclusive to hyperuniform fluids, but is characteristic of systems with a rapidly relaxing temperature field due to its non conservation. Indeed, it would be observed for a system without damping $\gamma$ (and thus non-hyperuniform) but with a non conserved temperature field ($G_T\neq 0$) due to energy change at collision; for example, with $\Delta \neq 0$, $\alpha < 1$, and $\gamma = 0$.

Note that at equilibrium, due to the fluctuation dissipation theorem, the static correlation functions at small $k$ are easily obtained because $\bm F(k, t = 0)$ is known. Therefore, the dynamic correlation functions can be found without having to define explicitly a noise \cite{hansen2013theory}.

\subsection{Numerical results}

\begin{figure*}[!ht]
\includegraphics[width=0.99\textwidth,clip=true]{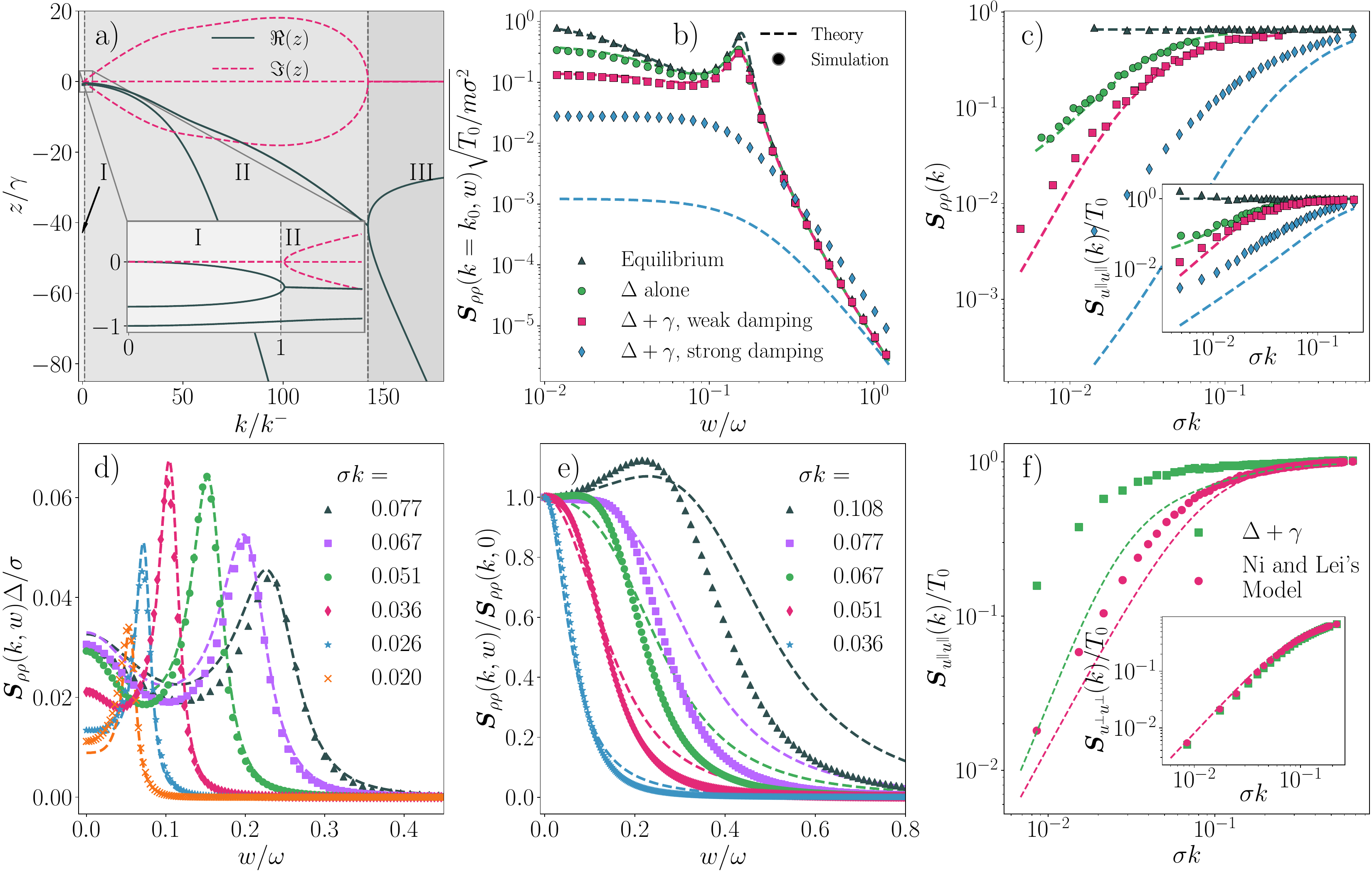}
\centering
\caption{a) Typical eigenvalues $z$ of the hydrodynamic matrix $\bm M(k)$ (Eq.~\eqref{eq: hydro matrix}) as a function of $k$. The trivial transverse mode is omitted. Dashed lines are the imaginary part of $z$ while solid ones are their real part. We see three typical regions, I and III where sound cannot propagate ($z$ are purely real) and II, at intermediate $k$, for which sound is present. Inset is a zoom on the region I, normalized by theoretical wavenumber at which we expect a bifurcation $k^-$. $\phi = 0.3$, $(\Delta/\gamma\sigma) = 1.5$. b) Theoretical (dashed lines) and measured (dots) dynamic structure factor $\bm S(k, w)$ as a function of $w$ for different systems. We provide only $w>0$ since the function is even. For each system the parameters are chosen to keep the kinetic temperature roughly constant, we also take $\phi = 0.1$ and $\sigma k_0=0.046$. ``Equilibrium'' is a system of hard disks. For each non-equilibrium systems, the temperature is fixed to the same value by varying $\Delta$ and $\gamma$ while keeping $\alpha = 0.95$.  ``$\Delta$ alone'' is a system without $\gamma$. ``$\Delta + \gamma$, weak damping'' is $\Delta/(\sigma\gamma)=3.4$. ``$\Delta + \gamma$, strong damping'' is $\Delta/(\sigma\gamma)=2.86$ and the system is starting to enter into the critical region. c) Structure factor for the 4 systems considered in b). Inset is the longitudinal velocity-velocity static correlation function. d) Evolution of the dynamic structure factor as $k$ is varied for the ``$\Delta + \gamma$, weak damping'' system. d)  Evolution of the dynamic structure factor as $k$ is varied for the ``$\Delta + \gamma$, strong damping'' system.  f) Comparison of the theory and the simulations between the model introduced in Ref.~\onlinecite{lei2019hydrodynamics} and our model for the longitudinal velocity-velocity static correlation function and the transverse one in the inset. In both systems, $\alpha = 1$ and $\phi = 0.05$. For our system $\Delta/(\sigma \gamma)=7.67$. For Lei and Ni's model, $\Delta E/m(\sigma\gamma)^2 = 1388$  These choices of parameters lead to approximately the same temperature for both systems. In all cases, $N$ varies between $5\times 10^4$ and $1.5\times 10^6$.}\label{fig:hydrodynamic}
\end{figure*}

We performed simulations of our system and determined the different correlation functions by computing $\bm F(\bm k, t)$. Given the isotropic nature of our system, we averaged over a range of wavevectors $\bm{k}$ with similar magnitudes $k=|\bm k|$ and fitted the resulting function to a sum of complex exponentials, from which we subsequently took the Fourier transform. This procedure is necessary to obtain a smooth, non-oscillating Fourier transform, as $\bm F$, being an autocorrelation function, is inherently noisy. However, we consistently verified that the fitting (used as a smoothing method) did not introduce any artifacts.

In Fig.~\ref{fig:hydrodynamic}b, we provide the dynamic structure factor as a function of $w$ for a given $k$ and temperature. We only present the correlation functions at $w>0$ since they are even. The theoretical prediction from Eq.~\eqref{eq: dynamic corr} is included as a dashed line. Additionally, in Fig.~\ref{fig:hydrodynamic}c, we present, in the main figure, the static structure factor and in the inset the longitudinal velocity-velocity static correlation for the same systems as those studied in panel b. To provide an overview of the peculiarities and similarities between the system of interest and typical fluids, we present these correlation functions for four different systems:
\begin{itemize}
    \item ``Equilibrium'' is a simple Equilibrium hard disk system. Its dynamic structure factor (panel b) exhibits three peaks \cite{landau1934structure} (only two are visible since we only show positive values of $w$): the Rayleigh peak at $k=0$,  related to heat diffusion and to the heat mode, and the Brillouin peaks at $w = \pm w_s(k_0)$, related to propagating sound waves. The structure factor (panel c) and static longitudinal velocity correlation (inset of panel c) are almost flat. The theory works flawlessly in this case \cite{hansen2013theory}. Note that, that at larger $k$, we expect a non-flat structure factor for example, with a shape given by the Ornstein–Zernike theory of liquids \cite{hansen2013theory}. 
    
    \item ``$\Delta$ alone'' is a non-equilibrium system with $\alpha<1$, $\Delta > 0$ but $\gamma = 0$ which was thoroughly studied in Ref.~\onlinecite{brito2013hydrodynamic}. Due to the non-conservation of energy in this system at collision (the particles still undergo a ballistic free flight with constant velocity), the Rayleigh peak in the dynamic structure factor begins to disappear. At lower $k$ or lower $\alpha$, it would not be visible at all, and only the Brillouin peak would be observable since sound can still propagate due to the conservation of momentum and density. This shows that even if the temperature field is technically a fast field because it is not conserved, it is still necessary to take it into account in the hydrodynamic description to describe qualitatively the observation since its relaxation time becomes comparable to the ones of the other fields at intermediate $k$. This regime was called quasi-elastic in Ref.~\onlinecite{brito2013hydrodynamic}. Moreover, while at this wavevector $\bm S_{\rho\rho}(k, w)$ has an equilibrium like shape with a Rayleigh peak, the corresponding static correlation functions are highly out of equilibrium and display a sharp decrease at low $k$. It should be emphasized that this system is not hyperuniform. Specifically, we can prove that without $\gamma$, contrary to the asymptotics found for $\gamma \neq 0$ in Eqs.~\eqref{eq:pp}, the structure factor reaches a finite value. A derivation of this fact is provided in Appendix \ref{sec: appendix adiabatic slaving delta model}.
    
    \item ``$\Delta+\gamma$, weak damping'' is the system  studied in the main part of this article. Due to the additional source of energy dissipation ($\gamma\neq0$) compared to the previous system, the Rayleigh peak has almost completely disappeared. We note that sound peaks are still visible. This indicates that with the chosen $k_0$, we fall into Region II of Fig.~\ref{fig:hydrodynamic}a. The structure factor and longitudinal velocity correlations are again well predicted, except at small $k$. We note that the $k_0$ used in panel b corresponds to a wavevector where the structure factor in panel c is well predicted. As a final remark, we find as expected the hyperuniformity $\bm S_{\rho\rho}( k)\sim k^2$ resulting from the interplay between the momentum conserving noise and the global damping predicted from the theory.
    
    \item ``$\Delta+\gamma$, strong damping'' is the same system as the previous one, except with significantly increased damping, placing us at the boundaries of the critical region. Here, the sound modes have completely disappeared due to strong non-conservation of momentum. Additionally, the theoretical predictions fail. This failure might be in part attributed to the small population of inactive particles not taken into account in the hydrodynamic slow modes and the slight non-Gaussianity of the velocity distribution. However, a more rigorous analysis indicates that the theoretical framework consistently exhibits deviations with the simulations when $\gamma\neq0$, although less pronounced, below a specific $k$ where $\gamma$ emerges as the predominant dissipative mechanism. This is why the density structure factor starts to deviate for the ``$\Delta+\gamma$, weak damping'' model below a given $k$. This is the inverse length at which the effect of $\gamma$ becomes strong compared to the effect of $\alpha$ and $\Delta$. In these cases, the velocity distribution is almost perfectly Gaussian and the discrepancies between simulation and theory cannot be attributed to our Gaussian assumptions. We note again the strong hyperuniformity of the structure factor and of the longitudinal static correlation function. However, the critical hyperuniform scaling $\bm S_{\rho\rho}\sim k^{0.5}$ is not observed at all, meaning that we are still far away from the transition. 
\end{itemize}

In the following, we aim to understand the limitations of our model and its region of validity.  In Fig.~\ref{fig:hydrodynamic}d, we present the evolution of the dynamic structure factor for the ``$\Delta+\gamma$, weak damping'' system as $k$ is varied. At large $k$, the theory works well except at $w=0$, this is likely due to the hydrodynamic approximation requiring a small $k$ since the same departure from the theory is observed for the ``equilibrium'' system (not shown). Additionally, although not easily visible in the structure factor Fig.~\ref{fig:hydrodynamic}c,  the theory slightly deviates for these wavevectors, resulting in disagreement for the dynamic structure factor. At intermediate $k$, the theory agrees perfectly with the simulations. Decreasing again the wavevector makes us fall into the region dominated by $\gamma$, where neither the structure factor nor the dynamic structure factor is well predicted. We also note that for the smallest $k$, we see the Brillouin's peak decreasing as we approach region I. 

In Fig.~\ref{fig:hydrodynamic}e, we give the evolution of the dynamic structure factor for the ``$\Delta+\gamma$, strong damping'' system as $k$ is varied. However, in this case, we know that the theory is completely wrong, partially because of the non-Gaussianity of the velocity distribution function but probably also due to additional effects not taken into account in the theory. We therefore normalize the dynamic structure factor by $\bm S_{\rho\rho}(k, 0)$, this helps us eliminate in part the issue induced by a wrong prediction of the structure factor. At equilibrium, we could completely factorize the dynamic structure factor into a part independent on the noise used thanks to the fluctuation dissipation theorem. Indeed, the quantity $\bm S_{\rho\rho}(k, w)/\bm S_{\rho\rho}(k)$ would be purely computable from $\bm M$ without the need to include a noise term. In this non-equilibrium system, $\bm S_{\rho\rho}(k, 0)$ is also related to $\bm S_{\rho T}(k)$ in a non-trivial way and the same argument cannot be made. This will however provide useful information. We see that, while the theory does not quantitatively predict the results from, the simulations, it captures well the change of region from II to I as $k$ is decreased. This leads us to conclude that the equilibrium noise we chose may be wrong, although the expressions for the transport coefficients are roughly correct.

 In light of Ref.~\onlinecite{lei2019hydrodynamics}, these results are quite surprising. In this study, a similar model was used where a constant energy $\Delta E$ is injected at collision instead of relying on a constant velocity injection $\Delta$ as in our case. Our systems are equivalent at the level of the density and momentum field but differ at the level of the temperature field due to different $G$:
 \begin{equation}
    \begin{split}
        G^{\text{Lei and Ni}}&=\omega(\phi, T)\Delta E - 2\gamma T\\
        G^{\Delta+\gamma}&=\omega(\phi, T)\langle m\Delta^2 + \alpha m(\mathbf v_{ij}\cdot \hat{\bm\sigma}_{ij})\Delta - \\ &~~~ m\dfrac{1-\alpha^2}{4}(\mathbf v_{ij}\cdot \hat{\bm\sigma}_{ij})^2 \rangle_{\text{coll}} - 2\gamma T.
    \end{split}
    \label{eq: comparison G}
 \end{equation}

 As clearly visible in these equations, our energy injection depends on the kinematic of the impact (and thus, on the temperature after the average on the collisions) while this is not the case in the system introduced in Ref.~\onlinecite{lei2019hydrodynamics}.
 
 Surprisingly, with only the density and velocity fields, Lei and Ni obtained very good agreement between their simulations and the theory in the case of strong damping (where the temperature is a fast field). This raises the question: why does our theory—which incorporates the temperature field in addition to the momentum and density fields—not work in our system, despite performing very well in the system studied by Lei and Ni? In Fig.~\ref{fig:hydrodynamic}f, we present a comparison of simulations between their model and ours at the same density, damping and coefficient of restitution (in both cases $\alpha = 1$ and all the dissipation is dealt by $\gamma$). We fixed $\Delta$ in our system and $\Delta E$ in theirs such that the temperature in both systems is the same. We also verified that the velocity distributions in both systems are similar and almost perfectly Gaussian (not shown). In the main panel, we provide measurement of the longitudinal static correlation function for both models alongside the predictions from the theory.  The theory works fairly well for the model by Lei and Ni up to a factor of around 1.5. However, it completely fails for our system. From Eq.~\eqref{eq:ss}, we know that according to the theory, the low $k$ longitudinal static velocity correlation function is proportional to the longitudinal viscosity $\eta_0^\parallel$, this suggests that the equilibrium viscosity coefficient used in the theory is not the right one. Especially in our model. This disagreement between the two models holds also for the structure factor (not shown). Interestingly, the models agree for the transverse velocity correlation function, as seen from the inset of Fig.~\ref{fig:hydrodynamic}f. Since for this observable the theory works well, we are led to assert that the shear viscosity of both systems is very close to the equilibrium one used in the theory. Indeed, the low $k$ transverse static velocity correlation is proportional to $\eta_0^\perp=\eta_0$. From this figure alone, we would thus be encouraged to think that the discrepancies between the two systems and between the theory and our system are solely due to the value of the longitudinal viscosity $\eta_0^\parallel = \eta_0 + \zeta_0$ used in the theory, and thus, the value of the bulk viscosity $\zeta_0$ since the measured $\eta_0$ from the transversal correlation function seems consistent with the equilibrium one used in the theory. This would imply a bulk viscosity an order of magnitude larger than the shear viscosity. 
 
 Nonetheless, from Eq.~\eqref{eq:viscosity from dynamic structure factor}, we also predict that the large $w$ tail of the dynamic structure factor is proportional to $\eta_0^\parallel$. But we see that this tail is very well predicted by the theory for the ``$\Delta+\gamma$, weak damping'' model in Fig.~\ref{fig:hydrodynamic}b, implying that from this observable we are using a correct value for the longitudinal viscosity. Moreover, we remark that according to the theory (Eq.~\eqref{eq: adiabatic slaving structure factor} and \eqref{eq: asymptotic w = 0 lei article}) both $\bm S_{\rho\rho}(k_0, w\to 0)$ and $\bm S_{\rho\rho}(k_0, w\to \infty)$ should be proportional to $\eta_0^{\parallel}$ which implies that, if the viscosity was the issue, we should observe discrepancies of similar order of magnitude at $w=0$ and $w\to\infty$, which is not the case. 
 
In short, following the theory with the equilibrium noises, two different measurements of the longitudinal viscosity lead to two different values. This hints at a potential flaw in the choice of noises used.

 Understanding exactly the root of the issue would require more work and is a very interesting question. It tells us that in systems where the energy is injected at collision, a slight difference in the method of energy injection can completely change the structural properties of the system. It also indicates that, the assumption of an equilibrium like noise is generally unjustified, or that the transport coefficients can be strongly affected by the nature of the collisions. We also note that only data at small densities were presented, but thanks to the Enskog's correction, the theory works very well at higher densities (densities up to $\phi  =0.5$ were tested, see Fig.~\ref{fig: adiabatic}).

In this section, we only analyzed the 2D system. A similar analysis could be performed for the realistic quasi-2D model. Analysis (not shown) indicate that the dynamic structure factor obtained in the realistic quasi-2D model is consistent with the picture given for the effective 2D model, however, there are differences concerning the structure factor. In the realistic quasi-2D model, the structure factor never shows a strong $\bm S_{\rho\rho}(k)\sim k^2$ hyperuniformity. This is explained by the lack of conservation of center of mass of the noise. In an idealized quasi-2D system, the global damping is due to tangential friction happening at each grain-plate collision. Thus, the damping is applied stochastically and discretely. Compared to the idealized global damping of the effective 2D model, this does not conserve the position of the center of mass and as was demonstrated in Ref.~\onlinecite{maire2024enhancing} forbids true hyperuniformity. In an experimental system, the roughness of the plate would also affect the dynamics and would be the most important effect. A particle would act closely to a Brownian particle and a global noise would be effectively applied \cite{puglisi2012structure, sarracino2010granular}, forbidding hyperuniformity \cite{maire2024enhancing}.

Finally, the dynamic structure factor of the system close to the transition, would merit further investigations. However, this is a topic for a separate study. In any case, the results are expected to diverge significantly from the predictions of a simple linear theory. Indeed, in the vicinity of the transition, even a Gaussian theory going beyond mean field is unable to predict the critical induced hyperuniformity \cite{ma2023theory}. Consequently, a renormalization group approach is necessary for a comprehensive understanding of the underlying physics \cite{ma2023theory}.

\section{Conclusion}

In this work, we conducted an in-depth investigation of the system introduced in Ref.~\onlinecite{maire2024interplay}. We demonstrated that an absorbing phase transition can occur in a granular system, with the nature of the transition determined by the presence or absence of synchronization. Building upon our previous findings, we analyzed in greater detail the structural properties of the system, both near the transition and within the active phase. We found that the critical point of the continuous transition exhibits a highly heterogeneous steady state, characterized by avalanche dynamics and non-trivial correlations, with critical exponents consistent with those of the conserved directed percolation universality class. In contrast, the discontinuous transition displays a more homogeneous state up to the onset of nucleation, which subsequently drives the system into a cooling regime reminiscent of free granular cooling and its associated instabilities. Furthermore, the active state of the corresponding two-dimensional effective model was shown to possess long-range correlations and to display hyperuniformity. Building on the theoretical framework established in Ref.~\onlinecite{maire2024interplay}, we derived from the microscopic degrees of freedom, the field theory of the conserved directed percolation for our particle system, enabling a quantitative prediction of the homogeneous, asynchronized state close and far from the transition. By incorporating synchronization at the mean-field level, we also quantitatively captured the discontinuous transition. Finally, by including fluctuations around the homogeneous state, our extended theory successfully accounts for all observed correlation functions of the non-equilibrium liquid, as well as its emergent hyperuniformity.

Future work will focus on a more detailed experimental analysis to further validate the universality of our findings. Numerical investigations are still needed to better understand the tricritical point and the nucleation pathway in this system.

\begin{acknowledgments}

This work has been done with the support of Investissements d'Avenir of LabEx PALM (Grant No. ANR-10-LABX-0039-PALM).  We thank Andrea Puglisi, Umberto Marini Bettolo Marconi and Lorenzo Caprini for fruitful scientific discussions concerning the fluctuating hydrodynamics of our model. We also thank Ran Ni for sharing with us its knowledge on the hyperuniformity in the metastable fluid and its relation to nucleation time. Finally, we are grateful to Yuta Kuroda for fruitful scientific discussions on hyperuniformity and fluctuating hydrodynamics.

\end{acknowledgments}

\bibliographystyle{apsrev4-1}
\bibliography{granular}

@article{10.21468/SciPostPhys.8.5.074,
    author = {Camille Aron and Claudio Chamon},
    doi = {10.21468/SciPostPhys.8.5.074},
    journal = {SciPost Phys.},
    pages = {074},
    publisher = {SciPost},
    title = {{Landau theory for non-equilibrium steady states}},
    url = {https://scipost.org/10.21468/SciPostPhys.8.5.074},
    volume = {8},
    year = {2020}
}

@article{acebron2005kuramoto,
    author = {Acebr{\'o}n, Juan A and Bonilla, Luis L and Vicente, Conrad J P{\'e}rez and Ritort, F{\'e}lix and Spigler, Renato},
    doi = {10.1103/RevModPhys.77.137},
    journal = {Reviews of modern physics},
    number = {1},
    pages = {137},
    publisher = {APS},
    title = {The Kuramoto model: A simple paradigm for synchronization phenomena},
    volume = {77},
    year = {2005}
}

@article{anand2025emergent,
    author = {Anand, Satyam and Zhang, Guanming and Martiniani, Stefano},
    journal = {arXiv preprint arXiv:2505.22933},
    title = {Emergent universal long-range structure in random-organizing systems},
    year = {2025}
}

@article{argentina2002van,
    author = {Argentina, Mederic and Clerc, MG and Soto, R},
    doi = {10.1007/978-3-540-39843-1_13},
    journal = {Physical review letters},
    number = {4},
    pages = {044301},
    publisher = {APS},
    title = {Van der Waals--like transition in fluidized granular matter},
    volume = {89},
    year = {2002}
}

@article{arnarson1998thermal,
    author = {Arnarson, Birgir {\"O} and Willits, Jeffrey T},
    doi = {10.1063/1.869658},
    journal = {Physics of Fluids},
    number = {6},
    pages = {1324--1328},
    publisher = {American Institute of Physics},
    title = {Thermal diffusion in binary mixtures of smooth, nearly elastic spheres with and without gravity},
    volume = {10},
    year = {1998}
}

@article{assaf2017wkb,
    author = {Assaf, Michael and Meerson, Baruch},
    doi = {10.1088/1751-8121/aa669a},
    journal = {Journal of Physics A: Mathematical and Theoretical},
    number = {26},
    pages = {263001},
    publisher = {IOP Publishing},
    title = {WKB theory of large deviations in stochastic populations},
    volume = {50},
    year = {2017}
}

@article{assis2009discontinuous,
    author = {Assis, Vladimir RV and Copelli, Mauro},
    doi = {10.1103/PhysRevE.80.061105},
    journal = {Physical Review E},
    number = {6},
    pages = {061105},
    publisher = {APS},
    title = {Discontinuous nonequilibrium phase transitions in a nonlinearly pulse-coupled excitable lattice model},
    volume = {80},
    year = {2009}
}

@article{bak1991self,
    author = {Bak, Per and Chen, Kan},
    doi = {10.1038/scientificamerican0191-46},
    journal = {Scientific American},
    number = {1},
    pages = {46--53},
    publisher = {JSTOR},
    title = {Self-organized criticality},
    volume = {264},
    year = {1991}
}

@article{bak1992self,
    author = {Bak, Per},
    doi = {10.1016/0378-4371(92)90503-I},
    journal = {Physica A: Statistical Mechanics and its Applications},
    number = {1-4},
    pages = {41--46},
    publisher = {Elsevier},
    title = {Self-organized criticality in non-conservative models},
    volume = {191},
    year = {1992}
}

@article{bertin2024cascade,
    author = {Bertin, Eric and Andrix, Alex and Le Godais, Ga{\"e}l},
    doi = {10.1007/s10955-024-03327-3},
    journal = {Journal of Statistical Physics},
    number = {9},
    pages = {1--15},
    publisher = {Springer},
    title = {A cascade model for the discontinuous absorbing phase transition between turbulent and laminar flows},
    volume = {191},
    year = {2024}
}

@article{bhaumik2021role,
    author = {Bhaumik, Himangsu and Foffi, Giuseppe and Sastry, Srikanth},
    doi = {10.1073/pnas.2100227118},
    journal = {Proceedings of the National Academy of Sciences},
    number = {16},
    pages = {e2100227118},
    publisher = {National Acad Sciences},
    title = {The role of annealing in determining the yielding behavior of glasses under cyclic shear deformation},
    volume = {118},
    year = {2021}
}

@article{bixon1969boltzmann,
    author = {Bixon, Mordechai and Zwanzig, Robert},
    doi = {10.1103/PhysRev.187.267},
    journal = {Physical review},
    number = {1},
    pages = {267},
    publisher = {APS},
    title = {Boltzmann-Langevin equation and hydrodynamic fluctuations},
    volume = {187},
    year = {1969}
}

@article{bixon1989hard,
    author = {Bixon, Mordechai and Dorfman, JR and Dufty, James W},
    doi = {10.1021/j100356a027},
    journal = {The Journal of Physical Chemistry},
    number = {19},
    pages = {7019--7022},
    publisher = {ACS Publications},
    title = {Hard sphere Langevin equations},
    volume = {93},
    year = {1989}
}

@article{bonachela2009self,
    author = {Bonachela, Juan A and Munoz, Miguel A},
    doi = {10.1088/1742-5468/2009/09/P09009},
    journal = {Journal of Statistical Mechanics: Theory and Experiment},
    number = {09},
    pages = {P09009},
    publisher = {IOP Publishing},
    title = {Self-organization without conservation: true or just apparent scale-invariance?},
    volume = {2009},
    year = {2009}
}

@article{bouchet2020boltzmann,
    author = {Bouchet, Freddy},
    doi = {10.1007/s10955-020-02588-y},
    journal = {Journal of Statistical Physics},
    number = {2},
    pages = {515--550},
    publisher = {Springer},
    title = {Is the Boltzmann equation reversible? A large deviation perspective on the irreversibility paradox},
    volume = {181},
    year = {2020}
}

@article{bray1994theory,
    author = {Bray, Alan J},
    doi = {10.1080/00018730110117433},
    journal = {Advances in Physics},
    number = {3},
    pages = {357--459},
    publisher = {Taylor \& Francis},
    title = {Theory of phase-ordering kinetics},
    volume = {43},
    year = {1994}
}

@article{brey1996model,
    author = {Brey, J Javier and Moreno, F and Dufty, James W},
    doi = {10.1103/PhysRevE.54.445},
    journal = {Physical Review E},
    number = {1},
    pages = {445},
    publisher = {APS},
    title = {Model kinetic equation for low-density granular flow},
    volume = {54},
    year = {1996}
}

@article{brey1998hydrodynamics,
    author = {Brey, J Javier and Dufty, James W and Kim, Chang Sub and Santos, Andr{\'e}s},
    doi = {10.1103/PhysRevE.58.4638},
    journal = {Physical Review E},
    number = {4},
    pages = {4638},
    publisher = {APS},
    title = {Hydrodynamics for granular flow at low density},
    volume = {58},
    year = {1998}
}

@article{brey2009fluctuating,
    author = {Brey, J Javier and Maynar, P and Garcia de Soria, MI},
    doi = {10.1103/PhysRevE.79.051305},
    journal = {Physical Review E—Statistical, Nonlinear, and Soft Matter Physics},
    number = {5},
    pages = {051305},
    publisher = {APS},
    title = {Fluctuating hydrodynamics for dilute granular gases},
    volume = {79},
    year = {2009}
}

@article{brey2013homogeneous,
    author = {Brey, J Javier and de Soria, MI Garc{\'\i}a and Maynar, P and Buz{\'o}n, V},
    doi = {10.1103/PhysRevE.88.062205},
    journal = {Physical Review E},
    number = {6},
    pages = {062205},
    publisher = {APS},
    title = {Homogeneous steady state of a confined granular gas},
    volume = {88},
    year = {2013}
}

@article{brey2015hydrodynamics,
    author = {Brey, J Javier and Buz{\'o}n, V and Maynar, P and Garc{\'\i}a de Soria, MI},
    doi = {10.1103/PhysRevE.91.052201},
    journal = {Physical Review E},
    number = {5},
    pages = {052201},
    publisher = {APS},
    title = {Hydrodynamics for a model of a confined quasi-two-dimensional granular gas},
    volume = {91},
    year = {2015}
}

@article{brey2019inhomogeneous,
    author = {Brey, J Javier and de Soria, MI Garc{\'\i}a and Maynar, P},
    doi = {10.1103/PhysRevE.100.052901},
    journal = {Physical Review E},
    number = {5},
    pages = {052901},
    publisher = {APS},
    title = {Inhomogeneous cooling state of a strongly confined granular gas at low density},
    volume = {100},
    year = {2019}
}

@article{brey2020kinetic,
    author = {Brey, J Javier and Maynar, P and de Soria, MI Garc{\'\i}a},
    doi = {10.1088/1742-5468/ab7124},
    journal = {Journal of Statistical Mechanics: Theory and Experiment},
    number = {3},
    pages = {034002},
    publisher = {IOP Publishing},
    title = {Kinetic model for a confined quasi-two-dimensional gas of inelastic hard spheres},
    volume = {2020},
    year = {2020}
}

@article{brey2020self,
    author = {Brey, J Javier and Garc{\'\i}a de Soria, MI and Maynar, P},
    doi = {10.1103/PhysRevE.101.012102},
    journal = {Physical Review E},
    number = {1},
    pages = {012102},
    publisher = {APS},
    title = {Self-diffusion in a quasi-two-dimensional gas of hard spheres},
    volume = {101},
    year = {2020}
}

@book{brilliantov2004kinetic,
    author = {Brilliantov, Nikolai V and P{\"o}schel, Thorsten},
    doi = {10.1093/acprof:oso/9780198530381.001.0001},
    publisher = {Oxford University Press, USA},
    title = {Kinetic theory of granular gases},
    year = {2004}
}

@article{brito2013hydrodynamic,
    author = {Brito, Ricardo and Risso, Dino and Soto, Rodrigo},
    doi = {10.1103/PhysRevE.87.022209},
    journal = {Physical Review E},
    number = {2},
    pages = {022209},
    publisher = {APS},
    title = {Hydrodynamic modes in a confined granular fluid},
    volume = {87},
    year = {2013}
}

@article{canet2011general,
    author = {Canet, L{\'e}onie and Chat{\'e}, Hugues and Delamotte, Bertrand},
    doi = {10.1088/1751-8113/44/49/495001},
    journal = {Journal of Physics A: Mathematical and Theoretical},
    number = {49},
    pages = {495001},
    publisher = {IOP Publishing},
    title = {General framework of the non-perturbative renormalization group for non-equilibrium steady states},
    volume = {44},
    year = {2011}
}

@book{cardy1996scaling,
    author = {Cardy, John},
    doi = {10.1017/CBO9781316036440},
    publisher = {Cambridge university press},
    title = {Scaling and renormalization in statistical physics},
    volume = {5},
    year = {1996}
}

@article{Castillo2012,
    author = {Castillo, Gustavo and Mujica, Nicol\'as and Soto, Rodrigo},
    doi = {10.1103/PhysRevLett.109.095701},
    issue = {9},
    journal = {Phys. Rev. Lett.},
    month = {Aug},
    numpages = {5},
    pages = {095701},
    publisher = {American Physical Society},
    title = {Fluctuations and Criticality of a Granular Solid-Liquid-Like Phase Transition},
    url = {https://link.aps.org/doi/10.1103/PhysRevLett.109.095701},
    volume = {109},
    year = {2012}
}

@article{castillo2012fluctuations,
    author = {Castillo, Gustavo and Mujica, Nicol{\'a}s and Soto, Rodrigo},
    doi = {10.1103/PhysRevLett.109.095701},
    journal = {Physical Review Letters},
    number = {9},
    pages = {095701},
    publisher = {APS},
    title = {Fluctuations and criticality of a granular solid-liquid-like phase transition},
    volume = {109},
    year = {2012}
}

@article{Castillo2015,
    author = {Castillo, Gustavo and Mujica, Nicol\'as and Soto, Rodrigo},
    doi = {10.1103/PhysRevE.91.012141},
    issue = {1},
    journal = {Phys. Rev. E},
    month = {Jan},
    numpages = {13},
    pages = {012141},
    publisher = {American Physical Society},
    title = {Universality and criticality of a second-order granular solid-liquid-like phase transition},
    url = {https://link.aps.org/doi/10.1103/PhysRevE.91.012141},
    volume = {91},
    year = {2015}
}

@article{castillo2015universality,
    author = {Castillo, Gustavo and Mujica, Nicol{\'a}s and Soto, Rodrigo},
    doi = {10.1103/PhysRevE.91.012141},
    journal = {Physical Review E},
    number = {1},
    pages = {012141},
    publisher = {APS},
    title = {Universality and criticality of a second-order granular solid-liquid-like phase transition},
    volume = {91},
    year = {2015}
}

@book{chaikin1995principles,
    author = {Chaikin, Paul M and Lubensky, Tom C and Witten, Thomas A},
    doi = {10.1063/1.2808258},
    publisher = {Cambridge university press Cambridge},
    title = {Principles of condensed matter physics},
    volume = {10},
    year = {1995}
}

@article{chang2020modelling,
    author = {Chang, Sheryl L and Harding, Nathan and Zachreson, Cameron and Cliff, Oliver M and Prokopenko, Mikhail},
    doi = {10.1038/s41467-020-19393-6},
    journal = {Nature communications},
    number = {1},
    pages = {5710},
    publisher = {Nature Publishing Group UK London},
    title = {Modelling transmission and control of the COVID-19 pandemic in Australia},
    volume = {11},
    year = {2020}
}

@article{chantry2017universal,
    author = {Chantry, Matthew and Tuckerman, Laurette S and Barkley, Dwight},
    doi = {10.1017/jfm.2017.405},
    journal = {Journal of Fluid Mechanics},
    pages = {R1},
    publisher = {Cambridge University Press},
    title = {Universal continuous transition to turbulence in a planar shear flow},
    volume = {824},
    year = {2017}
}

@article{chapman1912xii,
    author = {Chapman, Sydney},
    journal = {Philosophical Transactions of the Royal Society of London. Series A, Containing Papers of a Mathematical or Physical Character},
    number = {471-483},
    pages = {433--483},
    publisher = {The Royal Society London},
    title = {XII. The kinetic theory of a gas constituted of spherically symmetrical molecules},
    volume = {211},
    year = {1912}
}

@book{chapman1990mathematical,
    author = {Chapman, Sydney and Cowling, Thomas George},
    publisher = {Cambridge university press},
    title = {The mathematical theory of non-uniform gases: an account of the kinetic theory of viscosity, thermal conduction and diffusion in gases},
    year = {1990}
}

@article{Chastaing2015,
    author = {Chastaing, J.-Y. and Bertin, E. and Géminard, J.-C.},
    doi = {10.1119/1.4906418},
    issn = {0002-9505},
    journal = {American Journal of Physics},
    month = {06},
    number = {6},
    pages = {518-524},
    title = {{Dynamics of a bouncing ball}},
    volume = {83},
    year = {2015}
}

@article{chertkov2023characterizing,
    author = {Chertkov, Eli and Cheng, Zihan and Potter, Andrew C and Gopalakrishnan, Sarang and Gatterman, Thomas M and Gerber, Justin A and Gilmore, Kevin and Gresh, Dan and Hall, Alex and Hankin, Aaron and others},
    doi = {10.1038/s41567-023-02199-w},
    journal = {Nature physics},
    number = {12},
    pages = {1799--1804},
    publisher = {Nature Publishing Group UK London},
    title = {Characterizing a non-equilibrium phase transition on a quantum computer},
    volume = {19},
    year = {2023}
}

@misc{chiu2024theorynonequilibriummulticomponentcoexistence,
    archiveprefix = {arXiv},
    author = {Yu-Jen Chiu and Daniel Evans and Ahmad K. Omar},
    eprint = {2409.07620},
    primaryclass = {cond-mat.stat-mech},
    title = {Theory of Nonequilibrium Multicomponent Coexistence},
    url = {https://arxiv.org/abs/2409.07620},
    year = {2024}
}

@article{clar1997phase,
    author = {Clar, Siegfried and Schenk, Klaus and Schwabl, Franz},
    doi = {10.1103/PhysRevE.55.2174},
    journal = {Physical Review E},
    number = {3},
    pages = {2174},
    publisher = {APS},
    title = {Phase transitions in a forest-fire model},
    volume = {55},
    year = {1997}
}

@article{corte2008random,
    author = {Corte, Laurent and Chaikin, Paul M and Gollub, Jerry P and Pine, David J},
    doi = {10.1038/nphys891},
    journal = {Nature Physics},
    number = {5},
    pages = {420--424},
    publisher = {Nature Publishing Group UK London},
    title = {Random organization in periodically driven systems},
    volume = {4},
    year = {2008}
}

@article{Cundall79,
    author = {Cundall, P. A. and Strack, O. D. L.},
    doi = {10.1680/geot.1979.29.1.47},
    eprint = { 
https://doi.org/10.1680/geot.1979.29.1.47
},
    journal = {Géotechnique},
    number = {1},
    pages = {47-65},
    title = {A discrete numerical model for granular assemblies},
    url = { 
https://doi.org/10.1680/geot.1979.29.1.47
},
    volume = {29},
    year = {1979}
}

@book{de2006hydrodynamic,
    author = {De Zarate, Jose M Ortiz and Sengers, Jan V},
    publisher = {Elsevier},
    title = {Hydrodynamic fluctuations in fluids and fluid mixtures},
    year = {2006}
}

@article{de2023effects,
    author = {de Andrade, Marcelo Freitas and Figueiredo, Wagner},
    doi = {10.1016/j.physleta.2023.128863},
    journal = {Physics Letters A},
    pages = {128863},
    publisher = {Elsevier},
    title = {Effects of temporal disorder in the continuous phase transition of a catalytic reaction system},
    volume = {475},
    year = {2023}
}

@article{di2016self,
    author = {di Santo, Serena and Burioni, Raffaella and Vezzani, Alessandro and Munoz, Miguel A},
    doi = {10.1103/PhysRevLett.116.240601},
    journal = {Physical review letters},
    number = {24},
    pages = {240601},
    publisher = {APS},
    title = {Self-organized bistability associated with first-order phase transitions},
    volume = {116},
    year = {2016}
}

@article{Dickman1998,
    author = {Ronald Dickman and Alessandro Vespignani and Stefano Zapperi},
    doi = {10.1103/physreve.57.5095},
    journal = {Physical Review E},
    month = {May},
    number = {5},
    pages = {5095--5105},
    publisher = {American Physical Society ({APS})},
    title = {Self-organized criticality as an absorbing-state phase transition},
    url = {https://doi.org/10.1103/physreve.57.5095},
    volume = {57},
    year = {1998}
}

@article{dorfman1994generic,
    author = {Dorfman, JR and Kirkpatrick, TR and Sengers, JV},
    doi = {10.1146/annurev.pc.45.100194.001241},
    journal = {Annual Review of Physical Chemistry},
    number = {1},
    pages = {213--239},
    publisher = {Annual Reviews 4139 El Camino Way, PO Box 10139, Palo Alto, CA 94303-0139, USA},
    title = {Generic long-range correlations in molecular fluids},
    volume = {45},
    year = {1994}
}

@book{dorfman2021contemporary,
    author = {Dorfman, Jay Robert and van Beijeren, Henk and Kirkpatrick, Theodore Ross},
    doi = {10.1017/9781139025942},
    publisher = {Cambridge University Press},
    title = {Contemporary kinetic theory of matter},
    year = {2021}
}

@article{dornic2005integration,
    author = {Dornic, Ivan and Chat{\'e}, Hugues and Munoz, Miguel A},
    doi = {10.1103/PhysRevLett.94.100601},
    journal = {Physical review letters},
    number = {10},
    pages = {100601},
    publisher = {APS},
    title = {Integration of Langevin equations with multiplicative noise and the viability of field theories for absorbing phase transitions},
    volume = {94},
    year = {2005}
}

@article{duran2017general,
    author = {Dur{\'a}n-Olivencia, Miguel A and Yatsyshin, Peter and Goddard, Benjamin D and Kalliadasis, Serafim},
    doi = {10.1088/1367-2630/aa9041},
    journal = {New Journal of Physics},
    number = {12},
    pages = {123022},
    publisher = {IOP Publishing},
    title = {General framework for fluctuating dynamic density functional theory},
    volume = {19},
    year = {2017}
}

@article{elgart2004rare,
    author = {Elgart, Vlad and Kamenev, Alex},
    doi = {10.1103/PhysRevE.70.041106},
    journal = {Physical Review E},
    number = {4},
    pages = {041106},
    publisher = {APS},
    title = {Rare event statistics in reaction-diffusion systems},
    volume = {70},
    year = {2004}
}

@article{eltohfa2024simulations,
    author = {Eltohfa, Mohamed and Wang, Xinghan and Griffin, Colton M and Robicheaux, F},
    doi = {10.1103/PhysRevE.110.014114},
    journal = {Physical Review E},
    number = {1},
    pages = {014114},
    publisher = {APS},
    title = {Simulations of classical three-body thermalization in one dimension},
    volume = {110},
    year = {2024}
}

@book{enskog1917kinetische,
    author = {Enskog, David},
    publisher = {Almquist \& Wiksell},
    title = {Kinetische Theorie der Vorg{\"a}nge in m{\"a}ssig verd{\"u}nnten Gasen...},
    volume = {1},
    year = {1917}
}

@article{esipov1997granular,
    author = {Esipov, Sergei E and P{\"o}schel, Thorsten},
    doi = {10.1007/BF02183630},
    journal = {Journal of statistical physics},
    pages = {1385--1395},
    publisher = {Springer},
    title = {The granular phase diagram},
    volume = {86},
    year = {1997}
}

@article{everson1986chaotic,
    author = {Everson, RM},
    doi = {10.1016/0167-2789(86)90064-3},
    journal = {Physica D: Nonlinear Phenomena},
    number = {3},
    pages = {355--383},
    publisher = {Elsevier},
    title = {Chaotic dynamics of a bouncing ball},
    volume = {19},
    year = {1986}
}

@misc{feng2024theoryanomalousphasebehavior,
    archiveprefix = {arXiv},
    author = {Jiechao Feng and Ahmad K. Omar},
    eprint = {2407.08676},
    primaryclass = {cond-mat.soft},
    title = {Theory for the Anomalous Phase Behavior of Inertial Active Matter},
    url = {https://arxiv.org/abs/2407.08676},
    year = {2024}
}

@article{fiocco2013,
    author = {Fiocco, Davide and Foffi, Giuseppe and Sastry, Srikanth},
    doi = {10.1103/PhysRevE.88.020301},
    issue = {2},
    journal = {Phys. Rev. E},
    month = {Aug},
    numpages = {5},
    pages = {020301},
    publisher = {American Physical Society},
    title = {Oscillatory athermal quasistatic deformation of a model
glass},
    url = {https://link.aps.org/doi/10.1103/PhysRevE.88.020301},
    volume = {88},
    year = {2013}
}

@article{Fisher97,
    author = {Fisher, Daniel S. and Dahmen, Karin and Ramanathan, Sharad and Ben-Zion, Yehuda},
    doi = {10.1103/PhysRevLett.78.4885},
    issue = {25},
    journal = {Phys. Rev. Lett.},
    month = {Jun},
    numpages = {0},
    pages = {4885--4888},
    publisher = {American Physical Society},
    title = {Statistics of Earthquakes in Simple Models of Heterogeneous Faults},
    url = {https://link.aps.org/doi/10.1103/PhysRevLett.78.4885},
    volume = {78},
    year = {1997}
}

@article{franceschini2011transverse,
    author = {Franceschini, Alexandre and Filippidi, Emmanouela and Guazzelli, Elisabeth and Pine, David J},
    doi = {10.1103/PhysRevLett.107.250603},
    journal = {Physical review letters},
    number = {25},
    pages = {250603},
    publisher = {APS},
    title = {Transverse alignment of fibers in a periodically sheared suspension: an absorbing phase transition with a slowly varying control parameter},
    volume = {107},
    year = {2011}
}

@article{galliano2023two,
    author = {Galliano, Leonardo and Cates, Michael E and Berthier, Ludovic},
    doi = {10.1103/PhysRevLett.131.047101},
    journal = {Physical Review Letters},
    number = {4},
    pages = {047101},
    publisher = {APS},
    title = {Two-dimensional crystals far from equilibrium},
    volume = {131},
    year = {2023}
}

@article{garcia2024interactions,
    author = {Garcia Lorenzana, Giulia and Altieri, Ada and Biroli, Giulio},
    doi = {10.1103/PRXLife.2.013014},
    journal = {PRX Life},
    number = {1},
    pages = {013014},
    publisher = {APS},
    title = {Interactions and migration rescuing ecological diversity},
    volume = {2},
    year = {2024}
}

@book{gardiner2009stochastic,
    author = {Gardiner, Crispin},
    publisher = {Springer Berlin},
    title = {Stochastic methods},
    volume = {4},
    year = {2009}
}

@article{garrido1990long,
    author = {Garrido, Pedro L and Lebowitz, Joel L and Maes, Christian and Spohn, Herbert},
    doi = {10.1103/PhysRevA.42.1954},
    journal = {Physical Review A},
    number = {4},
    pages = {1954},
    publisher = {APS},
    title = {Long-range correlations for conservative dynamics},
    volume = {42},
    year = {1990}
}

@article{garzo2005instabilities,
    author = {Garz{\'o}, Vicente},
    doi = {10.1103/PhysRevE.72.021106},
    journal = {Physical Review E},
    number = {2},
    pages = {021106},
    publisher = {APS},
    title = {Instabilities in a free granular fluid described by the Enskog equation},
    volume = {72},
    year = {2005}
}

@article{garzo2018enskog,
    author = {Garz{\'o}, Vicente and Brito, Ricardo and Soto, Rodrigo},
    doi = {10.1103/PhysRevE.98.052904},
    journal = {Physical Review E},
    number = {5},
    pages = {052904},
    publisher = {APS},
    title = {Enskog kinetic theory for a model of a confined quasi-two-dimensional granular fluid},
    volume = {98},
    year = {2018}
}

@article{hinrichsen2000flowing,
  title={Flowing sand—a possible physical realization of Directed Percolation},
  author={Hinrichsen, Haye and Jim{\'e}nez-Dalmaroni, Andrea and Rozov, Yadin and Domany, Eytan},
  journal={Journal of Statistical Physics},
  volume={98},
  number={5},
  pages={1149--1168},
  year={2000},
  publisher={Springer},
  doi = {https://doi.org/10.1023/A:1018667712578}
}

@article{hinrichsen1999flowing,
  title={Flowing sand: A physical realization of directed percolation},
  author={Hinrichsen, Haye and Jim{\'e}nez-Dalmaroni, Andrea and Rozov, Yadin and Domany, Eytan},
  journal={Physical Review Letters},
  volume={83},
  number={24},
  pages={4999},
  year={1999},
  publisher={APS},
  doi = {https://doi.org/10.1103/PhysRevLett.83.4999}
}

@article{giusfredi2024localization,
    author = {Giusfredi, Michele and Iubini, Stefano and Politi, Paolo},
    doi = {10.1007/s10955-024-03324-6},
    journal = {Journal of Statistical Physics},
    number = {10},
    pages = {119},
    publisher = {Springer},
    title = {Localization in boundary-driven lattice models},
    volume = {191},
    year = {2024}
}

@book{goldenfeld2018lectures,
    author = {Goldenfeld, Nigel},
    doi = {10.1201/9780429493492},
    publisher = {CRC Press},
    title = {Lectures on phase transitions and the renormalization group},
    year = {2018}
}

@article{goldhirsch1993clustering,
    author = {Goldhirsch, ISAAC and Zanetti, G},
    doi = {10.1103/PhysRevLett.70.1619},
    journal = {Physical review letters},
    number = {11},
    pages = {1619},
    publisher = {APS},
    title = {Clustering instability in dissipative gases},
    volume = {70},
    year = {1993}
}

@article{gomez2022enskog,
    author = {G{\'o}mez Gonz{\'a}lez, Rub{\'e}n and Garz{\'o}, Vicente},
    doi = {10.1103/PhysRevE.106.064902},
    journal = {Physical Review E},
    number = {6},
    pages = {064902},
    publisher = {APS},
    title = {Enskog kinetic theory of binary granular suspensions: Heat flux and stability analysis of the homogeneous steady state},
    volume = {106},
    year = {2022}
}

@article{gradenigo2011fluctuating,
    author = {Gradenigo, Giacomo and Sarracino, Alessandro and Villamaina, Dario and Puglisi, Andrea},
    doi = {10.1088/1742-5468/2011/08/P08017},
    journal = {Journal of Statistical Mechanics: Theory and Experiment},
    number = {08},
    pages = {P08017},
    publisher = {IOP Publishing},
    title = {Fluctuating hydrodynamics and correlation lengths in a driven granular fluid},
    volume = {2011},
    year = {2011}
}

@article{grinstein1990conservation,
    author = {Grinstein, G and Lee, D-H and Sachdev, Subir},
    journal = {Physical review letters},
    number = {16},
    pages = {1927},
    publisher = {APS},
    title = {Conservation laws, anisotropy, and ‘‘self-organized criticality’’in noisy nonequilibrium systems},
    volume = {64},
    year = {1990}
}

@article{grinstein1991generic,
    author = {Grinstein, G},
    doi = {10.1063/1.348003},
    journal = {Journal of applied physics},
    number = {8},
    pages = {5441--5446},
    publisher = {American Institute of Physics},
    title = {Generic scale invariance in classical nonequilibrium systems},
    volume = {69},
    year = {1991}
}

@article{guzman2018critical,
    author = {Guzm{\'a}n, Marcelo and Soto, Rodrigo},
    doi = {10.1103/PhysRevE.97.012907},
    journal = {Physical Review E},
    number = {1},
    pages = {012907},
    publisher = {APS},
    title = {Critical phenomena in quasi-two-dimensional vibrated granular systems},
    volume = {97},
    year = {2018}
}

@book{hansen2013theory,
    author = {Hansen, Jean-Pierre and McDonald, Ian Ranald},
    publisher = {Academic press},
    title = {Theory of simple liquids: with applications to soft matter},
    year = {2013}
}

@misc{hazra2024hyperuniformitymasstransportprocesses,
    archiveprefix = {arXiv},
    author = {Animesh Hazra and Anirban Mukherjee and Punyabrata Pradhan},
    doi = {10.1088/1742-5468/ada88c},
    eprint = {2410.00613},
    primaryclass = {cond-mat.stat-mech},
    title = {Hyperuniformity in mass transport processes with center-of-mass conservation: Some exact results},
    url = {https://arxiv.org/abs/2410.00613},
    year = {2024}
}

@book{henkel2008non,
    author = {Henkel, Malte},
    doi = {10.1007/978-90-481-2869-3},
    publisher = {Springer},
    title = {Non-equilibrium phase transitions},
    year = {2008}
}

@article{hexner2017enhanced,
    author = {Hexner, Daniel and Chaikin, Paul M and Levine, Dov},
    doi = {10.1073/pnas.1619260114},
    journal = {Proceedings of the National Academy of Sciences},
    number = {17},
    pages = {4294--4299},
    publisher = {National Acad Sciences},
    title = {Enhanced hyperuniformity from random reorganization},
    volume = {114},
    year = {2017}
}

@article{hinrichsen2000non,
    author = {Hinrichsen, Haye},
    doi = {10.1080/00018730050198152},
    journal = {Advances in physics},
    number = {7},
    pages = {815--958},
    publisher = {Taylor \& Francis},
    title = {Non-equilibrium critical phenomena and phase transitions into absorbing states},
    volume = {49},
    year = {2000}
}

@article{hof2023directed,
    author = {Hof, Bj{\"o}rn},
    doi = {10.1038/s42254-022-00539-y},
    journal = {Nature Reviews Physics},
    number = {1},
    pages = {62--72},
    publisher = {Nature Publishing Group UK London},
    title = {Directed percolation and the transition to turbulence},
    volume = {5},
    year = {2023}
}

@article{honkonen2011ito,
    author = {Honkonen, Juha},
    journal = {arXiv preprint arXiv:1102.1581},
    title = {Ito and Stratonovich calculuses in stochastic field theory},
    year = {2011}
}

@article{Jagla2014,
    author = {Jagla, E. A. and Landes, Fran\ifmmode \mbox{\c{c}}\else \c{c}\fi{}ois P. and Rosso, Alberto},
    doi = {10.1103/PhysRevLett.112.174301},
    issue = {17},
    journal = {Phys. Rev. Lett.},
    month = {Apr},
    numpages = {5},
    pages = {174301},
    publisher = {American Physical Society},
    title = {Viscoelastic Effects in Avalanche Dynamics: A Key to Earthquake Statistics},
    url = {https://link.aps.org/doi/10.1103/PhysRevLett.112.174301},
    volume = {112},
    year = {2014}
}

@article{jarzynski1993universal,
    author = {Jarzynski, C and Swiatecki, WJ},
    doi = {10.1016/0375-9474(93)90327-T},
    journal = {Nuclear Physics A},
    number = {1},
    pages = {1--9},
    publisher = {Elsevier},
    title = {A universal asymptotic velocity distribution for independent particles in a time-dependent irregular container},
    volume = {552},
    year = {1993}
}

@article{jeanneret2014geometrically,
    author = {Jeanneret, Rapha{\"e}l and Bartolo, Denis},
    doi = {10.1038/ncomms4474},
    journal = {Nature Communications},
    number = {1},
    pages = {3474},
    publisher = {Nature Publishing Group UK London},
    title = {Geometrically protected reversibility in hydrodynamic Loschmidt-echo experiments},
    volume = {5},
    year = {2014}
}

@article{jocteur2025random,
    author = {Jocteur, Tristan and Nardini, Cesare and Bertin, Eric and Mari, Romain},
    journal = {arXiv preprint arXiv:2506.14330},
    title = {Random organization criticality with long-range hydrodynamic interactions},
    year = {2025}
}

@book{kardar2007statistical,
    author = {Kardar, Mehran},
    doi = {10.1017/CBO9780511815898},
    publisher = {Cambridge University Press},
    title = {Statistical physics of particles},
    year = {2007}
}

@article{Kawasaki2016,
    author = {Kawasaki, Takeshi and Berthier, Ludovic},
    doi = {10.1103/PhysRevE.94.022615},
    issue = {2},
    journal = {Phys. Rev. E},
    month = {Aug},
    numpages = {6},
    pages = {022615},
    publisher = {American Physical Society},
    title = {Macroscopic yielding in jammed solids is accompanied by a nonequilibrium first-order transition in particle trajectories},
    url = {https://link.aps.org/doi/10.1103/PhysRevE.94.022615},
    volume = {94},
    year = {2016}
}

@article{keim2014mechanical,
    author = {Keim, Nathan C and Arratia, Paulo E},
    doi = {10.1103/PhysRevLett.112.028302},
    journal = {Physical review letters},
    number = {2},
    pages = {028302},
    publisher = {APS},
    title = {Mechanical and microscopic properties of the reversible plastic regime in a 2D jammed material},
    volume = {112},
    year = {2014}
}

@misc{keta2024longrangeordertwodimensionalsystems,
    archiveprefix = {arXiv},
    author = {Yann-Edwin Keta and Silke Henkes},
    doi = {10.1039/D5SM00208G},
    eprint = {2410.14840},
    primaryclass = {cond-mat.soft},
    title = {Long-range order in two-dimensional systems with fluctuating active stresses},
    url = {https://arxiv.org/abs/2410.14840},
    year = {2024}
}

@article{khain2011hydrodynamics,
    author = {Khain, Evgeniy and Aranson, Igor S},
    doi = {10.1103/PhysRevE.84.031308},
    journal = {Physical Review E},
    number = {3},
    pages = {031308},
    publisher = {APS},
    title = {Hydrodynamics of a vibrated granular monolayer},
    volume = {84},
    year = {2011}
}

@inproceedings{kloss2010granular,
    author = {Kloss, Christoph and Goniva, Christoph},
    booktitle = {LAMMPS user workshop},
    title = {Granular Simulations in LAMMPS},
    year = {2010}
}

@article{knowltonsm14,
    author = {Knowlton, E. D. and Pine, D. J. and Cipelletti, L.},
    doi = {10.1039/C4SM00531G},
    issue = {36},
    journal = {Soft Matter},
    pages = {6931-6940},
    publisher = {The Royal Society of Chemistry},
    title = {A microscopic view of the yielding transition in
concentrated emulsions},
    url = {http://dx.doi.org/10.1039/C4SM00531G},
    volume = {10},
    year = {2014}
}

@article{Komatsu2015,
    author = {Komatsu, Yuta and Tanaka, Hajime},
    doi = {10.1103/PhysRevX.5.031025},
    issue = {3},
    journal = {Phys. Rev. X},
    month = {Aug},
    numpages = {17},
    pages = {031025},
    publisher = {American Physical Society},
    title = {Roles of Energy Dissipation in a Liquid-Solid Transition of Out-of-Equilibrium Systems},
    url = {https://link.aps.org/doi/10.1103/PhysRevX.5.031025},
    volume = {5},
    year = {2015}
}

@article{kundu2016long,
    author = {Kundu, Anupam and Hirschberg, Ori and Mukamel, David},
    doi = {10.1088/1742-5468/2016/03/033108},
    journal = {Journal of Statistical Mechanics: Theory and Experiment},
    number = {3},
    pages = {033108},
    publisher = {IOP Publishing},
    title = {Long range correlations in stochastic transport with energy and momentum conservation},
    volume = {2016},
    year = {2016}
}

@article{kuroda2023microscopic,
    author = {Kuroda, Yuta and Miyazaki, Kunimasa},
    doi = {10.1088/1742-5468/ad0639},
    journal = {Journal of Statistical Mechanics: Theory and Experiment},
    number = {10},
    pages = {103203},
    publisher = {IOP Publishing},
    title = {Microscopic theory for hyperuniformity in two-dimensional chiral active fluid},
    volume = {2023},
    year = {2023}
}

@article{kuroda2024long,
    author = {Kuroda, Yuta and Kawasaki, Takeshi and Miyazaki, Kunimasa},
    doi = {10.1103/PhysRevResearch.7.L012048},
    journal = {arXiv preprint arXiv:2402.19192},
    title = {Long-range translational order and hyperuniformity in two-dimensional chiral active crystal},
    year = {2024}
}

@misc{LammpsSiteGran,
    howpublished = {\url{https://docs.lammps.org/pair_granular.html}},
    year = {2024}
}

@article{landau1934structure,
    author = {Landau, L and Placzek, G},
    journal = {Phys. Z. Sowiet. Un},
    pages = {172},
    title = {Structure of the undisplaced scattering line},
    volume = {5},
    year = {1934}
}

@article{landau1980vol,
    author = {Landau, LD and Lifshitz, EM and Landau, LD and Lifschitz, EM},
    journal = {ZAMM Zeitschrift f},
    pages = {603},
    title = {Vol. 9—Statistical physics part 2. pdf},
    volume = {61},
    year = {1980}
}

@article{lasanta2015fluctuating,
    author = {Lasanta, Antonio and Manacorda, Alessandro and Prados, Antonio and Puglisi, Andrea},
    doi = {10.1088/1367-2630/17/8/083039},
    journal = {New Journal of Physics},
    number = {8},
    pages = {083039},
    publisher = {IOP Publishing},
    title = {Fluctuating hydrodynamics and mesoscopic effects of spatial correlations in dissipative systems with conserved momentum},
    volume = {17},
    year = {2015}
}

@article{LeDoussal2015,
    author = {Le Doussal, Pierre and Wiese, Kay J\"org},
    doi = {10.1103/PhysRevLett.114.110601},
    issue = {11},
    journal = {Phys. Rev. Lett.},
    month = {Mar},
    numpages = {5},
    pages = {110601},
    publisher = {American Physical Society},
    title = {Exact Mapping of the Stochastic Field Theory for Manna Sandpiles to Interfaces in Random Media},
    url = {https://link.aps.org/doi/10.1103/PhysRevLett.114.110601},
    volume = {114},
    year = {2015}
}

@article{lei2019hydrodynamics,
    author = {Lei, Qun-Li and Ni, Ran},
    doi = {10.1073/pnas.1911596116},
    journal = {Proceedings of the National Academy of Sciences},
    number = {46},
    pages = {22983--22989},
    publisher = {National Acad Sciences},
    title = {Hydrodynamics of random-organizing hyperuniform fluids},
    volume = {116},
    year = {2019}
}

@article{lei2021barrier,
    author = {Lei, Qun-Li and Hu, Hao and Ni, Ran},
    doi = {10.1103/PhysRevE.103.052607},
    journal = {Physical Review E},
    number = {5},
    pages = {052607},
    publisher = {APS},
    title = {Barrier-controlled nonequilibrium criticality in reactive particle systems},
    volume = {103},
    year = {2021}
}

@article{lei2023does,
    author = {Lei, Yusheng and Ni, Ran},
    doi = {10.1073/pnas.2312866120},
    journal = {Proceedings of the National Academy of Sciences},
    number = {48},
    pages = {e2312866120},
    publisher = {National Acad Sciences},
    title = {How does a hyperuniform fluid freeze?},
    volume = {120},
    year = {2023}
}

@article{lei2024non,
    author = {Lei, Yusheng and Ni, Ran},
    doi = {10.1088/1361-648X/ad83a0},
    journal = {arXiv preprint arXiv:2405.12818},
    title = {Non-equilibrium dynamic hyperuniform states},
    year = {2024}
}

@article{leishangthem2017yielding,
    author = {Leishangthem, Premkumar and Parmar, Anshul DS and Sastry, Srikanth},
    doi = {10.1038/ncomms14653},
    journal = {Nature communications},
    number = {1},
    pages = {14653},
    publisher = {Nature Publishing Group UK London},
    title = {The yielding transition in amorphous solids under oscillatory shear deformation},
    volume = {8},
    year = {2017}
}

@article{lemoult2016directed,
    author = {Lemoult, Gr{\'e}goire and Shi, Liang and Avila, Kerstin and Jalikop, Shreyas V and Avila, Marc and Hof, Bj{\"o}rn},
    doi = {10.1038/nphys3675},
    journal = {Nature Physics},
    number = {3},
    pages = {254--258},
    publisher = {Nature Publishing Group UK London},
    title = {Directed percolation phase transition to sustained turbulence in Couette flow},
    volume = {12},
    year = {2016}
}

@article{lubeck2004universal,
    author = {L{\"u}beck, Sven},
    doi = {10.1142/S0217979204027748},
    journal = {International Journal of Modern Physics B},
    number = {31n32},
    pages = {3977--4118},
    publisher = {World Scientific},
    title = {Universal scaling behavior of non-equilibrium phase transitions},
    volume = {18},
    year = {2004}
}

@article{Luck1993,
    author = {Luck, J. M. and Mehta, Anita},
    doi = {10.1103/PhysRevE.48.3988},
    issue = {5},
    journal = {Phys. Rev. E},
    month = {Nov},
    numpages = {0},
    pages = {3988--3997},
    publisher = {American Physical Society},
    title = {Bouncing ball with a finite restitution: Chattering, locking, and chaos},
    url = {https://link.aps.org/doi/10.1103/PhysRevE.48.3988},
    volume = {48},
    year = {1993}
}

@misc{ma2023theory,
    author = {Xiao Ma and Johannes Pausch and Michael E. Cates},
    eprint = {Arxiv preprint: 2310.17391},
    title = {Theory of Hyperuniformity at the Absorbing State Transition},
    year = {2023}
}

@article{maire2024enhancing,
    author = {Maire, R. and Plati, A.},
    doi = {10.1063/5.0217958},
    issn = {0021-9606},
    journal = {The Journal of Chemical Physics},
    month = {08},
    number = {5},
    pages = {054902},
    title = {{Enhancing (quasi-)long-range order in a two-dimensional driven crystal}},
    url = {https://doi.org/10.1063/5.0217958},
    volume = {161},
    year = {2024}
}

@article{pinto2025hydrodynamic,
  title={Hydrodynamic Equations for Active Brownian Particles in the High Persistence Regime},
  author={Pinto-Goldberg, Mart{\'\i}n and Soto, Rodrigo},
  journal={arXiv preprint arXiv:2506.17509},
  year={2025}
}

@article{maire2024interplay,
    author = {Maire, R and Plati, A and Stockinger, M and Trizac, E and Smallenburg, F and Foffi, G},
    doi = {10.1103/PhysRevLett.132.238202},
    journal = {Physical Review Letters},
    number = {23},
    pages = {238202},
    publisher = {APS},
    title = {Interplay between an absorbing phase transition and synchronization in a driven granular system},
    volume = {132},
    year = {2024}
}

@article{manacorda2017lattice,
    author = {Puglisi, A},
    journal = {Physical review letters},
    number = {20},
    pages = {208003},
    publisher = {APS},
    title = {Lattice model to derive the fluctuating hydrodynamics of active particles with inertia},
    volume = {119},
    year = {2017}
}

@book{manacorda2018lattice,
    author = {Manacorda, Alessandro},
    doi = {10.1007/978-3-319-95080-8},
    publisher = {Springer},
    title = {Lattice Models for Fluctuating Hydrodynamics in Granular and Active Matter},
    year = {2018}
}

@article{manna1990cascades,
    author = {Manna, SS and Kiss, L{\'a}szl{\'o} B and Kert{\'e}sz, J{\'a}nos},
    doi = {10.1007/BF01027312},
    journal = {Journal of statistical physics},
    pages = {923--932},
    publisher = {Springer},
    title = {Cascades and self-organized criticality},
    volume = {61},
    year = {1990}
}

@article{manna_sandpile_1999,
    author = {Manna, SS},
    journal = {Current Science},
    pages = {388--393},
    title = {Sandpile models of self-organized criticality},
    year = {1999}
}

@article{marcuzzi2016absorbing,
    author = {Marcuzzi, Matteo and Buchhold, Michael and Diehl, Sebastian and Lesanovsky, Igor},
    doi = {10.1103/PhysRevLett.116.245701},
    journal = {Physical review letters},
    number = {24},
    pages = {245701},
    publisher = {APS},
    title = {Absorbing state phase transition with competing quantum and classical fluctuations},
    volume = {116},
    year = {2016}
}

@article{mari2022absorbing,
    author = {Mari, Romain and Bertin, Eric and Nardini, Cesare},
    doi = {10.1103/PhysRevE.105.L032602},
    journal = {Physical Review E},
    number = {3},
    pages = {L032602},
    publisher = {APS},
    title = {Absorbing phase transitions in systems with mediated interactions},
    volume = {105},
    year = {2022}
}

@article{Martiniani2019,
    author = {Martiniani, Stefano and Chaikin, Paul M. and Levine, Dov},
    doi = {10.1103/PhysRevX.9.011031},
    issue = {1},
    journal = {Phys. Rev. X},
    month = {Feb},
    numpages = {13},
    pages = {011031},
    publisher = {American Physical Society},
    title = {Quantifying Hidden Order out of Equilibrium},
    url = {https://link.aps.org/doi/10.1103/PhysRevX.9.011031},
    volume = {9},
    year = {2019}
}

@article{Mata2021,
    author = {Ang{\'{e}}lica S. Mata},
    doi = {10.1063/5.0033130},
    journal = {Chaos: An Interdisciplinary Journal of Nonlinear Science},
    month = {January},
    number = {1},
    publisher = {{AIP} Publishing},
    title = {An overview of epidemic models with phase transitions to absorbing states running on top of complex networks},
    url = {https://doi.org/10.1063/5.0033130},
    volume = {31},
    year = {2021}
}

@article{maynar2009fluctuating,
    author = {Maynar, Pablo and Soria, MIG de and Trizac, Emmanuel},
    doi = {10.1140/epjst/e2010-01198-x},
    journal = {The European Physical Journal Special Topics},
    number = {1},
    pages = {123--139},
    publisher = {Springer},
    title = {Fluctuating hydrodynamics for driven granular gases},
    volume = {179},
    year = {2009}
}

@article{maynar2019homogeneous,
    author = {Maynar, P and de Soria, MI Garc{\'\i}a and Brey, J Javier},
    doi = {10.1088/1742-5468/ab3410},
    journal = {Journal of Statistical Mechanics: Theory and Experiment},
    number = {9},
    pages = {093205},
    publisher = {IOP Publishing},
    title = {Homogeneous dynamics in a vibrated granular monolayer},
    volume = {2019},
    year = {2019}
}

@article{maynar2019understanding,
    author = {Maynar, P and Garc{\'\i}a de Soria, MI and Brey, J Javier},
    doi = {10.1103/PhysRevE.99.032903},
    journal = {Physical Review E},
    number = {3},
    pages = {032903},
    publisher = {APS},
    title = {Understanding an instability in vibrated granular monolayers},
    volume = {99},
    year = {2019}
}

@article{mayo2023confined,
    author = {Mayo, M and Petit, JC and de Soria, MI Garc{\'\i}a and Maynar, P},
    doi = {10.1088/1742-5468/ad0828},
    journal = {Journal of Statistical Mechanics: Theory and Experiment},
    number = {12},
    pages = {123208},
    publisher = {IOP Publishing},
    title = {Confined granular gases under the influence of vibrating walls},
    volume = {2023},
    year = {2023}
}

@book{mazenko2006nonequilibrium,
    author = {Mazenko, Gene},
    doi = {10.1002/9783527618958},
    publisher = {John Wiley \& Sons},
    title = {Nonequilibrium statistical mechanics},
    year = {2006}
}

@article{mcnamara1993hydrodynamic,
    author = {McNamara, Sean},
    doi = {10.1063/1.858716},
    journal = {Physics of Fluids A: Fluid Dynamics},
    number = {12},
    pages = {3056--3070},
    publisher = {American Institute of Physics},
    title = {Hydrodynamic modes of a uniform granular medium},
    volume = {5},
    year = {1993}
}

@article{mcnamara1994inelastic,
    author = {McNamara, Sean and Young, William R},
    doi = {10.1103/PhysRevE.50.R28},
    journal = {Physical review E},
    number = {1},
    pages = {R28},
    publisher = {APS},
    title = {Inelastic collapse in two dimensions},
    volume = {50},
    year = {1994}
}

@article{meerson2011extinction,
    author = {Meerson, Baruch and Sasorov, Pavel V},
    doi = {10.1103/PhysRevE.83.011129},
    journal = {Physical Review E},
    number = {1},
    pages = {011129},
    publisher = {APS},
    title = {Extinction rates of established spatial populations},
    volume = {83},
    year = {2011}
}

@article{Metha1990,
    author = {Mehta, Anita and Luck, J. M.},
    doi = {10.1103/PhysRevLett.65.393},
    issue = {4},
    journal = {Phys. Rev. Lett.},
    month = {Jul},
    numpages = {0},
    pages = {393--396},
    publisher = {American Physical Society},
    title = {Novel temporal behavior of a nonlinear dynamical system: The completely inelastic bouncing ball},
    url = {https://link.aps.org/doi/10.1103/PhysRevLett.65.393},
    volume = {65},
    year = {1990}
}

@article{mindlin1949compliance,
    author = {Mindlin, R. D.},
    doi = {10.1115/1.4009973},
    issn = {0021-8936},
    journal = {Journal of Applied Mechanics},
    month = {04},
    number = {3},
    pages = {259-268},
    title = {{Compliance of Elastic Bodies in Contact}},
    url = {https://doi.org/10.1115/1.4009973},
    volume = {16},
    year = {2021}
}

@article{mitrano2011instabilities,
    author = {Mitrano, Peter P and Dahl, Steven R and Cromer, Daniel J and Pacella, Michael S and Hrenya, Christine M},
    doi = {10.1063/1.3633012},
    journal = {Physics of Fluids},
    number = {9},
    publisher = {AIP Publishing},
    title = {Instabilities in the homogeneous cooling of a granular gas: A quantitative assessment of kinetic-theory predictions},
    volume = {23},
    year = {2011}
}

@article{Mobius2014,
    author = {Möbius, Ronny and Heussinger, Claus},
    doi = {10.1039/C4SM00178H},
    issue = {27},
    journal = {Soft Matter},
    pages = {4806-4812},
    publisher = {The Royal Society of Chemistry},
    title = {(Ir)reversibility in dense granular systems driven by oscillating forces},
    url = {http://dx.doi.org/10.1039/C4SM00178H},
    volume = {10},
    year = {2014}
}

@article{movsko1995thermalization,
    author = {Mo{\v{s}}ko, M and Cambel, V},
    journal = {Acta Physica Polonica A},
    number = {1},
    pages = {157--160},
    publisher = {Polska Akademia Nauk. Instytut Fizyki PAN},
    title = {Thermalization of One-Dimensional Electron Gas by Many-Body Coulomb Scattering},
    volume = {87},
    year = {1995}
}

@article{mujica2016dynamics,
    author = {Mujica, Nicolas and Soto, Rodrigo},
    doi = {10.1007/978-3-319-27965-7_32},
    journal = {Recent advances in fluid dynamics with environmental applications},
    pages = {445--463},
    publisher = {Springer},
    title = {Dynamics of noncohesive confined granular media},
    year = {2016}
}

@article{mukherjee2024anomalous,
    author = {Mukherjee, Anirban and Tapader, Dhiraj and Hazra, Animesh and Pradhan, Punyabrata},
    doi = {10.1103/PhysRevE.110.024119},
    journal = {Physical Review E},
    number = {2},
    pages = {024119},
    publisher = {APS},
    title = {Anomalous relaxation and hyperuniform fluctuations in center-of-mass conserving systems with broken time-reversal symmetry},
    volume = {110},
    year = {2024}
}

@article{mulero2009equation,
    author = {Mulero, A and Cachadina, I and Solana, JR},
    doi = {10.1080/00268970902942250},
    journal = {Molecular Physics},
    number = {14},
    pages = {1457--1465},
    publisher = {Taylor \& Francis},
    title = {The equation of state of the hard-disc fluid revisited},
    volume = {107},
    year = {2009}
}

@article{nagamanasa2014experimental,
    author = {Nagamanasa, K Hima and Gokhale, Shreyas and Sood, AK and Ganapathy, Rajesh},
    doi = {10.1103/PhysRevE.89.062308},
    journal = {Physical Review E},
    number = {6},
    pages = {062308},
    publisher = {APS},
    title = {Experimental signatures of a nonequilibrium phase transition governing the yielding of a soft glass},
    volume = {89},
    year = {2014}
}

@article{neel2014dynamics,
    author = {N{\'e}el, Baptiste and Rondini, Ignacio and Turzillo, Alex and Mujica, Nicol{\'a}s and Soto, Rodrigo},
    doi = {10.1103/PhysRevE.89.042206},
    journal = {Physical Review E},
    number = {4},
    pages = {042206},
    publisher = {APS},
    title = {Dynamics of a first-order transition to an absorbing state},
    volume = {89},
    year = {2014}
}

@article{Ness2020,
    author = {Ness, Christopher and Cates, Michael E.},
    doi = {10.1103/PhysRevLett.124.088004},
    issue = {8},
    journal = {Phys. Rev. Lett.},
    month = {Feb},
    numpages = {6},
    pages = {088004},
    publisher = {American Physical Society},
    title = {Absorbing-State Transitions in Granular Materials Close to Jamming},
    url = {https://link.aps.org/doi/10.1103/PhysRevLett.124.088004},
    volume = {124},
    year = {2020}
}

@book{nise2020control,
    author = {Nise, Norman S},
    publisher = {John Wiley \& Sons},
    title = {Control systems engineering},
    year = {2020}
}

@article{odor2004universality,
    author = {{\'O}dor, G{\'e}za},
    journal = {Reviews of modern physics},
    number = {3},
    pages = {663--724},
    publisher = {APS},
    title = {Universality classes in nonequilibrium lattice systems},
    volume = {76},
    year = {2004}
}

@article{Olafsen98,
    author = {Olafsen, J. S. and Urbach, J. S.},
    doi = {10.1103/PhysRevLett.81.4369},
    issue = {20},
    journal = {Phys. Rev. Lett.},
    month = {Nov},
    numpages = {0},
    pages = {4369--4372},
    publisher = {American Physical Society},
    title = {Clustering, Order, and Collapse in a Driven Granular Monolayer},
    url = {https://link.aps.org/doi/10.1103/PhysRevLett.81.4369},
    volume = {81},
    year = {1998}
}

@article{omar2023mechanical,
    author = {Omar, Ahmad K and Row, Hyeongjoo and Mallory, Stewart A and Brady, John F},
    doi = {10.1073/pnas.2219900120},
    journal = {Proceedings of the National Academy of Sciences},
    number = {18},
    pages = {e2219900120},
    publisher = {National Acad Sciences},
    title = {Mechanical theory of nonequilibrium coexistence and motility-induced phase separation},
    volume = {120},
    year = {2023}
}

@article{onorato2023wave,
    author = {Onorato, Miguel and Lvov, Yuri V and Dematteis, Giovanni and Chibbaro, Sergio},
    doi = {10.1016/j.physrep.2023.09.006},
    journal = {Physics Reports},
    pages = {1--36},
    publisher = {Elsevier},
    title = {Wave Turbulence and thermalization in one-dimensional chains},
    volume = {1040},
    year = {2023}
}

@article{Osullivan2009,
    author = {Catherine O'Sullivan and Liang Cui},
    doi = {https://doi.org/10.1016/j.powtec.2009.03.003},
    issn = {0032-5910},
    journal = {Powder Technology},
    note = {Special Issue: Discrete Element Methods: The 4th International conference on Discrete Element Methods},
    number = {3},
    pages = {289-302},
    title = {Micromechanics of granular material response during load reversals: Combined DEM and experimental study},
    url = {https://www.sciencedirect.com/science/article/pii/S0032591009001661},
    volume = {193},
    year = {2009}
}

@article{pagonabarraga2001randomly,
    author = {Pagonabarraga, I and Trizac, E and Van Noije, TPC and Ernst, MH},
    doi = {10.1103/PhysRevE.65.011303},
    journal = {Physical Review E},
    number = {1},
    pages = {011303},
    publisher = {APS},
    title = {Randomly driven granular fluids: Collisional statistics and short scale structure},
    volume = {65},
    year = {2001}
}

@article{pal2015diffusion,
    author = {Pal, Arnab},
    doi = {10.1103/PhysRevE.91.012113},
    journal = {Physical Review E},
    number = {1},
    pages = {012113},
    publisher = {APS},
    title = {Diffusion in a potential landscape with stochastic resetting},
    volume = {91},
    year = {2015}
}

@article{pascual2005criticality,
    author = {Pascual, Mercedes and Guichard, Fr{\'e}d{\'e}ric},
    doi = {10.1016/j.tree.2004.11.012},
    journal = {Trends in ecology \& evolution},
    number = {2},
    pages = {88--95},
    publisher = {Elsevier},
    title = {Criticality and disturbance in spatial ecological systems},
    volume = {20},
    year = {2005}
}

@article{paul2017ballistic,
    author = {Paul, Subhajit and Das, Subir K},
    doi = {10.1103/PhysRevE.96.012105},
    journal = {Physical Review E},
    number = {1},
    pages = {012105},
    publisher = {APS},
    title = {Ballistic aggregation in systems of inelastic particles: Cluster growth, structure, and aging},
    volume = {96},
    year = {2017}
}

@article{PhysRevB.109.L020304,
    author = {O'Dea, Nicholas and Morningstar, Alan and Gopalakrishnan, Sarang and Khemani, Vedika},
    doi = {10.1103/PhysRevB.109.L020304},
    issue = {2},
    journal = {Phys. Rev. B},
    month = {Jan},
    numpages = {6},
    pages = {L020304},
    publisher = {American Physical Society},
    title = {Entanglement and absorbing-state transitions in interactive quantum dynamics},
    url = {https://link.aps.org/doi/10.1103/PhysRevB.109.L020304},
    volume = {109},
    year = {2024}
}

@article{PhysRevE.93.012110,
    author = {de Oliveira, Marcelo M. and Alves, Sidiney G. and Ferreira, Silvio C.},
    doi = {10.1103/PhysRevE.93.012110},
    issue = {1},
    journal = {Phys. Rev. E},
    month = {Jan},
    numpages = {7},
    pages = {012110},
    publisher = {American Physical Society},
    title = {Continuous and discontinuous absorbing-state phase transitions on Voronoi-Delaunay random lattices},
    url = {https://link.aps.org/doi/10.1103/PhysRevE.93.012110},
    volume = {93},
    year = {2016}
}

@article{PhysRevLett.114.110602,
    author = {Hexner, Daniel and Levine, Dov},
    doi = {10.1103/PhysRevLett.114.110602},
    issue = {11},
    journal = {Phys. Rev. Lett.},
    month = {Mar},
    numpages = {5},
    pages = {110602},
    publisher = {American Physical Society},
    title = {Hyperuniformity of Critical Absorbing States},
    url = {https://link.aps.org/doi/10.1103/PhysRevLett.114.110602},
    volume = {114},
    year = {2015}
}

@article{PhysRevLett.125.148001,
    author = {Wilken, Sam and Guerra, Rodrigo E. and Pine, David J. and Chaikin, Paul M.},
    doi = {10.1103/PhysRevLett.125.148001},
    issue = {14},
    journal = {Phys. Rev. Lett.},
    month = {Sep},
    numpages = {5},
    pages = {148001},
    publisher = {American Physical Society},
    title = {Hyperuniform Structures Formed by Shearing Colloidal Suspensions},
    url = {https://link.aps.org/doi/10.1103/PhysRevLett.125.148001},
    volume = {125},
    year = {2020}
}

@article{PhysRevLett.127.038002,
    author = {Wilken, Sam and Guerra, Rodrigo E. and Levine, Dov and Chaikin, Paul M.},
    doi = {10.1103/PhysRevLett.127.038002},
    issue = {3},
    journal = {Phys. Rev. Lett.},
    month = {Jul},
    numpages = {6},
    pages = {038002},
    publisher = {American Physical Society},
    title = {Random Close Packing as a Dynamical Phase Transition},
    url = {https://link.aps.org/doi/10.1103/PhysRevLett.127.038002},
    volume = {127},
    year = {2021}
}

@article{PhysRevLett.130.120402,
    author = {Sierant, Piotr and Turkeshi, Xhek},
    doi = {10.1103/PhysRevLett.130.120402},
    issue = {12},
    journal = {Phys. Rev. Lett.},
    month = {Mar},
    numpages = {8},
    pages = {120402},
    publisher = {American Physical Society},
    title = {Controlling Entanglement at Absorbing State Phase Transitions in Random Circuits},
    url = {https://link.aps.org/doi/10.1103/PhysRevLett.130.120402},
    volume = {130},
    year = {2023}
}

@article{PhysRevLett.130.207102,
    author = {Guislain, Laura and Bertin, Eric},
    doi = {10.1103/PhysRevLett.130.207102},
    issue = {20},
    journal = {Phys. Rev. Lett.},
    month = {May},
    numpages = {6},
    pages = {207102},
    publisher = {American Physical Society},
    title = {Nonequilibrium Phase Transition to Temporal Oscillations in Mean-Field Spin Models},
    url = {https://link.aps.org/doi/10.1103/PhysRevLett.130.207102},
    volume = {130},
    year = {2023}
}

@article{PhysRevLett.131.238202,
    author = {Wilken, Sam and Guo, Ashley Z. and Levine, Dov and Chaikin, Paul M.},
    doi = {10.1103/PhysRevLett.131.238202},
    issue = {23},
    journal = {Phys. Rev. Lett.},
    month = {Dec},
    numpages = {6},
    pages = {238202},
    publisher = {American Physical Society},
    title = {Dynamical Approach to the Jamming Problem},
    url = {https://link.aps.org/doi/10.1103/PhysRevLett.131.238202},
    volume = {131},
    year = {2023}
}

@article{PhysRevLett.132.264002,
    author = {Gom\'e, S\'ebastien and Rivi\`ere, Ali\'enor and Tuckerman, Laurette S. and Barkley, Dwight},
    doi = {10.1103/PhysRevLett.132.264002},
    issue = {26},
    journal = {Phys. Rev. Lett.},
    month = {Jun},
    numpages = {6},
    pages = {264002},
    publisher = {American Physical Society},
    title = {Phase Transition to Turbulence via Moving Fronts},
    url = {https://link.aps.org/doi/10.1103/PhysRevLett.132.264002},
    volume = {132},
    year = {2024}
}

@article{PhysRevLett.132.268203,
    author = {Jocteur, Tristan and Figueiredo, Shana and Martens, Kirsten and Bertin, Eric and Mari, Romain},
    doi = {10.1103/PhysRevLett.132.268203},
    issue = {26},
    journal = {Phys. Rev. Lett.},
    month = {Jun},
    numpages = {6},
    pages = {268203},
    publisher = {American Physical Society},
    title = {Yielding Is an Absorbing Phase Transition with Vanishing Critical Fluctuations},
    url = {https://link.aps.org/doi/10.1103/PhysRevLett.132.268203},
    volume = {132},
    year = {2024}
}

@article{PhysRevLett.89.190602,
    author = {van Wijland, Fr\'ed\'eric},
    issue = {19},
    journal = {Phys. Rev. Lett.},
    month = {Oct},
    numpages = {4},
    pages = {190602},
    publisher = {American Physical Society},
    title = {Universality Class of Nonequilibrium Phase Transitions with Infinitely Many Absorbing States},
    url = {https://link.aps.org/doi/10.1103/PhysRevLett.89.190602},
    volume = {89},
    year = {2002}
}

@article{PhysRevMaterials.8.104002,
    author = {Mkhonta, S. K. and Huang, Zhi-Feng and Elder, K. R.},
    doi = {10.1103/PhysRevMaterials.8.104002},
    issue = {10},
    journal = {Phys. Rev. Mater.},
    month = {Oct},
    numpages = {8},
    pages = {104002},
    publisher = {American Physical Society},
    title = {Liquid-substrate fluctuation effects on crystal growth and disordered hyperuniformity of two-dimensional materials},
    url = {https://link.aps.org/doi/10.1103/PhysRevMaterials.8.104002},
    volume = {8},
    year = {2024}
}

@article{PhysRevResearch.2.013318,
    author = {Buend\'{\i}a, Victor and di Santo, Serena and Villegas, Pablo and Burioni, Raffaella and Mu\~noz, Miguel A.},
    doi = {10.1103/PhysRevResearch.2.013318},
    issue = {1},
    journal = {Phys. Rev. Res.},
    month = {Mar},
    numpages = {10},
    pages = {013318},
    publisher = {American Physical Society},
    title = {Self-organized bistability and its possible relevance for brain dynamics},
    url = {https://link.aps.org/doi/10.1103/PhysRevResearch.2.013318},
    volume = {2},
    year = {2020}
}

@article{PhysRevResearch.2.043390,
    author = {Aron, Camille and Kulkarni, Manas},
    doi = {10.1103/PhysRevResearch.2.043390},
    issue = {4},
    journal = {Phys. Rev. Res.},
    month = {Dec},
    numpages = {10},
    pages = {043390},
    publisher = {American Physical Society},
    title = {Nonanalytic nonequilibrium field theory: Stochastic reheating of the Ising model},
    url = {https://link.aps.org/doi/10.1103/PhysRevResearch.2.043390},
    volume = {2},
    year = {2020}
}

@article{PhysRevResearch.3.013238,
    author = {Jo, Minjae and Lee, Jongshin and Choi, K. and Kahng, B.},
    doi = {10.1103/PhysRevResearch.3.013238},
    issue = {1},
    journal = {Phys. Rev. Res.},
    month = {Mar},
    numpages = {15},
    pages = {013238},
    publisher = {American Physical Society},
    title = {Absorbing phase transition with a continuously varying exponent in a quantum contact process: A neural network approach},
    url = {https://link.aps.org/doi/10.1103/PhysRevResearch.3.013238},
    volume = {3},
    year = {2021}
}

@article{Pieraski1983,
    author = {P. Piera{\'{n}}ski},
    doi = {10.1051/jphys:01983004405057300},
    journal = {Journal de Physique},
    number = {5},
    pages = {573--578},
    publisher = {{EDP} Sciences},
    title = {Jumping particle model. Period doubling cascade in an experimental system},
    url = {https://doi.org/10.1051/jphys:01983004405057300},
    volume = {44},
    year = {1983}
}

@article{Pieraski1985,
    author = {Pi. Piera{\'{n}}ski and Z. Kowalik and M. Franaszek},
    doi = {10.1051/jphys:01985004605068100},
    journal = {Journal de Physique},
    number = {5},
    pages = {681--686},
    publisher = {{EDP} Sciences},
    title = {Jumping particle model. A study of the phase space of a non-linear dynamical system below its transition to chaos},
    url = {https://doi.org/10.1051/jphys:01985004605068100},
    volume = {46},
    year = {1985}
}

@article{pine2005chaos,
    author = {Pine, David J and Gollub, Jerry P and Brady, John F and Leshansky, Alexander M},
    doi = {10.1038/nature04380},
    journal = {Nature},
    number = {7070},
    pages = {997--1000},
    publisher = {Nature Publishing Group UK London},
    title = {Chaos and threshold for irreversibility in sheared suspensions},
    volume = {438},
    year = {2005}
}

@article{pires2023tricritical,
    author = {Pires, Marcelo A and Sampaio Filho, Cesar IN and Herrmann, Hans J and Andrade Jr, Jos{\'e} S},
    doi = {10.1016/j.chaos.2023.113761},
    journal = {Chaos, Solitons \& Fractals},
    pages = {113761},
    publisher = {Elsevier},
    title = {Tricritical behavior in epidemic dynamics with vaccination},
    volume = {174},
    year = {2023}
}

@book{pitaevskii2012physical,
    author = {Pitaevskii, Lev P and Lifshitz, EM},
    publisher = {Butterworth-Heinemann},
    title = {Physical Kinetics: Volume 10},
    volume = {10},
    year = {2012}
}

@article{Plati2019,
    author = {Plati, A. and Baldassarri, A. and Gnoli, A. and Gradenigo, G. and Puglisi, A.},
    doi = {10.1103/PhysRevLett.123.038002},
    issue = {3},
    journal = {Phys. Rev. Lett.},
    month = {Jul},
    numpages = {5},
    pages = {038002},
    publisher = {American Physical Society},
    title = {Dynamical Collective Memory in Fluidized Granular Materials},
    url = {https://link.aps.org/doi/10.1103/PhysRevLett.123.038002},
    volume = {123},
    year = {2019}
}

@article{plati2021long,
    author = {Plati, Andrea and Puglisi, Andrea},
    doi = {10.1038/s41598-021-93091-1},
    journal = {Scientific Reports},
    number = {1},
    pages = {14206},
    publisher = {Nature Publishing Group UK London},
    title = {Long range correlations and slow time scales in a boundary driven granular model},
    volume = {11},
    year = {2021}
}

@article{Plati2022,
    author = {Plati, A. and Puglisi, A.},
    doi = {10.1103/PhysRevLett.128.208001},
    issue = {20},
    journal = {Phys. Rev. Lett.},
    month = {May},
    numpages = {6},
    pages = {208001},
    publisher = {American Physical Society},
    title = {Collective Drifts in Vibrated Granular Packings: The Interplay of Friction and Structure},
    url = {https://link.aps.org/doi/10.1103/PhysRevLett.128.208001},
    volume = {128},
    year = {2022}
}

@article{Plimpton2022,
    author = {Aidan P. Thompson and H. Metin Aktulga and Richard Berger and Dan S. Bolintineanu and W. Michael Brown and Paul S. Crozier and Pieter J. {in 't Veld} and Axel Kohlmeyer and Stan G. Moore and Trung Dac Nguyen and Ray Shan and Mark J. Stevens and Julien Tranchida and Christian Trott and Steven J. Plimpton},
    doi = {https://doi.org/10.1016/j.cpc.2021.108171},
    issn = {0010-4655},
    journal = {Computer Physics Communications},
    pages = {108171},
    title = {LAMMPS - a flexible simulation tool for particle-based materials modeling at the atomic, meso, and continuum scales},
    url = {https://www.sciencedirect.com/science/article/pii/S0010465521002836},
    volume = {271},
    year = {2022}
}

@book{PoeschelBook,
    address = {Berlin},
    author = {T. P\"oschel and T. Schwager},
    publisher = {Springer},
    title = {Computational Granular Dynamics},
    year = {2005}
}

@book{PopovBook,
    address = {Berlin},
    author = {V. L. Popov},
    doi = {10.1007/978-3-642-10803-7},
    publisher = {Springer-Verlag},
    title = {Contact Mechanics and Friction},
    year = {2010}
}

@article{pruessner2013average,
    author = {Pruessner, Gunnar},
    doi = {10.1142/S0217979213500094},
    journal = {International Journal of Modern Physics B},
    number = {05},
    pages = {1350009},
    publisher = {World Scientific},
    title = {The average avalanche size in the manna model and other models of self-organized criticality},
    volume = {27},
    year = {2013}
}

@misc{ptaszynski2025nonanalyticlandaufunctionalsshaping,
    archiveprefix = {arXiv},
    author = {Krzysztof Ptaszynski and Massimiliano Esposito},
    doi = {10.1103/PhysRevE.111.044142},
    eprint = {2502.06226},
    primaryclass = {cond-mat.stat-mech},
    title = {Nonanalytic Landau functionals shaping the finite-size scaling of fluctuations and response functions in and out of equilibrium},
    url = {https://arxiv.org/abs/2502.06226},
    year = {2025}
}

@article{puglisi2012structure,
    author = {Puglisi, Andrea and Gnoli, Andrea and Gradenigo, Giacomo and Sarracino, Alessandro and Villamaina, Dario},
    doi = {10.1063/1.3673876},
    journal = {The Journal of chemical physics},
    number = {1},
    publisher = {AIP Publishing},
    title = {Structure factors in granular experiments with homogeneous fluidization},
    volume = {136},
    year = {2012}
}

@article{regev2013onset,
    author = {Regev, Ido and Lookman, Turab and Reichhardt, Charles},
    doi = {10.1103/PhysRevE.88.062401},
    journal = {Physical Review E},
    number = {6},
    pages = {062401},
    publisher = {APS},
    title = {Onset of irreversibility and chaos in amorphous solids
under periodic shear},
    volume = {88},
    year = {2013}
}

@article{reichhardt2023reversible,
    author = {Reichhardt, Charles and Regev, Ido and Dahmen, K and Okuma, Satoshi and Reichhardt, Cynthia Jane Olson},
    doi = {10.1103/PhysRevResearch.5.021001},
    journal = {Physical Review Research},
    number = {2},
    pages = {021001},
    publisher = {APS},
    title = {Reversible to irreversible transitions in periodic driven many-body systems and future directions for classical and quantum systems},
    volume = {5},
    year = {2023}
}

@article{Reis2006,
    author = {Reis, P. M. and Ingale, R. A. and Shattuck, M. D.},
    doi = {10.1103/PhysRevLett.96.258001},
    issue = {25},
    journal = {Phys. Rev. Lett.},
    month = {Jun},
    numpages = {4},
    pages = {258001},
    publisher = {American Physical Society},
    title = {Crystallization of a Quasi-Two-Dimensional Granular Fluid},
    url = {https://link.aps.org/doi/10.1103/PhysRevLett.96.258001},
    volume = {96},
    year = {2006}
}

@article{risso2018effective,
    author = {Risso, Dino and Soto, Rodrigo and Guzm{\'a}n, Marcelo},
    doi = {10.1103/PhysRevE.98.022901},
    journal = {Physical Review E},
    number = {2},
    pages = {022901},
    publisher = {APS},
    title = {Effective two-dimensional model for granular matter with phase separation},
    volume = {98},
    year = {2018}
}

@article{rivas2011sudden,
    author = {Rivas, Nicol{\'a}s and Ponce, Suomi and Gallet, Basile and Risso, Dino and Soto, Rodrigo and Cordero, Patricio and Mujica, Nicol{\'a}s},
    doi = {10.1103/PhysRevLett.106.088001},
    journal = {Physical review letters},
    number = {8},
    pages = {088001},
    publisher = {APS},
    title = {Sudden chain energy transfer events in vibrated granular media},
    volume = {106},
    year = {2011}
}

@article{Rossi2000,
    author = {Rossi, Michela and Pastor-Satorras, Romualdo and Vespignani, Alessandro},
    doi = {10.1103/PhysRevLett.85.1803},
    issue = {9},
    journal = {Phys. Rev. Lett.},
    month = {Aug},
    numpages = {0},
    pages = {1803--1806},
    publisher = {American Physical Society},
    title = {Universality Class of Absorbing Phase Transitions with a Conserved Field},
    url = {https://link.aps.org/doi/10.1103/PhysRevLett.85.1803},
    volume = {85},
    year = {2000}
}

@article{salvalaglio2024persistent,
    author = {Salvalaglio, Marco and Skinner, Dominic J and Dunkel, J{\"o}rn and Voigt, Axel},
    doi = {10.1103/PhysRevResearch.6.023107},
    journal = {Physical Review Research},
    number = {2},
    pages = {023107},
    publisher = {APS},
    title = {Persistent homology and topological statistics of hyperuniform point clouds},
    volume = {6},
    year = {2024}
}

@article{sarracino2010granular,
    author = {Sarracino, Alessandro and Villamaina, Dario and Costantini, Giulio and Puglisi, Andrea},
    doi = {10.1088/1742-5468/2010/04/P04013},
    journal = {Journal of Statistical Mechanics: Theory and Experiment},
    number = {04},
    pages = {P04013},
    publisher = {IOP Publishing},
    title = {Granular brownian motion},
    volume = {2010},
    year = {2010}
}

@article{sarracino2010irreversible,
    author = {Sarracino, Alessandro and Villamaina, Dario and Gradenigo, Giacomo and Puglisi, Andrea},
    doi = {10.1209/0295-5075/92/34001},
    journal = {Europhysics Letters},
    number = {3},
    pages = {34001},
    publisher = {IOP Publishing},
    title = {Irreversible dynamics of a massive intruder in dense granular fluids},
    volume = {92},
    year = {2010}
}

@phdthesis{schindler2023phase,
    author = {Schindler, Thomas},
    school = {Friedrich-Alexander-Universit{\"a}t Erlangen-N{\"u}rnberg (FAU)},
    title = {Phase behavior and perturbation response of quasi-2D driven granulates employing molecular dynamics simulations},
    year = {2023}
}

@article{schlogl1972chemical,
    author = {Schl{\"o}gl, Friedrich},
    journal = {Zeitschrift f{\"u}r physik},
    number = {2},
    pages = {147--161},
    publisher = {Springer},
    title = {Chemical reaction models for non-equilibrium phase transitions},
    volume = {253},
    year = {1972}
}

@article{simha2002hydrodynamic,
    author = {Simha, R Aditi and Ramaswamy, Sriram},
    journal = {Physical review letters},
    number = {5},
    pages = {058101},
    publisher = {APS},
    title = {Hydrodynamic fluctuations and instabilities in ordered suspensions of self-propelled particles},
    volume = {89},
    year = {2002}
}

@article{singh2020resetting,
    author = {Singh, RK and Metzler, R and Sandev, T},
    doi = {10.1088/1751-8121/abc83a},
    journal = {Journal of Physics A: Mathematical and Theoretical},
    number = {50},
    pages = {505003},
    publisher = {IOP Publishing},
    title = {Resetting dynamics in a confining potential},
    volume = {53},
    year = {2020}
}

@article{smallenburg2022efficient,
    author = {Smallenburg, Frank},
    doi = {10.1140/epje/s10189-022-00180-8},
    journal = {The European Physical Journal E},
    number = {3},
    pages = {22},
    publisher = {Springer},
    title = {Efficient event-driven simulations of hard spheres},
    volume = {45},
    year = {2022}
}

@article{solon2015pressure,
    author = {Solon, Alexandre P and Stenhammar, Joakim and Wittkowski, Raphael and Kardar, Mehran and Kafri, Yariv and Cates, Michael E and Tailleur, Julien},
    doi = {10.1103/PhysRevLett.114.198301},
    journal = {Physical review letters},
    number = {19},
    pages = {198301},
    publisher = {APS},
    title = {Pressure and phase equilibria in interacting active Brownian spheres},
    volume = {114},
    year = {2015}
}

@article{solon2018generalized,
    author = {Solon, Alexandre P and Stenhammar, Joakim and Cates, Michael E and Kafri, Yariv and Tailleur, Julien},
    doi = {10.1088/1367-2630/aaccdd},
    journal = {New Journal of Physics},
    number = {7},
    pages = {075001},
    publisher = {IOP Publishing},
    title = {Generalized thermodynamics of motility-induced phase separation: phase equilibria, Laplace pressure, and change of ensembles},
    volume = {20},
    year = {2018}
}

@article{spohn1983long,
    author = {Spohn, Herbert},
    doi = {10.1088/0305-4470/16/18/029},
    journal = {Journal of Physics A: Mathematical and General},
    number = {18},
    pages = {4275},
    publisher = {IOP Publishing},
    title = {Long range correlations for stochastic lattice gases in a non-equilibrium steady state},
    volume = {16},
    year = {1983}
}

@article{spohn2014nonlinear,
    author = {Spohn, Herbert},
    doi = {10.1007/s10955-014-0933-y},
    journal = {Journal of Statistical Physics},
    pages = {1191--1227},
    publisher = {Springer},
    title = {Nonlinear fluctuating hydrodynamics for anharmonic chains},
    volume = {154},
    year = {2014}
}

@article{spohn2016fluctuating,
    author = {Spohn, Herbert},
    doi = {10.1007/978-3-319-29261-8_3},
    journal = {Thermal Transport in Low Dimensions: From Statistical Physics to Nanoscale Heat Transfer},
    pages = {107--158},
    publisher = {Springer},
    title = {Fluctuating hydrodynamics approach to equilibrium time correlations for anharmonic chains},
    year = {2016}
}

@article{takeuchi2007directed,
    author = {Takeuchi, Kazumasa A and Kuroda, Masafumi and Chat{\'e}, Hugues and Sano, Masaki},
    doi = {10.1103/PhysRevLett.99.234503},
    journal = {Physical review letters},
    number = {23},
    pages = {234503},
    publisher = {APS},
    title = {Directed percolation criticality in turbulent liquid crystals},
    volume = {99},
    year = {2007}
}

@article{tjhung2015hyperuniform,
    author = {Tjhung, Elsen and Berthier, Ludovic},
    doi = {10.1103/PhysRevLett.114.148301},
    journal = {Physical review letters},
    number = {14},
    pages = {148301},
    publisher = {APS},
    title = {Hyperuniform density fluctuations and diverging dynamic correlations in periodically driven colloidal suspensions},
    volume = {114},
    year = {2015}
}

@article{tjhung2016criticality,
    author = {Tjhung, Elsen and Berthier, Ludovic},
    doi = {10.1088/1742-5468/2016/03/033501},
    journal = {Journal of Statistical Mechanics: Theory and Experiment},
    number = {3},
    pages = {033501},
    publisher = {IOP Publishing},
    title = {Criticality and correlated dynamics at the irreversibility transition in periodically driven colloidal suspensions},
    volume = {2016},
    year = {2016}
}

@article{torquato2016hyperuniformity,
    author = {Torquato, Salvatore},
    doi = {10.1103/PhysRevE.94.022122},
    journal = {Physical Review E},
    number = {2},
    pages = {022122},
    publisher = {APS},
    title = {Hyperuniformity and its generalizations},
    volume = {94},
    year = {2016}
}

@article{torquato2018hyperuniform,
    author = {Torquato, Salvatore},
    doi = {10.1016/j.physrep.2018.03.001},
    journal = {Physics Reports},
    pages = {1--95},
    publisher = {Elsevier},
    title = {Hyperuniform states of matter},
    volume = {745},
    year = {2018}
}

@book{van1992stochastic,
    author = {Van Kampen, Nicolaas Godfried},
    doi = {10.1063/1.2915501},
    publisher = {Elsevier},
    title = {Stochastic processes in physics and chemistry},
    volume = {1},
    year = {1992}
}

@article{van1998ring,
    author = {Van Noije, TPC and Ernst, MH and Brito, Ricardo},
    journal = {Physica A: Statistical Mechanics and its Applications},
    number = {1-2},
    pages = {266--283},
    publisher = {Elsevier},
    title = {Ring kinetic theory for an idealized granular gas},
    volume = {251},
    year = {1998}
}

@article{van1998velocity,
    author = {Van Noije, TPC and Ernst, MH},
    doi = {10.1007/s100350050009},
    journal = {Granular Matter},
    number = {2},
    pages = {57--64},
    publisher = {Springer},
    title = {Velocity distributions in homogeneous granular fluids: the free and the heated case},
    volume = {1},
    year = {1998}
}

@article{van1998wilson,
    author = {Van Wijland, F and Oerding, K and Hilhorst, HJ},
    journal = {Physica A: Statistical Mechanics and its Applications},
    number = {1-2},
    pages = {179--201},
    publisher = {Elsevier},
    title = {Wilson renormalization of a reaction--diffusion process},
    volume = {251},
    year = {1998}
}

@article{van1999randomly,
    author = {Van Noije, TPC and Ernst, MH and Trizac, Emmanuel and Pagonabarraga, I},
    doi = {10.1103/PhysRevE.59.4326},
    journal = {Physical Review E},
    number = {4},
    pages = {4326},
    publisher = {APS},
    title = {Randomly driven granular fluids: Large-scale structure},
    volume = {59},
    year = {1999}
}

@incollection{van2001kinetic,
    author = {van Noije, Twan PC and Ernst, Matthieu H},
    booktitle = {Granular Gases},
    doi = {10.1007/3-540-44506-4_1},
    pages = {3--30},
    publisher = {Springer},
    title = {Kinetic theory of granular gases},
    year = {2001}
}

@article{vazquez2001fluctuating,
    adsurl = {https://ui.adsabs.harvard.edu/abs/2001JNET...26..279V},
    author = {{V{\'a}zquez}, F. and {Haro}, M. L{\'o}pez de},
    doi = {10.1515/JNETDY.2001.020},
    journal = {Journal of Non Equilibrium Thermodynamics},
    month = {September},
    number = {3},
    pages = {279-290},
    title = {{Fluctuating Hydrodynamics and Irreversible Thermodynamics}},
    volume = {26},
    year = {2001}
}

@article{vellela2009stochastic,
    author = {Vellela, Melissa and Qian, Hong},
    doi = {10.1098/rsif.2008.0476},
    journal = {Journal of The Royal Society Interface},
    number = {39},
    pages = {925--940},
    publisher = {The Royal Society},
    title = {Stochastic dynamics and non-equilibrium thermodynamics of a bistable chemical system: the Schl{\"o}gl model revisited},
    volume = {6},
    year = {2009}
}

@article{visco2008non,
    author = {Visco, Paolo and van Wijland, Fr{\'e}d{\'e}ric and Trizac, Emmanuel},
    doi = {10.1021/jp800333h},
    journal = {The Journal of Physical Chemistry B},
    number = {17},
    pages = {5412--5415},
    publisher = {ACS Publications},
    title = {Non-Poissonian statistics in a low-density fluid},
    volume = {112},
    year = {2008}
}

@article{Weijs2015,
    author = {Weijs, Joost H. and Jeanneret, Rapha\"el and Dreyfus, R\'emi and Bartolo, Denis},
    doi = {10.1103/PhysRevLett.115.108301},
    issue = {10},
    journal = {Phys. Rev. Lett.},
    month = {Sep},
    numpages = {5},
    pages = {108301},
    publisher = {American Physical Society},
    title = {Emergent Hyperuniformity in Periodically Driven Emulsions},
    url = {https://link.aps.org/doi/10.1103/PhysRevLett.115.108301},
    volume = {115},
    year = {2015}
}

@article{wiese2024hyperuniformity,
    author = {Wiese, Kay J{\"o}rg},
    doi = {10.1103/PhysRevLett.133.067103},
    journal = {Physical Review Letters},
    number = {6},
    pages = {067103},
    publisher = {APS},
    title = {Hyperuniformity in the manna model, conserved directed percolation and depinning},
    volume = {133},
    year = {2024}
}

@article{yu2020velocity,
    author = {Yu, Peidong and Schr{\"o}ter, Matthias and Sperl, Matthias},
    doi = {10.1103/PhysRevLett.124.208007},
    journal = {Physical Review Letters},
    number = {20},
    pages = {208007},
    publisher = {APS},
    title = {Velocity distribution of a homogeneously cooling granular gas},
    volume = {124},
    year = {2020}
}

@article{zakine2023minimum,
    author = {Zakine, Ruben and Vanden-Eijnden, Eric},
    doi = {10.1103/PhysRevX.13.041044},
    journal = {Physical Review X},
    number = {4},
    pages = {041044},
    publisher = {APS},
    title = {Minimum-Action Method for Nonequilibrium Phase Transitions},
    volume = {13},
    year = {2023}
}

@article{Zhang2005,
    author = {Zhang, H. P. and Makse, H. A.},
    doi = {10.1103/PhysRevE.72.011301},
    issue = {1},
    journal = {Phys. Rev. E},
    month = {Jul},
    numpages = {12},
    pages = {011301},
    publisher = {American Physical Society},
    title = {Jamming transition in emulsions and granular materials},
    url = {https://link.aps.org/doi/10.1103/PhysRevE.72.011301},
    volume = {72},
    year = {2005}
}

@article{ziff1986kinetic,
    author = {Ziff, Robert M and Gulari, Erdagon and Barshad, Yoav},
    doi = {10.1103/PhysRevLett.56.2553},
    journal = {Physical review letters},
    number = {24},
    pages = {2553},
    publisher = {APS},
    title = {Kinetic phase transitions in an irreversible surface-reaction model},
    volume = {56},
    year = {1986}
}

\appendix

\section{Argument and potential derivation of the overpopulated tail of the velocity distribution}
\label{sec: appendix brilliantov}

We follow Refs.~\cite{brilliantov2004kinetic, van1998velocity, esipov1997granular} to derivate an overpopulated tail in granular medium. We adapt their argument in order to potentially explain the overpopulated tail observed close to the continuous transition in Fig.~\ref{fig:Fig9}. We note however that the physical mechanism leading to this overpopulated tail is different in the simple granular cooling and in our case.

We start from the Boltzmann equation in our system Eq.~\eqref{eq: boltzmann} and focus on large velocities $v_i\gg \Delta$. At high velocities, the only terms that will contribute to the probability change are the viscous damping drift $\gamma\partial/\partial\bm v \cdot (\bm v f)$ and the collisional gain term (the first term in Eq.~\eqref{eq: collisional part boltzmann}). The collisional loss term is negligible, as particles primarily cool down due to drag rather than collisions. Hence, in the steady state, assuming homogeneity, we can write down the following approximated Boltzmann equation:
\begin{equation}
    \begin{split}
    -\gamma\dfrac{\partial }{\partial \bm v_i}\cdot (\bm v_if)\simeq& \chi \sigma\int \dd \bm v_j\int \dd\hat{\bm\sigma}_{ij}\Theta(\bm {v}_{ij}\cdot \hat{ \bm\sigma}_{ij} - 2\Delta)\times\\&(\bm {v}_{ij}\cdot \hat{ \bm\sigma}_{ij} - 2\Delta)\dfrac{f(\bm v_j'')f(\bm v_i'')}{\alpha^2}
    \end{split}
    \label{eq: approximation boltzmann}
\end{equation}
We proceed by noting that since the considered particle $i$ has a large velocity $v_i$, it will on average collide with particles of smaller velocity. Consequently, $\bm{v}_{ij}\simeq \bm v_i$ and $\bm{v}_{ij}\cdot \hat{\bm\sigma}_{ij}-2\Delta\simeq \bm{v}_{i}\cdot \hat{\bm\sigma}_{ij}$. For ease of theoretical computation and to deal with the post collisional velocities, with focus on 1D systems where we do not expect any differences for the velocity distribution in simulations. We further simplify the problem by setting $\alpha = 1$ since the same overpopulated tail is also obtained in an elastic system \cite{lei2019hydrodynamics}. These assumptions for positive $v_i$ lead to:
\begin{equation}
    -\gamma\dfrac{\partial }{\partial v_i}( v_if)\simeq 2\chi \sigma(v_if(v_i)),
\end{equation}
which admits as solution:
\begin{equation}
    f(v)=Z e^{\displaystyle{-\beta v}-\log(v)}
    \label{eq: tail}
\end{equation}
with $\beta=2\chi\sigma/\gamma$ and $Z$ an integration constant.
Eq.~\eqref{eq: tail} demonstrates the potential development of an exponential overpopulated tail. However, a rigorous numerical and theoretical analysis are necessary to validate our approximations. Notably, the assumption of homogeneity is uncontrolled, as we saw that the population of active particle has a drastic effect on the global temperature of the system. Moreover, 1D systems are known to exhibit peculiar thermalization \cite{spohn2014nonlinear, spohn2016fluctuating, eltohfa2024simulations, movsko1995thermalization, onorato2023wave}. For example, a system of 1D hard elastic particles would never reach equilibrium because the collision rule would simply exchange the velocities of the colliding particles. A Gaussian velocity distribution would never be reached and the initial distribution would be conserved. With our non-equilibrium collision rule, a 1D fluid can equilibrate. However, there might still be some differences compared to the $2D$ case, particularly regarding the efficiency of the energy spreading.

Verifying our proposition is not the main concern of the paper, we thus left this investigation for further studies.

\section{\texorpdfstring{Critical hyperuniform scaling for $\bm{S}_{u^\parallel u^\parallel}(k)$ and $\bm{S}_{u^\perp u^\perp}(k)$}{Critical hyperuniform scaling for S parallel and S perp}}
\label{sec: new hyperuniform scaling}

\begin{figure}[!ht]
\centering
\includegraphics[width=0.99\columnwidth,clip=true]{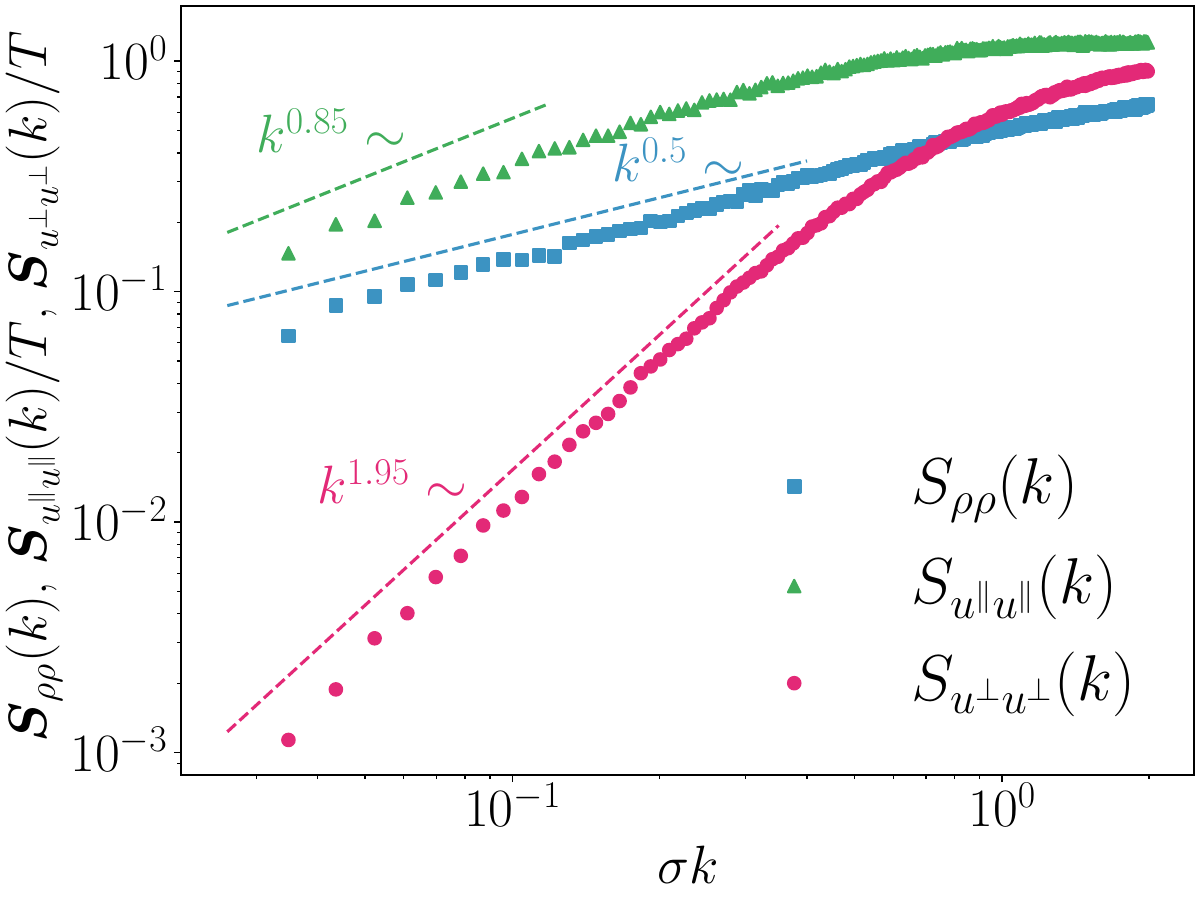}
\caption{$\bm S_{\rho\rho}(k)$, $\bm S_{u^\parallel u^\parallel}(k)$ and $\bm S_{u^\perp u^\perp}(k)$ as a function of $k$ close to the critical point. $N = 10^5$, $\alpha = 0.95$, $\Delta/(\sigma\gamma)=1.5$ and $\phi = 0.1512$.}\label{fig: critical vel struc}
\end{figure}

In Fig.~\ref{fig: critical vel struc},  we present the well-established hyperuniform scaling of the structure factor $\bm S_{\rho\rho}(k)$ at the critical point. More unexpectedly, we also observe hyperuniform scaling for the longitudinal and transverse velocity correlation functions. Specifically, the longitudinal velocity correlation function decays as a power law $\sim k^{0.85}$ over an appreciable range of $k$. One might have anticipated that, near the critical packing fraction $\phi_c$, the only relevant fields would be the number of active particles and the density, leading to trivial velocity correlations. However, this is not the case. Remarkably, the transverse velocity field also exhibits hyperuniform scaling, decaying approximately as $\sim k^{1.95}$, which is consistent with the expected $\sim k^2$ scaling induced by the interplay between a global damping and a noise conserving the center of mass of the system as described in Sec.~\ref{sec: hydro}. In the linear regime — expected to break down near the transition — this scaling is indeed anticipated for the transverse velocity field.

\section{Hyperuniform and clustered metastable fluids}
\label{sec: appendix ran ni}

\begin{figure}[!ht]
\centering
\includegraphics[width=0.99\columnwidth,clip=true]{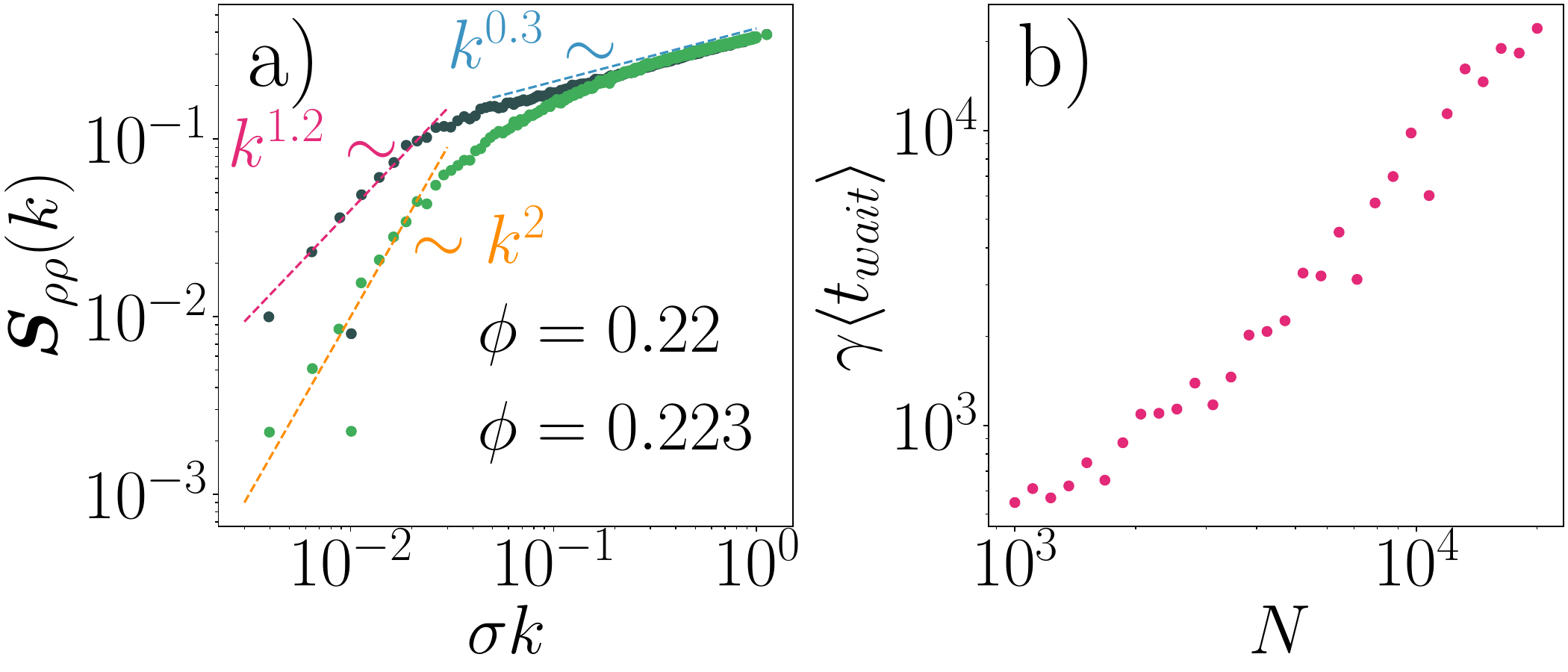}
\caption{a) Structure factor as a function of $k$ for our system with synchronization but with $\alpha = 1$ at $\phi= 0.22$, $N =10^6$ (very close to transition packing fraction)  and $\phi = 0.223$, $N=4\times 10^5$ (almost stable). b) Average waiting time for the metastable state ($\phi = 0.22$) to reach the absorbing state as a function of the number of particle in the system. The average is performed over 20 independent runs. $1/(\gamma\tau_s) = 4.33$ and $\Delta/(\gamma\sigma)=1.33$.}\label{fig: ranni}
\end{figure}

The model presented in Ref.~\cite{lei2021barrier, lei2023does} is similar to ours. While it does not contain synchronization, the collision dynamics is effectively similar to ours. In their model, if two particles collide with an energy smaller than some threshold, then the particles undergo elastic collision, otherwise they gain a finite amount of energy. A viscous drag is also applied during the free flight. In our model, particles that do not have a lot of energy are, on average, those which traveled for a long time  and are slowed down due to drag. Hence, they are likely to be synchronized and will therefore undergo collisions with $\Delta = 0$. This might mimic the effect of the energy-activated active collision of the models studied by Lei et al. Thus, the rough phenomenology of both models is similar even though Refs. \onlinecite{lei2021barrier, lei2023does} do not consider dissipative collisions, and drag is \textit{de facto} their only source of dissipation. Interestingly, they find that the pathway of the metastable fluid to the absorbing state is seemingly not driven by a nucleation process but by large wavelength fluctuations which are suppressed due to the non-equilibrium nature of the system. Indeed, they find that the metastable fluid has suppressed density fluctuations. On this basis, they argue that the non-equilibrium equivalent of the equilibrium metastable state becomes kinetically stable in the thermodynamic limit, due to the suppression of large wavelength fluctuations. The shrinking of the metastable region with system size is also observed using the minimal field theory of the conserved directed percolation universality class, modified to obtain a first order transition \cite{di2016self, PhysRevResearch.2.013318}. This is at odds with our results, as well as what was found in Ref.~\onlinecite{neel2014dynamics}. We believe the discrepancy is caused by the presence of dissipative collisions in our system that induces the formation of clusters, which enhance long wavelength fluctuations and allow for the regular nucleation pathway to the transition.
Indeed, in contrast to Ref.~\onlinecite{lei2023does}, where dissipation arises solely from viscous drag, our model with $\alpha < 1$ includes dissipative collisions, which are responsible for driving the system to the absorbing state. The purely dissipative collisions between synchronized particles cause clustering, therefore inducing a peak at a wavevector $k^*$ in the structure factor of the metastable fluid. These clusters appear to aid the system in reaching an absorbing state via a mechanism akin to equilibrium nucleation. Indeed, the system size $L^*$ at which a sharp drop in waiting time is found in Fig.~\ref{fig: waiting Time} corresponds roughly to $2\pi/k^*$. However, it remains highly unclear why such phenomena facilitate the transition.

We now turn our analysis to a metastable fluid with a discontinuous transition but with $\alpha=1$. In Fig.~\ref{fig: ranni}a), we present the structure factor in the metastable regime (close to the discontinuous transition) in our model, but with $\alpha = 1$ where collisions are elastic, and therefore clustering does not occur. In contrast to the metastable fluid with $\alpha\neq 1$ (Fig.~\ref{fig:Fig9}), we do not observe any peak in the structure factor. Instead, a hyperuniform scaling close to $k^{1.2}$ (but not incompatible with $k^2$ scaling at lower $k$) is visible for the metastable fluid ($\phi = 0.22$), in agreement with results from Ref.~\onlinecite{lei2023does}. We also note that a very long $k^{0.3}$ scaling is observed close to the metastable fluid. Additionally, the average waiting time $\langle t_{wait}\rangle$ increases monotonically with system size (or $N$) as given in Fig.~\ref{fig: ranni}b). This is in contrast with the non-monotonic behavior observed in Fig.~\ref{fig: waiting Time}. Thus, when dissipative collisions are absent, our system behaves similarly to the one studied in Ref.~\onlinecite{lei2023does}, allowing for a hyperuniform scaling to emerge in the metastable fluid. This scaling seemingly prohibits nucleation in the limit $N\to \infty$.

\vspace{1em}

\section{Tricritical point and continuous transition with synchronization}
\label{sec: appendix tricritical}

\begin{figure}
    \centering
    \includegraphics[width=0.99\columnwidth]{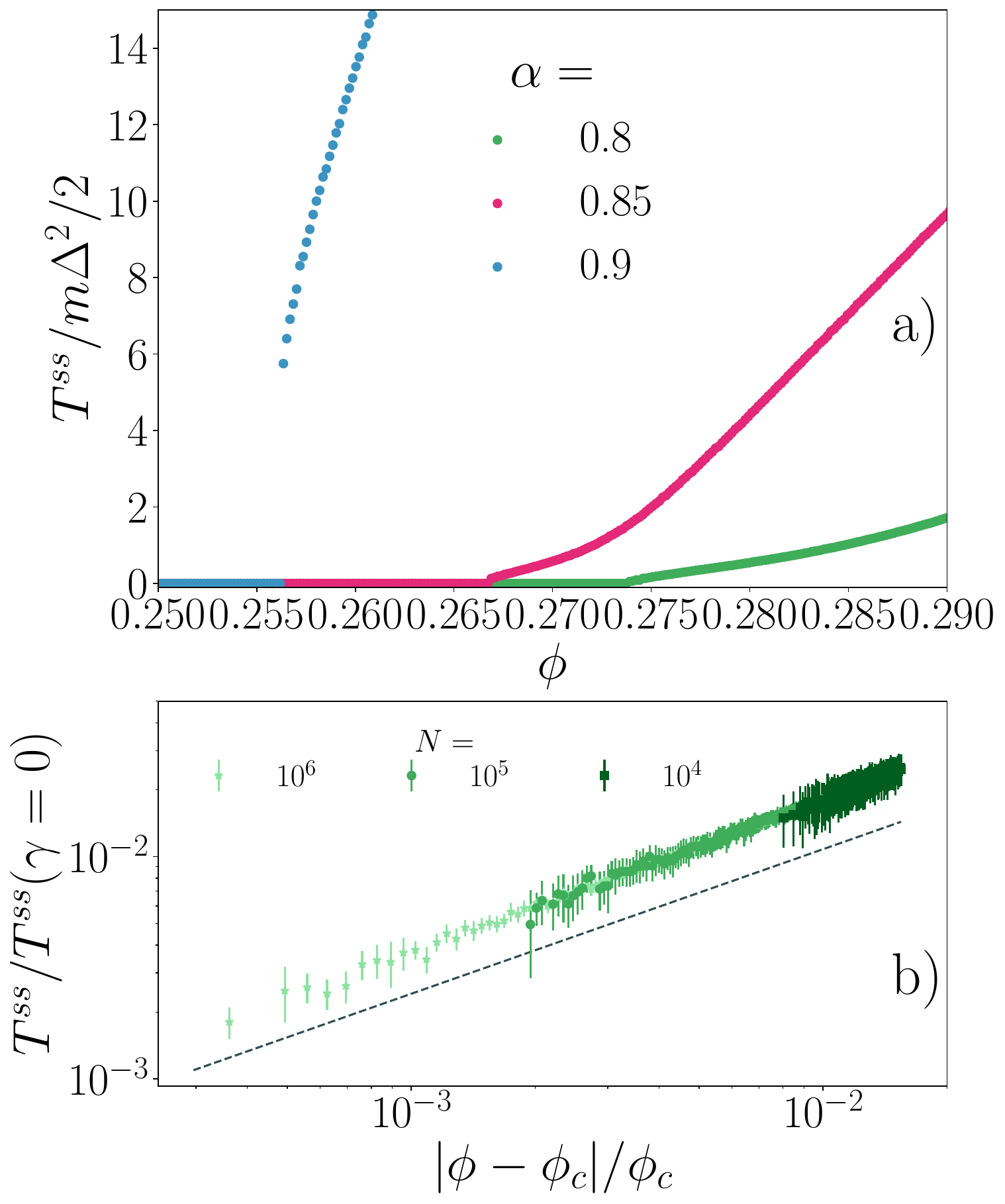}
    \caption{a) Transition from the discontinuous to continuous transition with synchronization as $\alpha$ is decreased: $N = 10000$, $\Delta/(\gamma\sigma) = 2.5$ and $1/\gamma\tau_s=25$. b) Is a close-up on the transition for $\alpha = 0.8$ with a scaling analysis showing that the last active steady state $\Delta T_{ss}$ goes to 0 with system size, proving the continuity of the transition. The dashed line is the expected power law with the exponent from the conserved directed percolation universality class.}
    \label{fig: tricritical}
\end{figure}

As discussed in the main text, even with synchronization, we can observe a continuous transition because the synchronization time competes with a finite mean collision time in the active region. Thus, at a large but finite synchronization time $\tau_s$, significantly longer than the minimum mean collision time in active regions, synchronization will not influence the transition, which remains continuous. A less straightforward example of a continuous transition at finite $\tau_s$ occurs when $\alpha \ll 1$ as illustrated by Fig.~\ref{fig: tricritical}a. Indeed, we observe that by decreasing $\alpha$, the transition becomes continuous. The continuity of the transition is made explicit in Fig.~\ref{fig: tricritical}b where a scale analysis is performed for $\alpha = 0.8$ and shows that the temperature of the last active steady state tends to approach 0 with increasing system size. While the theory cannot predict quantitatively this transition, it still gives a qualitative answer. As evidenced from Eq.~\eqref{eq: K52}, in the case of small $\alpha$, a discontinuous transition is theoretically predicted to occur at a small $T_a$. However, the population of inactive particles will reach 0 before the theoretical discontinuous transition happens, leading to a continuous transition instead. Similar transitions have been investigated in sandpiles belonging to the conserved directed percolation universality class \cite{di2016self}. In those systems, the mean field approximation has been shown to quantitatively fail considerably around the passage from the discontinuous to continuous transition, notably, it has been found that the continuous transition for a range of parameters, is fluctuation induced.

This behavior implies the existence of a non-equilibrium tricritical point in our system. Such a phenomenon has been documented and thoroughly investigated in a closely related system in Ref.~\onlinecite{lei2021barrier}. It would be interesting to observe a tricritical point in the realistic system as well. It will require further exploration, for example through the use of very inelastic beads such as plastic ones. Note that a tricritical point as been observed as well for the liquid-solid phase transition of a quasi-2D granular gas \cite{guzman2018critical, castillo2012fluctuations, castillo2015universality}.

\section{Expression for the transport coefficients and numerical values}
\label{sec: appendix transport}

As mentioned in the main text, we use the simple Enskog's equilibrium transport coefficients. They should give reasonable results in the active state. Moreover, while our simulations are mostly done at $\phi = 0.1$, the theory works well at higher densities (see Fig.~\ref{fig: adiabatic}, where $\phi = 0.5$ for instance). Here, we provide a list of the expressions used. They are derived assuming molecular chaos and a Gaussian velocity distribution. We will also use the equilibrium contact value of the pair correlation function \cite{mulero2009equation}:
\begin{equation}
    \chi(\phi) = \dfrac{1 - 7\phi/16 - \phi^3/20}{(1-\phi)^2}.
\end{equation}
The pressure is given by \cite{garzo2018enskog}:
\begin{equation}
    p = \rho T(1 + (1+\alpha)\phi \chi(\phi))+8 \sqrt{\pi m}\phi\rho \chi(\phi)\Delta\sqrt{T}.
\end{equation}
The transport coefficients we used are the equilibrium ones from Enskog's theory \cite{gradenigo2011fluctuating}:
\begin{equation}
    \begin{split}
     \eta &= \frac{1}{2\sigma}\sqrt{\frac{mT}{\pi}}\left(\chi^{-1} + 
   2\phi + (1 + 8/\pi)\chi \phi^2\right)\\
   \zeta &= \dfrac{8\phi^2}{\sigma\pi}\chi\sqrt{\frac{mT}{\pi}}\\
   \kappa &= \dfrac{2}{\sigma}\sqrt{\frac{T}{m\pi}}\left(\chi^{-1} + 3\phi + (9/4 + 4/\pi)\chi\phi^2\right)\\
   \mu &= 0\\
   \nu &= 0
    \end{split}
\end{equation}
For the transport coefficients taking into account $\Delta$ and $\alpha$, see Ref.~\onlinecite{garzo2018enskog}. The transport coefficient with $\Delta+\gamma$ are not known and should be derived from the Chapman-Enskog procedure.

\section{\texorpdfstring{Adiabatic slaving of the temperature field: $\bm{M}^{(r)}$ and theoretical correlation functions}{Adiabatic slaving of the temperature field: M(r) and theoretical correlation functions}}
\label{sec: appendix adiabatic slaving delta model}

As already stated in the main text, we can find exact theoretical expressions for the eigenvalues of $\bm M$ or the dynamic and static correlation functions through Eqs.~\eqref{eq: static corr} and \eqref{eq: dynamic corr}. However, their expressions are not particularly useful since they are third order polynomial roots or made of ratio of polynomials of degree up to $k^6$. Therefore, it becomes necessary to perform approximations to make the computations more tractable and the results clearer. One way is to integrate the temperature field by assuming that it varies quicker than any other field. This assumption is exact at $k=0$ since the others field are conserved and hence, have a relaxation timescale diverging with $1/k$. However, at larger $k$, this assumption will not be as good. Moreover, when $\gamma\neq 0$, the momentum field is also a fast field at small $k$. In what follow, we will only integrate the temperature field following Ref.~\onlinecite{brito2013hydrodynamic}, in order to obtain a hydrodynamic description with only the velocity and density fields. 

From Eq.~\eqref{eq: hydro matrix}, the evolution for the temperature field reads (we again drop the explicit evaluation of the derivative at the homogeneous state and the small transport coefficients $\mu_0$ and $\nu_0$):

\begin{equation}
    \delta \dot T(t) =-\Gamma\delta T(t) + \mathcal S(t),
\end{equation}
with:
\begin{align}
    \Gamma(k) &= -G_T + \dfrac{\kappa_0}{n_0 }k^2, \\
     \mathcal S(t) &= G_\rho \delta \rho -i k\dfrac{p_0}{n_0 }\delta u_\parallel+ \Xi_T(t).
\end{align}
where, $\Xi_T$ the noise on the temperature field defined in Eq.~\eqref{eq: noise correlation}. 
A very crude approximation would be to set $\delta \dot T = 0$. However, we can go further by formally solving this equation, taking the disturbance due to $\delta \rho$ and $\delta u_\parallel$ as sources $S \mathcal (t)$ for $\delta T$ which is damped at a frequency $\Gamma$. Its solution is:

\begin{equation}
    \delta T(t) = \int_{-\infty}^t \dd t'e^{\Gamma (t'-t) } \mathcal S(t'),
    \label{eq: solution temperature field}
\end{equation}
with $\Gamma = -G_T + \kappa_0k^2/n_0>0$ is an inverse relaxation time for the temperature field and where we assumed $\delta T(-\infty) =0$ without loss of generality since we are only concerned about the long time limit. We could then replace this solution for $\delta T(t)$ in the hydrodynamic equation given by $\dot{\bm \Psi} = \bm M \bm \Psi$. We would have eliminated the temperature field exactly, at the price of having to deal with a non-Markovian system with an exponential kernel. At that point, it is possible to indeed follow this route, integrate by parts the equations and eventually perform an expansion on the resulting equation for the density and velocity. However, we follow Ref.~\onlinecite{brito2013hydrodynamic} and immediately perform the adiabatic expansion on Eq.~\eqref{eq: solution temperature field} by performing successive integration by parts to expand the integral in powers of $\Gamma^{-1}$. $\Gamma$ is assumed to be large compared to the other relaxation scale thanks to its finiteness at $k = 0$, allowing us to truncate the expansion without significant loss of physical accuracy. The expansion is performed as follows:
\begin{equation}
    \begin{split}
\delta T(t) &= \int_{-\infty}^t \dd t'e^{\Gamma (t'-t) } \mathcal S(t')\\
            &= \left[\dfrac{e^{\Gamma (t'-t)} \mathcal S(t')}{\Gamma}\right]_{-\infty}^{t}-\dfrac{1}{\Gamma}\int_{-\infty}^t \dd t'e^{\Gamma (t'-t) } \mathcal S'(t')\\
            &= \sum_{n=0}^N \dfrac{(-1)^n}{\Gamma^{n+1}}\dfrac{\partial^{n} \mathcal S(t)}{\partial t^{n}}+\\
            &~\sum_{m=N+1}^\infty \dfrac{(-1)^m}{\Gamma^{m}}\int_{-\infty}^{t}dt'e^{\Gamma (t'-t) }\dfrac{\partial^{m} \mathcal S(t)}{\partial t^{m}}\\
            &= \dfrac{ \mathcal S(t)}{\Gamma} - \dfrac{\dot { \mathcal S}(t)}{\Gamma^2} + \mathcal{O}(\Gamma^{-3})
    \end{split}
    \label{eq: copy past brito}
\end{equation}

In what follow, we will retain only the first two terms (up to the first time derivative) because the remaining terms contribute at orders higher than $k^2$ in the hydrodynamic equations. By doing this, we transformed the memory dependence into a derivative dependence, effectively rendering our equation Markovian. Naturally, an infinite series of derivatives is non-local; only the truncation at a finite order is Markovian. It is important to note that the first term of the expansion corresponds to $\delta \dot{T} = 0$, indicating that the second term, proportional to $\dot{S}$, introduces a genuinely non-instantaneous contribution.

By taking the derivative of $ \mathcal S$ and inserting in it the hydrodynamic equations for $\delta \rho$ and $\delta u_{\parallel}$ we obtain:

\begin{equation}
    \dot { \mathcal S}(t)=-i(\rho_0 G_\rho+\gamma p_0/n_0) k \delta u_\parallel(t)+ \dot \Xi_T(t) + \mathcal{O}(k^2),
\end{equation}
where only terms up to first order in $k$ are kept because higher term would contribute to order $k^{n>2}$ in the hydrodynamic equations. The formal derivative of the white Gaussian noise has to be understood as a weak derivative in the sense of distributions. It has the following correlation:

\begin{equation}
    \langle \dot{\Xi}_T(k, t)\dot{\Xi}_T(k', t') \rangle=-\dfrac{2\langle \Xi_T(k, t)\Xi_T(k', t') \rangle}{(t-t')^2},
\end{equation}
or equivalently, in Fourier space:
\begin{equation}
    \langle \dot{\Xi}_T(k, w)\dot{\Xi}_T(k', w') \rangle=w^2\langle \Xi_T(k, w)\Xi_T(k', w') \rangle.
\end{equation}
Note that these relations hold only when $\Xi$ is a Gaussian white noise process.

Setting the noise to 0, and replacing all $\delta T$ in the first two rows of $\bm M$ by the expression found in Eq.~\eqref{eq: copy past brito}, we find $\bm M^{(r)}$:
\begin{widetext}
\begin{equation}
    \bm M^{(r)}(k) = \begin{pmatrix}
    0 & -i\rho_0  k\\
    -\dfrac{i k}{\rho_0}\left.\left(p_\rho - \dfrac{G_\rho p_T}{G_T}\right)\right|_{\rho_0, T_0} ~& ~-\gamma - \left.\left(\dfrac{\eta_0^\parallel }{\rho_0}-\dfrac{(G_T - \gamma)(p_0 + \nu_0)p_T}{G_T^2\rho_0 n_0}-\dfrac{G_\rho p_T}{G_T^2}\right)\right|_{\rho_0, T_0}  k^2 
    \end{pmatrix}.
    \label{eq: hydro matrix reduced}
    \end{equation}
\end{widetext}

This reduced hydrodynamics model includes correction to the simple model introduced by Lei and Ni in Ref.~\onlinecite{lei2019hydrodynamics} to describe a hyperuniform fluid. The first stability criterion is again apparent here in the velocity-density coupling. Similar systems, with an effective $\Delta$, were found to be unstable due to the positivity of this term \cite{risso2018effective, argentina2002van}. 
At equilibrium, spinodal instabilities related to phase transitions are also captured by the term $\bm M_{u^{\parallel}\rho }$ in a hydrodynamic theory, since the compressibility $1/(\rho p_\rho)$ becomes negative. However, such instability analysis on the hydrodynamic matrix cannot capture the binodals, metastability and coexistence densities, as these require thermodynamic information or additional constraints \cite{chiu2024theorynonequilibriummulticomponentcoexistence, solon2015pressure, solon2018generalized,omar2023mechanical, feng2024theoryanomalousphasebehavior}.

We now turn to correlation computation. Due to the peculiar correlation of the derivative of the white noise, it is not very practical to use the relations used in the main text to find $\bm S(\bm k, w)$ and $\bm S(\bm k)$ (Eqs.~\eqref{eq: dynamic corr} and \eqref{eq: static corr}). Moreover, the adiabatic expansion in all its glory starts to become heavy to carry since second derivative of noises arise. In order to easily compute $\bm S(\bm k, w)$ and $\bm S(\bm k)$, we chose to ignore the time derivative contribution $\dot S(t)$ in the expansion of $\delta T(t)$ which amounts to simply setting $\delta \dot{T} = 0$. This is inconsistent but leads to good results. For simplicity and for illustrative purposes, we will work in the case $\gamma=0$ which implies $G_\rho=0$. This allows us to nicely predict the observed shape of the structure factor for the system ``$\Delta$ alone'' in Fig.~\ref{fig:hydrodynamic}. 
\begin{figure}[t]
\centering
\includegraphics[width=0.99\columnwidth,clip=true]{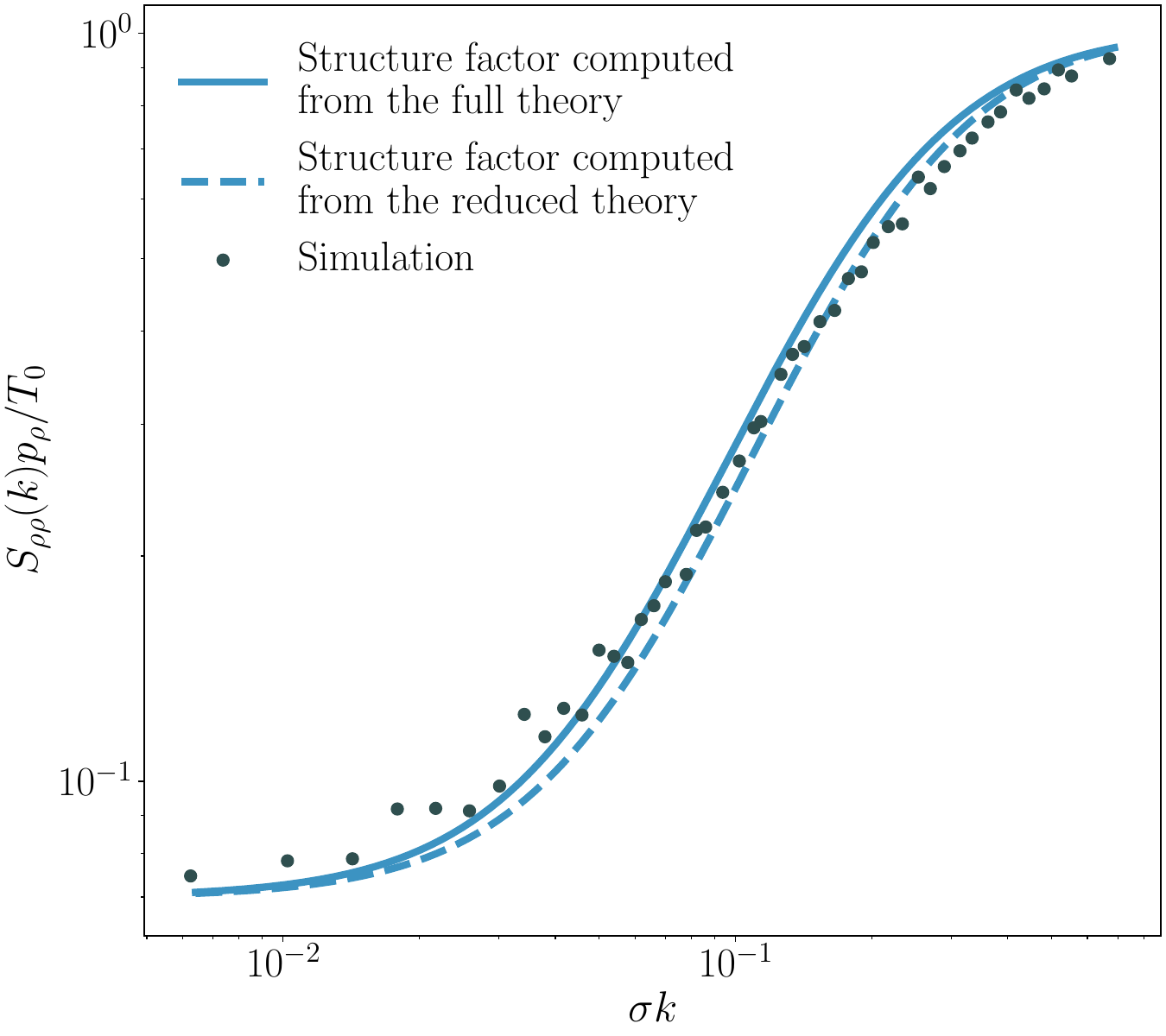}
\caption{Comparison between the structure factor found in the simulations of the simple $\Delta$ model without damping, the full theory (obtained from the full hydrodynamic matrix) and the adiabatic slaving (Eq.~\eqref{eq: adiabatic slaving structure factor}). $\alpha = 0.9$, $\phi = 0.5$ and $N = 2.5\times 10^6$.}\label{fig: adiabatic}
\end{figure}
To continue, we take the derivative of $\delta \dot u_{\parallel}$:

\begin{equation}
    \delta \ddot{u}_{\parallel}(t)=-p_\rho k^2\delta u_\parallel(t)- \frac{\eta_0^\parallel}{\rho_0}k^2\delta \dot u_\parallel(t)- ik\frac{p_T}{\rho_0} \frac{S(t)}{\Gamma} + \dot{\Xi}_{u_\parallel}(t).
\end{equation}

Taking the Fourier transform of this equation and computing $\bm S_{{u_\parallel}{u_\parallel}}(k, w)=\langle {u_\parallel}(k, w){u_\parallel}(-k, -w)\rangle$ leads to:
\begin{equation}
    \bm S_{{u_\parallel}{u_\parallel}}(k, w)=\dfrac{\mu_{u_{\parallel}} + \mu_{T}}{(k^2p_\rho-w^2)^2+w^2k^4\left(\frac{p_0 p_T}{\Gamma\rho_0n_0} - \frac{\eta^\parallel_0}{\rho_0}\right)^2},
    \label{eq: dynamic correlation slaving}
\end{equation}
with $\mu_{u_{\parallel}}$ and $\mu_{T}$ related to the variance of the time derivative of the velocity and temperature field, respectively:

\begin{equation}
    \begin{split}
    \mu_{u_{\parallel}}  &= 2k^2w^2T\eta_0^{\parallel}/\rho^2,\\
    \mu_{T}              &= 2k^4p_T^2T^2w^2\kappa/(\Gamma^2\rho^4).
    \end{split}
\end{equation}

We can integrate between $-\infty$ and $\infty$ and divide by $2\pi$ Eq.~\eqref{eq: dynamic correlation slaving} to find the static correlation function. This gives the simple result:

\begin{equation}
    \begin{split}
    \bm S_{u^\parallel u^\parallel}(k)\dfrac{\rho_0}{T_0} &=\bm  S_{\rho\rho}(k)\dfrac{p_\rho}{\rho_0T_0}\\ 
    &= \dfrac{\kappa_0 k^2}{\kappa_0 k^2+G_T\rho_0}+\dfrac{G_T\eta_0^\parallel\rho_0}{p_T^2T_0+\eta_0^\parallel(\kappa_0 k^2 + G_T\rho_0)},
    \end{split}
\end{equation}
where the density structure factor was found using the density field hydrodynamic equation. 

As with $\gamma\neq 0$, at large $k$ we recover the equilibrium result. Nonetheless, when $G_T\neq 0$, we see that the system has a transient hyperuniform scaling (first term) until reaching a constant value given by:

\begin{equation}
    \bm S_{\rho\rho}(0)/\rho_0 = \dfrac{T_0}{P_\rho}\left(1-\dfrac{1}{1+\dfrac{G_T \eta_0^\parallel \rho}{p_T^2 T_0}}\right).
    \label{eq: adiabatic slaving structure factor}
\end{equation}

Interestingly, the equilibrium limit $ G_T \to 0 $, which occurs, for example, when $\alpha \to 1$, is singular. Indeed,  as we approach this singular equilibrium limit $\alpha \to 1$, the value of $S(0)$ decreases to $0$, moving us further away from the true equilibrium result proportional to the temperature and the compressibility. However, as this limit is approached, increasingly smaller values of $ k $ are required for the transient hyperuniform scaling to manifest. This implies that infinitely close to the equilibrium limit, an infinitely large system is required to observe the non-equilibrium behavior of the structure factor. One might argue that our theoretical predictions become unreliable near equilibrium because the parameter $ T$ does not vary rapidly. However, it is important to note that there will always exist a value of $ k $ for which the conserved fields $\rho$ and $\bm u$  evolve more slowly than the non-conserved temperature field. Therefore, the asymptotic behavior remains reliable and, indeed, exact.

A comparison between simulations and theory is given in Fig.~\ref{fig: adiabatic} for the structure factor. The reduced theory (the one developed in this section, with the adiabatic slaving) works very well. Note here that $\alpha = 0.9$ and $\phi = 0.5$. It is satisfying that the theory still works at such high densities. At lower $\alpha$ however, we would not expect such good agreement. Note as well that we chose a reasonably small $\alpha$ and high densities because they both facilitate the observation of the transient hyperuniform scaling.

\newpage

\end{document}